\documentclass[a4paper]{article}

\usepackage{amsmath,amsfonts,amssymb}
\usepackage{hyperref}
\usepackage{amsthm}
\usepackage{enumerate}
\usepackage{bbm}
\usepackage{graphicx}
\usepackage{lscape,epsf}
\usepackage[vcentermath]{youngtab}
\usepackage{rotating}
\usepackage{float}
\usepackage{subcaption}
\usepackage{color}

\def\ds{\displaystyle}
\def\bea{\begin{array}{c}}
\def\ea{\end{array}}
\def\be{\begin{equation}\bea\ds}
\def\ee{\ea\end{equation}}
\def\bee{\begin{equation}\begin{array}{rcl}\ds}
\def\eee{\end{array}\end{equation}}

\def\nb{{\bf n}}
\def\bb{{\bf b}}
\def\tb{{\bf t}}

\def\htau{\hat{\tau}}
\def\hk{\hat{\kappa}}

\def\eb[#1]{{\bf e}_{#1}}

\title{Chern-Simons-Higgs Model as a Theory of Protein Molecules}

\author{Dmitry Melnikov
	and Alyson B.~F.~Neves
	}

\begin{document}

\maketitle

\vspace{-5cm}
\hfill{ITEP-TH-12/19}
\vspace{5cm}

\begin{abstract}
In this paper we discuss a one-dimensional Abelian Higgs model with Chern-Simons interaction as an effective theory of one-dimensional curves embedded in three-dimensional space. We demonstrate how this effective model is compatible with the geometry of protein molecules. Using standard field theory techniques we analyze phenomenologically interesting static configurations of the model and discuss their stability. This simple model predicts some characteristic relations for the geometry of secondary structure motifs of proteins, and we show how this is consistent with the experimental data. After using the data to universally fix basic local geometric parameters, such as the curvature and torsion of the helical motifs, we are left with a single free parameter. We explain how this parameter controls the abundance and shape of the principal motifs (alpha helices, beta strands and loops connecting them).
\end{abstract}

\section{Introduction}

\begin{figure}[htb]
\begin{minipage}{0.45\linewidth}
\includegraphics[width=\linewidth]{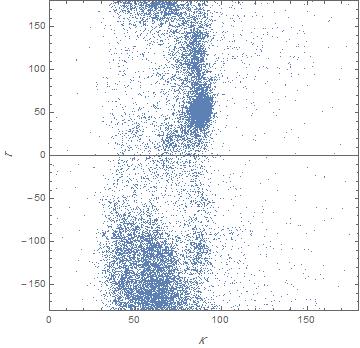}
\end{minipage}
\hfill{
\begin{minipage}{0.45\linewidth}
\includegraphics[width=\linewidth]{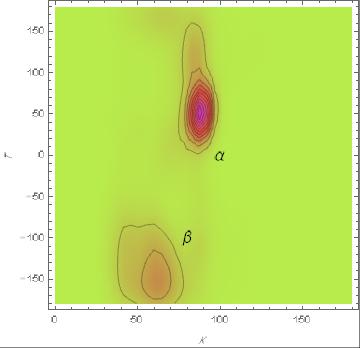}
\end{minipage}
}
\caption{Correlation plots of curvature and torsion angles calculated with respect to positions of $C_\alpha$ atoms in the backbones of a hundred selected proteins. These plots contain similar information to that of the Ramachandran plots.}
\label{fig:ramachandran}
\end{figure}

Proteins are very complex objects, but in the meantime there is an impressive underlying regularity beyond their structure. This regularity is captured in part by the famous Ramachandran plots that map the correlation of the dihedral (torsion) angles of consecutive bonds in the protein backbone (see for example~\cite{Ramachandran:1963,Ramachandran:1965,Lovell:2003}). The example on figure~\ref{fig:ramachandran} shows an analogous data of about hundred of proteins, illustrating the localization of the pairs of angles around two regions.\footnote{In figure~\ref{fig:ramachandran} we rather plot correlation of the curvature and torsion angles of the chain of backbone $C_\alpha$ atoms. This coordinate system was proposed in~\cite{Hu:2011wg} (see also~\cite{Hinsen:2013,Peng:2014,Melnikov:2019len}).}

These two most densely populated regions, labeled $\alpha$ and $\beta$, correspond to the most common secondary structure elements in proteins: (right) alpha helices and beta strands. To be more precise, less localized $\beta$ region also contains contribution of structural motifs (loops) connecting the helices. Proteins may also contain structures that are equivalent to left-handed helices, which are usually rare (they appear if the protein contains a large number of glycine amino acids). In the latter case one would see a localization of points at the position corresponding to the reflection of the locus of the alpha helices with respect to the $\kappa$ axis.

In this paper we would like to discuss the basic theory behind the Ramachandran plots. At the microscopic level the $\alpha$ region of the plots can be explained by the existence of hydrogen bonds beyond the next-to-nearest amino acids, so that the global energy minimum corresponds to a state with a non-zero curvature of the amino acid chain (helical configuration). The chirality of the helix is inferred from the chirality of the amino acids. Alpha helix configuration is favored by proteins containing certain amino acids (alanine, glutamate, lysine, leucine, methionine) and by more hydrophobic environments (cellular membranes).  However, the same hydrogen bonds can make other configurations possible, creating attractive forces between subchains and forming two-dimensional sheets, so that amino acids in the subchains assume the $\beta$ configurations in terms of the Ramachandran plots. Amino acids like thereonine, valine and isoleucine are more prone to participate in such configurations. Other amino acids, like proline and glycine, tend to break the most common structures. Although many details of the chemical structure and its effect on the geometry of the molecules is well understood, the very large number of participating degrees of freedom makes the prediction of the protein shape from a given chain of amino acids a formidable task. Here we would like to explore a different, yet technically simpler approach. We will write an effective model and test its compatibility with the basic features of the secondary structures of proteins.

In general, the idea of effective field theory approach is to focus on the universal long-distance features of the system, ignoring the many details of its microscopic structure. From the long-distance point of view, protein is a long one-dimensional continuous curve with certain self-interaction properties. One question is, what is the universal theory of such objects. A certain class of effective theories, similar to the elastic rod description of the DNA molecules (\emph{e.g.}~\cite{Klapper:1996}), was proposed in~\cite{Danielsson:2009qm,Chernodub:2010xz}. Standard approach to constructing an effective model consists of identifying (gauge) symmetries and effective degrees of freedom (mean fields) and writing all relevant interaction terms parameterized by phenomenologically determined interaction constants. We note that methods of the renormalization group of statistical field theory are implicitly present in the protein studies. Indeed, depending on the characteristic scale, at which we want to look at the protein, and its function, there exist various representations of the protein molecules. In particular, one common graphical representation of a protein, visualizing its secondary structures and motifs, images the corresponding chains of amino acids as continuous curves.

Continuous curves in three dimensions are parameterized by a pair of functions of a single parameter. It was suggested in~\cite{Danielsson:2009qm} to use curvature and torsion as two mean fields, describing the effective dynamics of curves. It was argued in that paper that there is a natural gauge symmetry associated with those fields. Therefore, the gauge invariance can be used as one of the guiding principles in the construction of the effective energy functionals (Hamiltonians) of the theory. The natural analog of such theory in high energy physics is the Abelian Higgs model, reduced to one dimension. The most general situation was discussed in~\cite{Hu:2013,Gordeli:2015aya}.

As in high energy physics it is natural to consider ground states of the Higgs model with constant fields, which spontaneously break the gauge symmetry. In terms of protein applications, the ground states are configurations of curves with constant curvature and torsion, \emph{i.e.} helices. There is a subtlety related to the way a non-trivial torsion appears in this model. Since it plays the role of a gauge field, in one dimension it enters through a version of the Chern-Simons gauge field, which is chiral, providing the necessary chirality to the helices. This otherwise trivial field becomes physical through the Higgs mechanism of spontaneous symmetry breaking. As an additional result of this breaking, the model possesses soliton-like solutions, which appear as loops breaking long helical pieces into finite domains~\cite{Chernodub:2010xz}. The discrete (lattice) models of solitons fitting protein structures in the Protein Data Bank~\cite{pdb}~(PDB) have been extensively studied in papers~\cite{Molkenthin:2011,Hu:2011,Krokhotin:2012,Niemi:2014,Molochkov:2017jmv}. The conclusion of those papers is that using a relatively small number of parameters describing soliton solutions it is possible to achieve a good, potentially sub-angstrom accuracy in the description of the configurations of a vast majority of structures appearing in the PDB. The large reduction of the number of parameters describing a protein, as compared with all atom or even coarse grained approach, give a big advantage in the simulation of the protein dynamics and design of new proteins.

Here we would like to make a step back and have a more general discussion of the continuous Abelian-Chern-Simons-Higgs model from the point of view of its consistency with protein physics. The model, as we will define it, will depend on only four, minimally required, phenomenological parameters. We would like to see how far this minimal model can be pushed to reproduce the variety of features observed in real proteins. 

We will find that this model can possess two types of vacua, which can be compared with alpha helices and beta strands. The model will predict a certain relation between the curvature and torsion of the helical structures, which we show to be consistent with the protein data. Depending on the structure of the vacua we will show that the model can have solitons that interpolate between two helices, two strands, or between helices and strands. In the continuous model the second type of solitons is unstable, but it might stabilize when the continuous model is replaced by its discrete version. When the relation between the curvature and the torsion is fixed, as well as the overall curvature scale of the helices, the model effectively remains with a single parameter. We show that this parameter is connected to the propensity of the curve to form beta strand-like configurations. We also estimate the stability of such configurations in our model. 

We note that many elements of the discussion below can be found to be either explicitly or implicitly present in the numerical simulations of the discrete version of the model. However, the continuous case shows them differently, by stressing the importance of the first and highlighting the existence of the second. For example, the relation between the curvature and torsion of helices also exists in the discrete case, but in the continuous case its role is more fundamental as we shall claim below.

This paper is mostly based on techniques of classical and quantum field theory and contains some details of the calculations. We summarized the main conclusions obtained in our study in a companion paper~\cite{Melnikov:2019}. The shorter version is intended to be less technical and suitable for a broader range of specialists.

The remainder of this paper is organized as follows. In section~\ref{sec:model} we present the one-dimensional Abelian-Chern-Simons Higgs (ACSH) model. We introduce a subtle topological charge associated with this model, which implies that it can be treated in either the canonical or grand canonical ensemble. In section~\ref{sec:GC} we discuss the static configurations in the grand canonical ensemble. We discuss ground states corresponding to helices and solitons corresponding to loops connecting them. In section~\ref{sec:sols} we pass to the canonical ensemble. In the canonical ensemble the model can have two vacua, with zero and non-zero curvature: one global and one local. Apart from stable kink-like solitons, the canonical picture can have unstable sphaleron-like solutions, describing phase transitions from the local to the global vacuum. In section~\ref{sec:stability} we analyze the stability of the local vacua against classical thermal fluctuations and against quantum tunneling processes. In section~\ref{sec:proteins} we test the model against the real protein data. First, we check the geometry of typical PDB helices and test them against the predicted torsion-curvature relation. This allows us to fix three of the four phenomenological parameters of the model. We show that our model disfavors the situation, in which the non-zero curvature minimum is local. The most interesting case is when this minimum is global, while the minimum at zero curvature is local, but nearly degenerate with the global minimum. We estimate the stability of the zero-curvature minimum and find that it depends on the remaining parameter of the model in the desired way: the minimum is less stable when the solitons are short and more stable when they are long. In section~\ref{sec:conclusions} we summarize our findings and discuss their interpretation in terms of the protein geometry.

\section{The model}
\label{sec:model}

In three dimensions curves are parameterized by a pair of functions of a single parameter. In the description of Frenet~\cite{Frenet:1852}, the functions are curvature $\kappa$ and torsion $\tau$. We will consider them as functions of the arclength parameter $s$, which measures the length of the curve. Functions $\kappa(s)$ and $\tau(s)$ are parameters of an evolution matrix of the frame -- the right triplet of vectors ${\bf e}^a(s)=(\nb(s),\bb(s),\tb(s))$ (normal, binormal and tangent). From the point of view of the curve itself, the frame description is redundant, since all the same information is already contained in the tangent vector $\tb$.

The redundancy can be cast in the form of a $U(1)$ gauge symmetry (see~\cite{Danielsson:2009qm,Hu:2011}), which reflects the invariance of the equations with respect to a rotation of the frame around the tangent vector $\tb$. Hence gauge transformations connect two identical curves with different local choices of framing. Function $\kappa(s)$ in the gauge theory presentation naturally generalizes to a complex quantity. Its phase reflects the relative angle of the local frame with respect to the original Frenet frame. Consequently, functions $\kappa(s)$ and $\tau(s)$ transform under the $U(1)$ rotations of the frame:
\begin{eqnarray}
  \kappa(s) & \to & \hk(s)={e}^{i\alpha(s)}\kappa(s)\,, \label{gaugeK} \\
  \tau(s) & \to & \htau(s)=\tau(s) + \alpha'(s)\,, \label{gaugeT}
\end{eqnarray}
what represents a typical action of gauge symmetry transformations on an Abelian Higgs multiplet (see \emph{e.g.}~\cite{Niemi:2014} for additional details). Having identified the effective degrees of freedom and symmetries one can proceed to constructing an effective gauge theory functional .

In principle, energy functionals for one-dimensional objects are widely used in the theory of elastic rods~\cite{TheoryOfElasticRods}. This theory has related biological applications, notably in the study of the DNA conformations~\cite{Fuller:1978,Tsuru:1986,Benham:1989,Schlick:1992,Klapper:1996,Vologodskii:1997}. However, the classification of general functionals based on gauge symmetry and effective theory approach, to our knowledge, is a relatively new and independent development of this theory~\cite{Danielsson:2009qm,Chernodub:2010xz}. In this work we would like to focus on a particular minimal choice, which can be called Abelian-Chern-Simons-Higgs  model. It is introduced by the following energy functional for a one-dimensional curve in three dimensions:
\be
\label{AHfunctional}
E \ = \ \int\limits_0^L ds\ \frac{1}{2}\left(|\nabla\hk|^2 - m^2|\hk|^2 + \lambda|\hk|^4 \right) - F\int\limits_0^L ds\ \htau\,,
\ee
where the dependence of the ``gauge covariant'' curvature $\hk$ and torsion $\htau$ on arclength parameter $s$ is implied. $L$ is the curve's length and $\nabla$ is the one-dimensional covariant derivative defined with respect to the gauge field $\htau$,
\be
\nabla \ = \ \frac{d}{ds} - i\htau\,.
\ee
The model depends on three parameters $m^2>0$, $\lambda>0$ and $F$. We will assume in this functional that $\hk$ is dimensionless and $\htau$ has dimension of energy in natural units.

The last term in energy functional~(\ref{AHfunctional}) is a one-dimensional version of the Chern-Simons (CS) action. This term is very important. The parameter $F$, in particular, sets a scale for the torsion of the curve. $F$ can be either positive, or negative, corresponding to right and left chirality (helicity) respectively. It is straightforward to generalize the CS term to a more conventional three-dimensional CS term, which can be thought as mediating statistical interactions between curves~\cite{Gordeli:2015aya}.

Functional $E$ is invariant under gauge transformations~(\ref{gaugeK}) and~(\ref{gaugeT}) everywhere, except the endpoints of the curve, $s=0$ and $s=L$. A generic transformation at the endpoints induces a shift of the functional,
\be
E \ \to \ E - F(\alpha(L)-\alpha(0))\,.
\ee
In CS theory, presence of a boundary restricts gauge transformations from arbitrary local functions to functions satisfying certain boundary conditions. Here we are only allowed to use those gauge transformations, satisfying
\be
\label{gaugecondition}
\Delta\alpha\ = \ \alpha(L) - \alpha(0) \ = \ 0\,.
\ee
Note that phase angle $\alpha$ is defined modulo $2\pi$. Naively, this can be thought as of a trivial relative rotation of the frame at one of the endpoints. However, condition~(\ref{gaugecondition}) is violated by such a rotation, and two configurations different by a relative $2\pi$ rotation of the frames at the endpoints correspond to two physically distinct choices. Respectively, gauge transformations can be split in two classes: small transformations, which keep $\Delta\alpha=0$ -- these are true gauge transformations, and large transformations. The latter change $\Delta\alpha\to\Delta\alpha+2\pi n$ for some integer $n$.

The large gauge transformations are singular. If one considers the case of a fixed relative orientation of the frames at the endpoints, there is clearly no regular deformation of the orientation of the frame along the curve that would transform one choice of the relative orientation to a choice different by a full $2\pi$ rotation. The associated topological invariant of smooth deformations is the self-linking number of the curve. In the physical model, the rotations of the endpoints change the energy functional. One can think of the boundary degrees of freedom, that define different boundary conditions, that is different mutual orientations of the endpoint frames, as of non-local ``entangled'' edge modes, by analogy with three-dimensional applications of CS theory. To better understand the role of the edge modes (boundary conditions) we analyze the equations of motion obtained from the minimization of the energy functional.

Minimization of~(\ref{AHfunctional}) gives two independent equations, which can be cast in the form of equations for $|\hk|$ and $\eta={\rm Arg}\,\hk$:
\begin{eqnarray}
|\hk|'' + m^2|\hk| - 2\lambda |\hk|^3 - (\htau-\eta')^2|\hk| & = & 0\,,\\
|\hk|^2(\htau-\eta') & = & F\,.
\end{eqnarray}
Since the combination $\htau-\eta'$ is gauge invariant (in the high energy physics jargon the gauge field eats the phase of $\kappa$), we can rewrite the equations in explicitly gauge invariant form replacing $|\hk|\to \kappa$ and $\htau-\eta'\to\tau$. Physical, gauge independent $\kappa$ and $\tau$ coincide with those of Frenet, which can be defined as a particular choice of a gauge, with real $\hk$.

We notice that even though $\tau$, after eating $\eta$, has become a non-pure-gauge degree of freedom, it is still non-dynamical,
\be
\label{torsion}
\tau \ = \ \frac{F}{\kappa^2}\,,
\ee
and can be integrated out from the equations leaving
\be
\label{eom}
\kappa'' + m^2\kappa - 2\lambda\kappa^3 - \frac{F^2}{\kappa^3} \ = \ 0\,,
\ee
Therefore, we arrive at an effective energy functional in terms of $\kappa$ only,
\be
\label{EffAction}
E \ = \ \int ds\ \frac{1}{2}\left({\kappa'}^2 - m^2\kappa^2 + \lambda\kappa^4 - \frac{F^2}{\kappa^2}\right) - F\Delta\eta\,.
\ee
In the original theory any two values of $\hk$ related by a phase rotation~(\ref{gaugeK}) are equivalent. What matters physically is the modulus $\kappa=|\hk|$. Hence, in general, we will think of $\kappa$ as a positive quantity, while $\tau$ can be both positive and negative. After $\htau$ eats the phase of $\hk$ (Higgs mechanism), there is still a residual $\mathbb{Z}_2$ symmetry in the system $\kappa\to -\kappa$, which can be broken spontaneously. In terms of curves, it means that there is no natural local definition of the sign of curvature, but as long as such choice is made at one point of a smooth curve, it becomes uniquely defined everywhere.

The role of the CS term in effective functional~(\ref{EffAction}) is twofold. First, it produces a singular ``self-energy" potential term. Second, there is a piece, which came from a ``physical" part of the non-physical phase $\Delta\eta$. The gauge invariant energy functional depends on the mutual orientation of the frames at the endpoints. We remind that $\eta$ is defined as a relative angle with respect to the Frenet frame at the given point.

It is natural to associate $\Delta\eta$ with a (topological) charge of the curve. Consequently, the parameter $F$ is analogous to the chemical potential in the grand canonical ensemble. As usual, one can choose different boundary conditions. The trivial choice would be to fix $\Delta\eta$, that is to work in the canonical ensemble. Alternatively, the chemical potential $F$ can be fixed. In the second case the equilibrium topological charge is found by minimizing the energy with respect to $F$:
\be
\label{topcharge}
\Delta\eta \ = \ -\int\limits_0^L ds\,\frac{F}{\kappa^2} \ = \ -\int\limits_0^L ds\ \tau \,.
\ee

In topology of one-dimensional curves the integral of torsion (up to a normalization) is called \emph{twist number} $Tw$. In general, it is a non-integral number and it is not a topological invariant. A theorem of Calugareanu (Calugareanu-White-Fuller~\cite{Calugareanu,White,Fuller}) relates $Tw$ with the self-linking number of the curve: $Lk=Tw+Wr$, where $Wr$ is another quantity called \emph{writhe}, which computes the number of self-intersections of the curve seen in a selected projection on a 2D plane, averaged over all choices of the projection plane~\cite{Dennis}. In~\cite{Melnikov:2019len} twist, writhe and self-linking number were discussed in the context of proteins. It was shown, in particular, how a discrete analog of the Calugareanu's theorem can be used count the structural motifs in the molecules.

We see that the chemical potential, introduced by $F$, controls the chirality of the molecule. Apart from chemical composition, the chemical potential can be introduced by external agents, such as properties of the medium (\emph{e.g.} acidity). Solvents like water compete for the hydrogen bonds, making proteins less prone to form helices. It would also be interesting to further understand the physical manifestation of the ``edge modes'', or the topological charge $\Delta\eta$.

Substituting equilibrium value in~(\ref{EffAction}) one finds an energy functional, which is bounded from below
\be
\label{EGC}
E \ = \ \int ds\ \frac{1}{2}\left({\kappa'}^2 - m^2\kappa^2 + \lambda\kappa^4 + \frac{F^2}{\kappa^2}\right) \,.
\ee

In summary, we derived an effective theory~(\ref{EffAction}) with equations of motion~(\ref{eom}), which should describe static (equilibrium) configurations of curves. Interaction with gauge field $\htau$ leads to a divergent self-energy-like term in the Hamiltonian. The effective theory should be supplemented by a choice of appropriate boundary conditions for the additional parameter $\Delta\eta$ associated with edge modes (edge frames). Choosing the canonical, or grand canonical ensembles, yields opposite signs of the singular term in the potential.

\section{Field configurations in the grand canonical ensemble}
\label{sec:GC}

\begin{figure}[htb]
  \begin{minipage}{0.45\linewidth}
   \includegraphics[width=\linewidth]{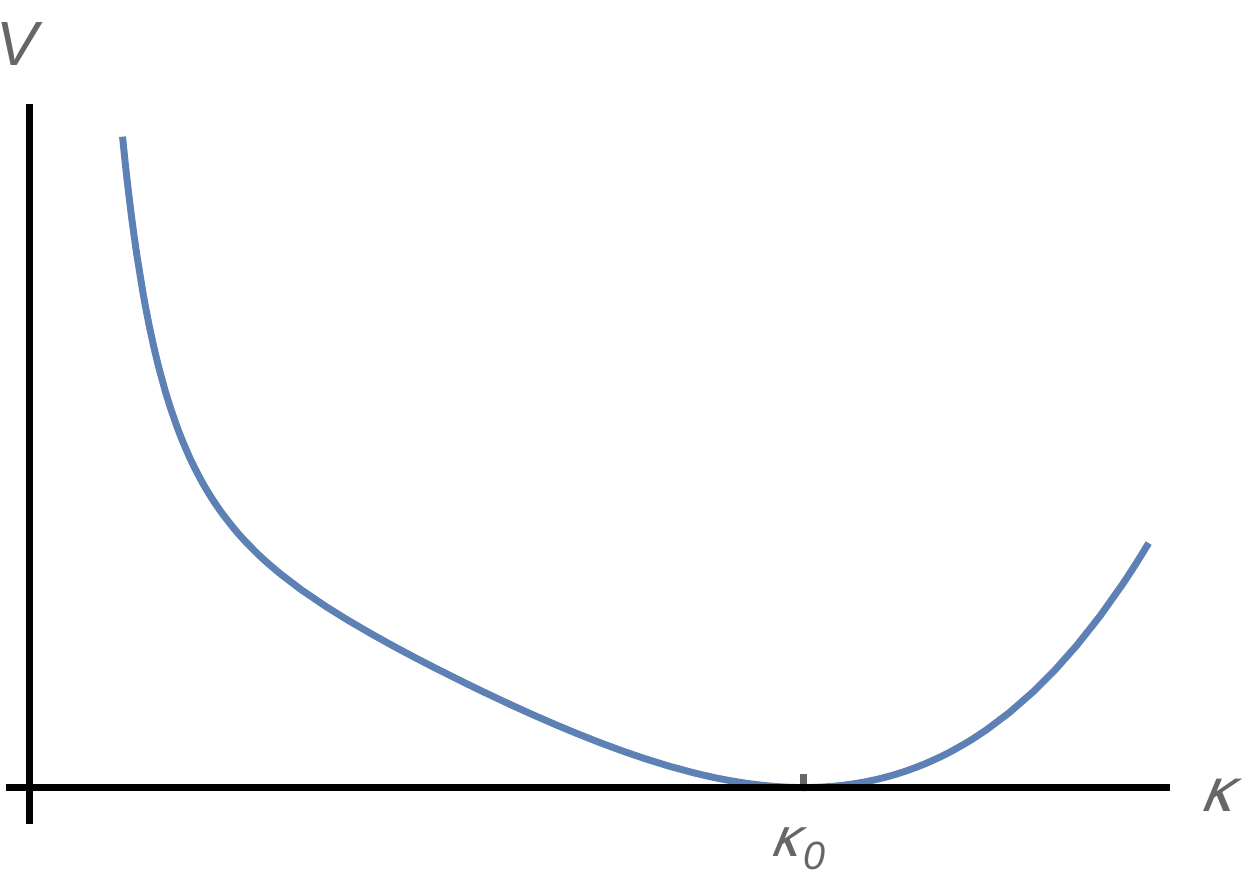}
  \end{minipage}
\hfill{
\begin{minipage}{0.45\linewidth}
   \includegraphics[width=\linewidth]{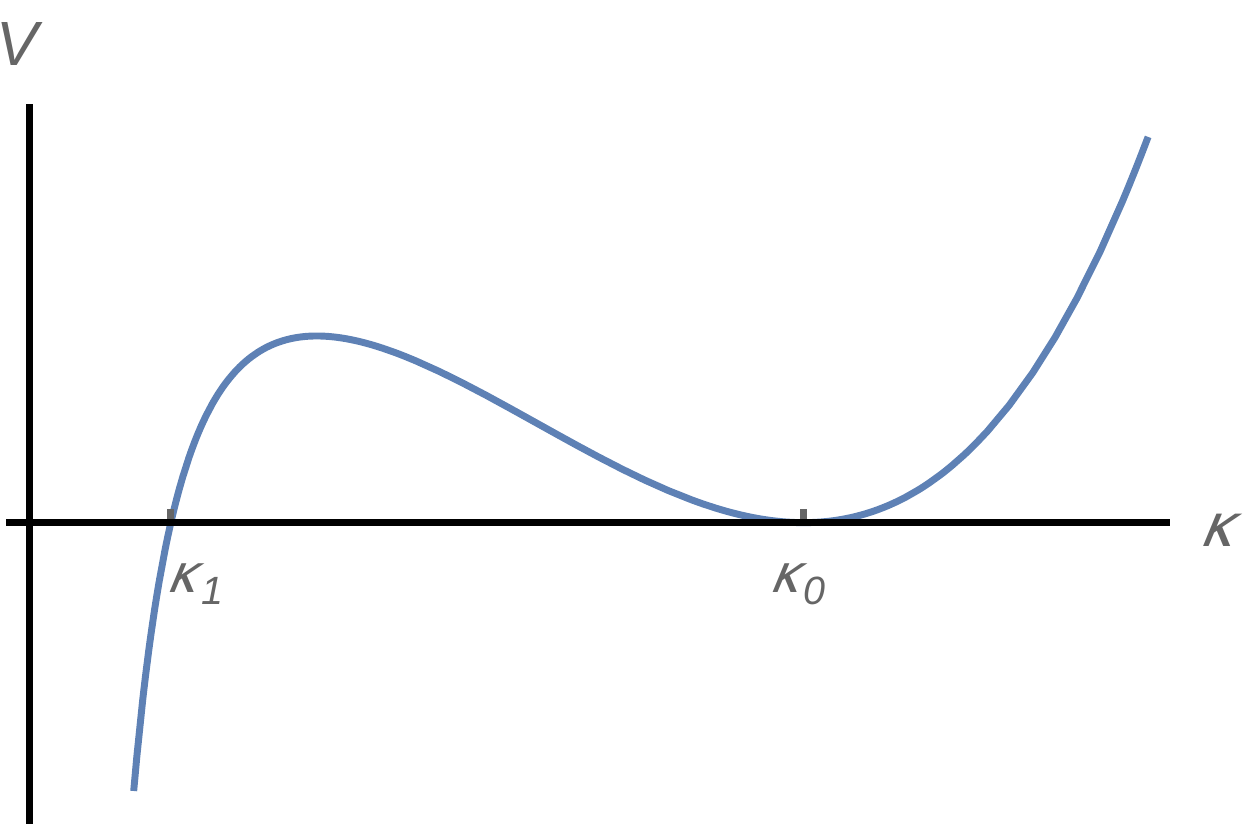}
  \end{minipage}
}
  \caption{(\emph{Left}) Potential energy in the grand canonical ensemble theory~(\ref{EGC}). (\emph{Right}) Potential energy in the canonical ensemble~(\ref{EffAction}) with phenomenologically interesting choices of parameters.}
  \label{fig:potential}
\end{figure}

We start our analysis of the equilibrium configurations of theory~(\ref{EffAction}) from the grand canonical picture~(\ref{EGC}). The potential energy term, \emph{i.e.} the part of~(\ref{EGC}) that does not contain derivatives, in this case is shown on figure~\ref{fig:potential}~(left). 

It is convenient to introduce the following parameterization of the potential, which would also apply to the study of the canonical ensemble:
\be
\label{potential}
V(\kappa) \ = \ \frac{\lambda(\kappa^2-\kappa_0^2)^2(\kappa^2-\kappa_1^2)}{2\kappa^2}\,.
\ee
In the grand canonical case we assume $\kappa_1^2<0$, so that the potential has a single minimum at $\kappa_0>0$, as in figure~\ref{fig:potential} (left). We also remind that we always consider $\lambda>0$ and $m^2>0$. In general, there is no nice formula expressing $\kappa_0$ and $\kappa_1$ as a function of $F$ and $m^2$, since the two sets of parameters are essentially related through a solution of a cubic equation:
\be
m^2 \ = \ \lambda(\kappa_1^2+2\kappa_0^2)\,, \qquad F^2 \ =\ \mp{\lambda}\kappa_0^4\kappa_1^2\,,
\ee
where the sign in the definition of $F^2$ is negative for the grand canonical and positive for the canonical ensemble in order to be consistent with the signs in~(\ref{EffAction}) and~(\ref{EGC}).

\subsection{Ground state configuration}
\label{sec:helix}

Energy minimum at $\kappa=\kappa_0$ corresponds to a uniform static field configuration with constant curvature $\kappa_0$ and torsion
\be
\label{torsion0}
\tau_0 \ = \ \frac{F}{\kappa_0^2}\,.
\ee
A curve with constant curvature and torsion is a helix, which can be seen, for example, by solving the Frenet equations:
\begin{eqnarray}
x(s) & = & - \frac{\Lambda\kappa_0}{\tau_0^2+\Lambda^2\kappa_0^2}\, \cos\left(s\sqrt{\tau_0^2+\Lambda^2\kappa_0^2}\right)\,, \label{xhelix}\\
y(s) & = & - \frac{\Lambda\kappa_0}{\tau_0^2+\Lambda^2\kappa_0^2}\, \sin\left(s\sqrt{\tau_0^2+\Lambda^2\kappa_0^2}\right)\,, \label{yhelix}\\
z(s) & = & \frac{\tau_0 s}{\sqrt{\tau_0^2+\Lambda^2\kappa_0^2}} \,. \label{zhelix}
\end{eqnarray}
In the energy functional we preferred to keep $\kappa$ dimensionless, in order not to overload it with unnecessary dimensionful constants. The price to pay however, is the necessity to introduce it in the physical solutions, in which curvature must have the same dimension as $\tau$ (inverse distance, or energy in natural units). Hence in equations~(\ref{xhelix})-(\ref{zhelix}) we ``dressed" the dimensionless curvature with some scale $\Lambda$. For example, it is convenient to select $\Lambda$ to be a characteristic scale of the protein molecule, $\Lambda=1 {\rm \AA}^{-1}$. In the following we will mostly assume this value and not write $\Lambda$ explicitly. However, in section~\ref{sec:proteins} we will need to relate $\Lambda$ to an energy scale to make phenomenological predictions. Otherwise, $\Lambda$ should be kept in mind when the physical dimensions of different quantities are compared.

\begin{figure}[htb]
\centering
\begin{minipage}{0.45\linewidth}
 \includegraphics[width=\linewidth]{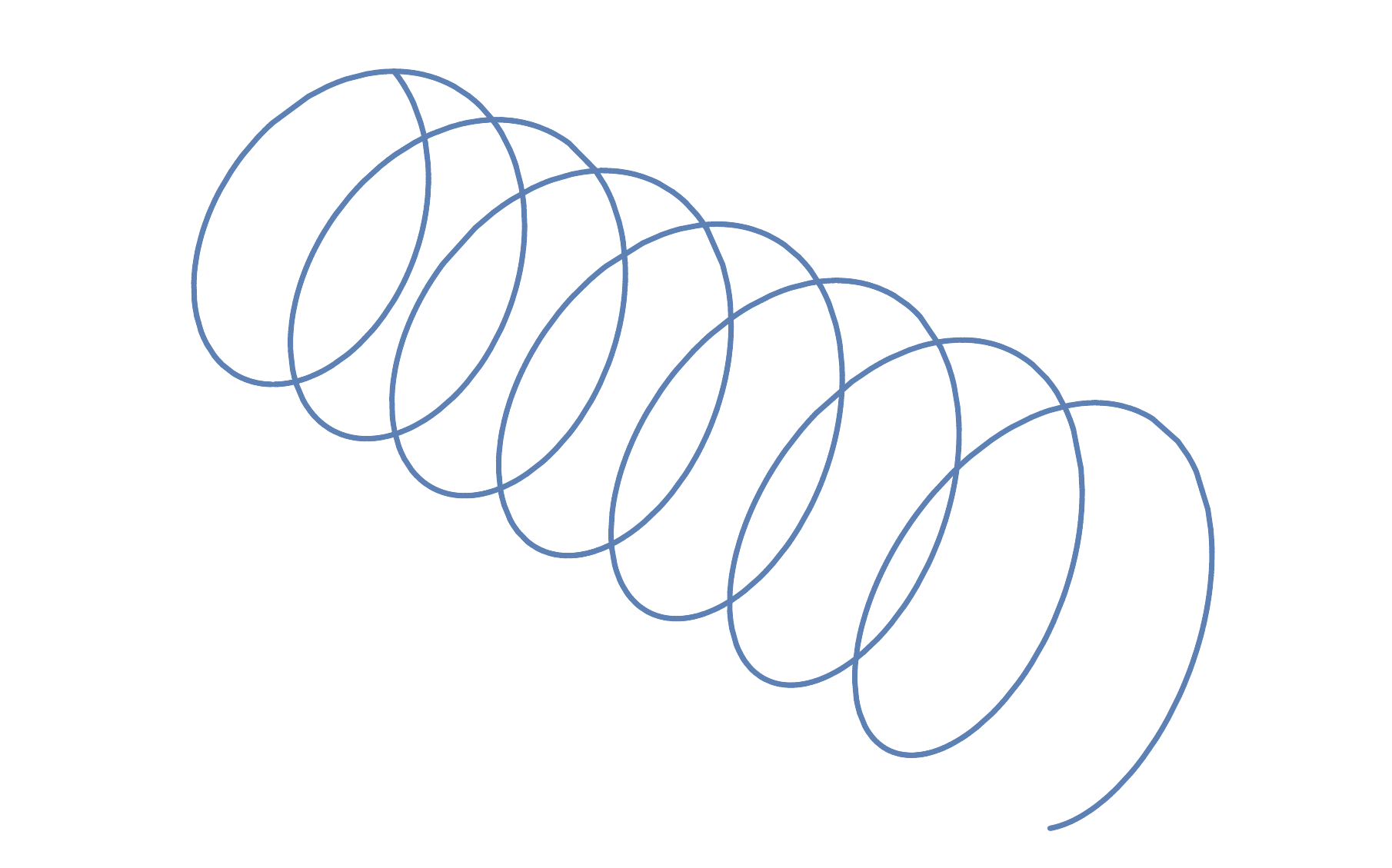}
\end{minipage}
\hfill{
\begin{minipage}{0.45\linewidth}
\includegraphics[width=\linewidth]{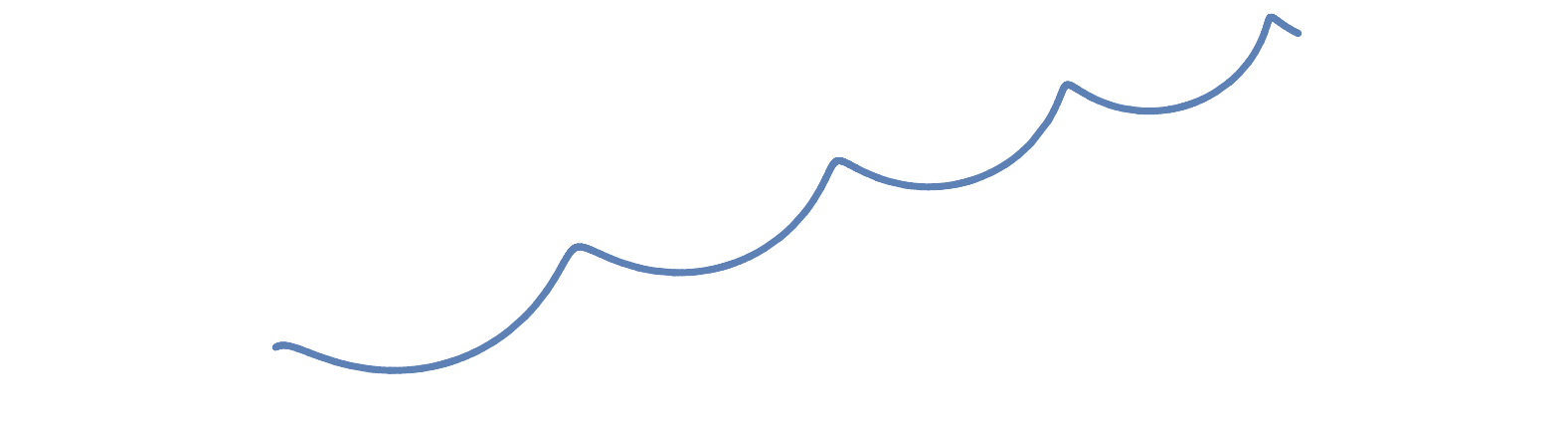}
\end{minipage}
}
\caption{Static minimum energy solutions of the model are helices. The ratio of the pitch to the radius distinguishes $\alpha$ (left) and $\beta$ (right) helices.}
\label{fig:helices}
\end{figure}

Solution~(\ref{xhelix})-(\ref{zhelix}) describes a helix, whose axis is parallel to the $z$-axis. It can be characterized by two parameters: radius $A$ and pitch $T$,
\be
\label{RadiusPitch}
A \ = \ \frac{\Lambda\kappa_0}{\tau_0^2+\Lambda^2\kappa_0^2}\,, \qquad T\ = \ \frac{2\pi\tau_0}{\tau_0^2+\Lambda^2\kappa_0^2}\,.
\ee
We remind that the pitch is the distance between the loops of the helix, \emph{i.e.} distance in the $z$-direction of the points with the same value of $x$ and $y$. Consequently, the dimensionless ratio
\be
\label{gamma}
\gamma \ = \ \frac{T}{A} \ = \ \frac{2\pi\tau_0}{\Lambda\kappa_0}\,,
\ee
shows how much the helix is ``stretched". On figure~\ref{fig:helices} we show two typical examples. We will call alpha-helix the solution with $\gamma\ll 1$. The opposite case $\gamma\gg 1$ will be referred to as beta-helix. It can also be seen from the solution, that positive (negative) $F$ corresponds to right (left) helices. When $F$ vanishes, the curve collapses to a circle.

We further note that the potential shown on figure~\ref{fig:potential} (left), singular at $\kappa=0$, does not allow other interesting static configurations. This happens in particular, that there is an infinite penalty to pay for a straight piece of the curve. In a physical approach to this problem one should say that the central singularity should be regularized. It is standard in the effective field theory approach that the theory is valid up to a certain cutoff scale. An efficient way to introduce a cutoff in the present model is to add a mass term for the field $\tau$ to the energy functional.
\be
E \ \to \ E + \int ds\ \frac{1}{2}\epsilon^2\tau^2\,.
\ee
This gauge invariant extension of the original theory is typically called a Proca mass term.

In this case relation~(\ref{torsion}) between the curvature and torsion gets modified,
\be
\label{regtorsion}
\tau \ = \ \frac{F}{\kappa^2+\epsilon^2}\,,
\ee
and the new effective functional has the form
\be
\label{Emod}
E \ = \ \int ds\ \frac{1}{2}\left({\kappa'}^2 - m^2\kappa^2 + \lambda\kappa^4 + \frac{F^2}{\kappa^2+\epsilon^2}\right)\,.
\ee
It is extremized on solutions to
\be
\label{eomReg}
\kappa'' + m^2\kappa - 2\lambda\kappa^3 + \frac{2F^2\kappa}{(\kappa^2+\epsilon^2)^2} \ = \ 0\,.
\ee

Again it is convenient to pass to a new set of parameters, so that the regularized potential is
\be
\label{regpot}
V(\kappa) \ = \ \frac{\lambda(\kappa^2-\kappa_0^2)^2(\kappa^2+\kappa_1^2)}{2(\kappa^2+\epsilon^2)}\,.
\ee
As compared to equation~(\ref{potential}) we changed the sign in front of $\kappa_1^2$ in order for $\kappa_1$ to be real. The new definitions of parameters $F$ and $m^2$ are
\be
\label{FmRelReg}
m^2 \ = \ \lambda(2\kappa_0^2-\kappa_1^2+\epsilon^2)\,, \qquad F^2 \ =\ {\lambda}(\kappa_0^2+\epsilon^2)^2(\kappa_1^2-\epsilon^2)\,.
\ee

\begin{figure}[htb]
\centering
\begin{minipage}{0.45\linewidth}
 \includegraphics[width=\linewidth]{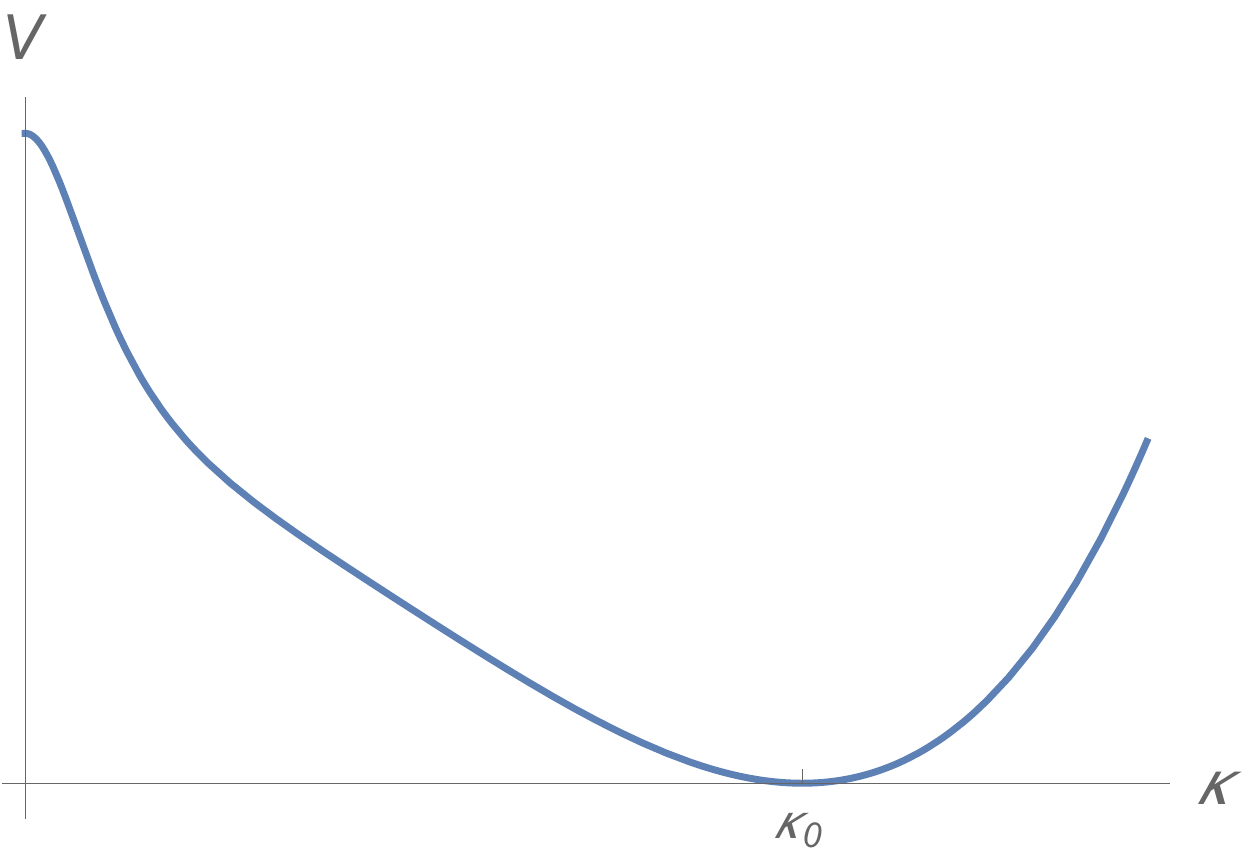}
\end{minipage}
\hfill{
\begin{minipage}{0.45\linewidth}
\includegraphics[width=\linewidth]{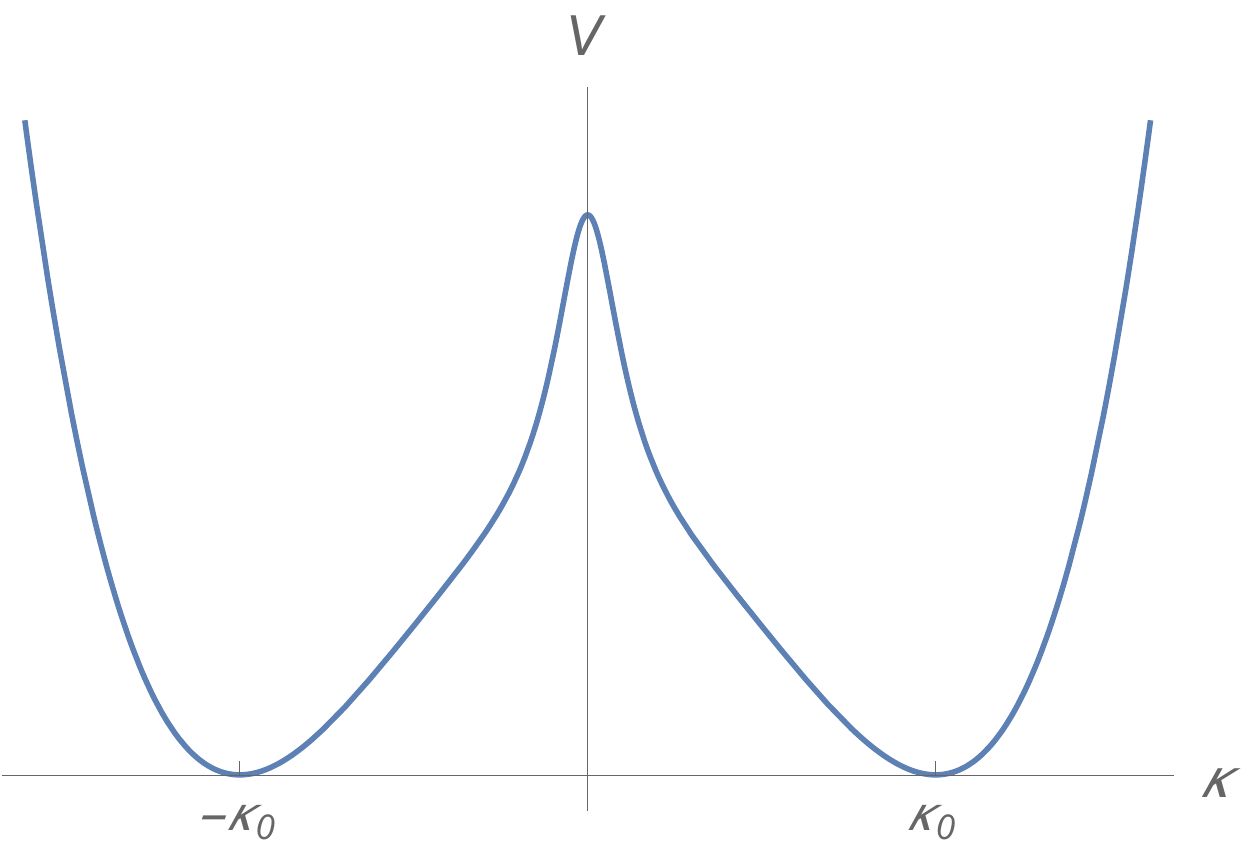}
\end{minipage}
}
\caption{(\emph{Left}) Regularized potential in the grand canonical ensemble for $\epsilon^2/\kappa_1^2<1$. (\emph{Right}) The same potential extended to negative values of $\kappa$.}
\label{fig:regpotGC}
\end{figure}

As one can notice from equation~(\ref{FmRelReg}), for $F^2$ to be positive, one has to assume $\kappa_1^2>\epsilon^2$. Otherwise, it will not be compatible with the grand canonical picture. This restricts the potential to have the form shown on the left of figure~\ref{fig:regpotGC}. We remind that we can always use the $\mathbb{Z}_2$ symmetry to extend the potential to negative values of $\kappa$, as shown on the right of the same figure.  If $\epsilon^2=\kappa_1^2$ (\ref{regpot}) becomes a standard double well potential.

\subsection{Solitons}

\begin{figure}[t]
\begin{minipage}{0.4\linewidth}
\centering
\includegraphics[width=0.8\linewidth,clip=true,trim=0pt 90pt 0pt 70pt]{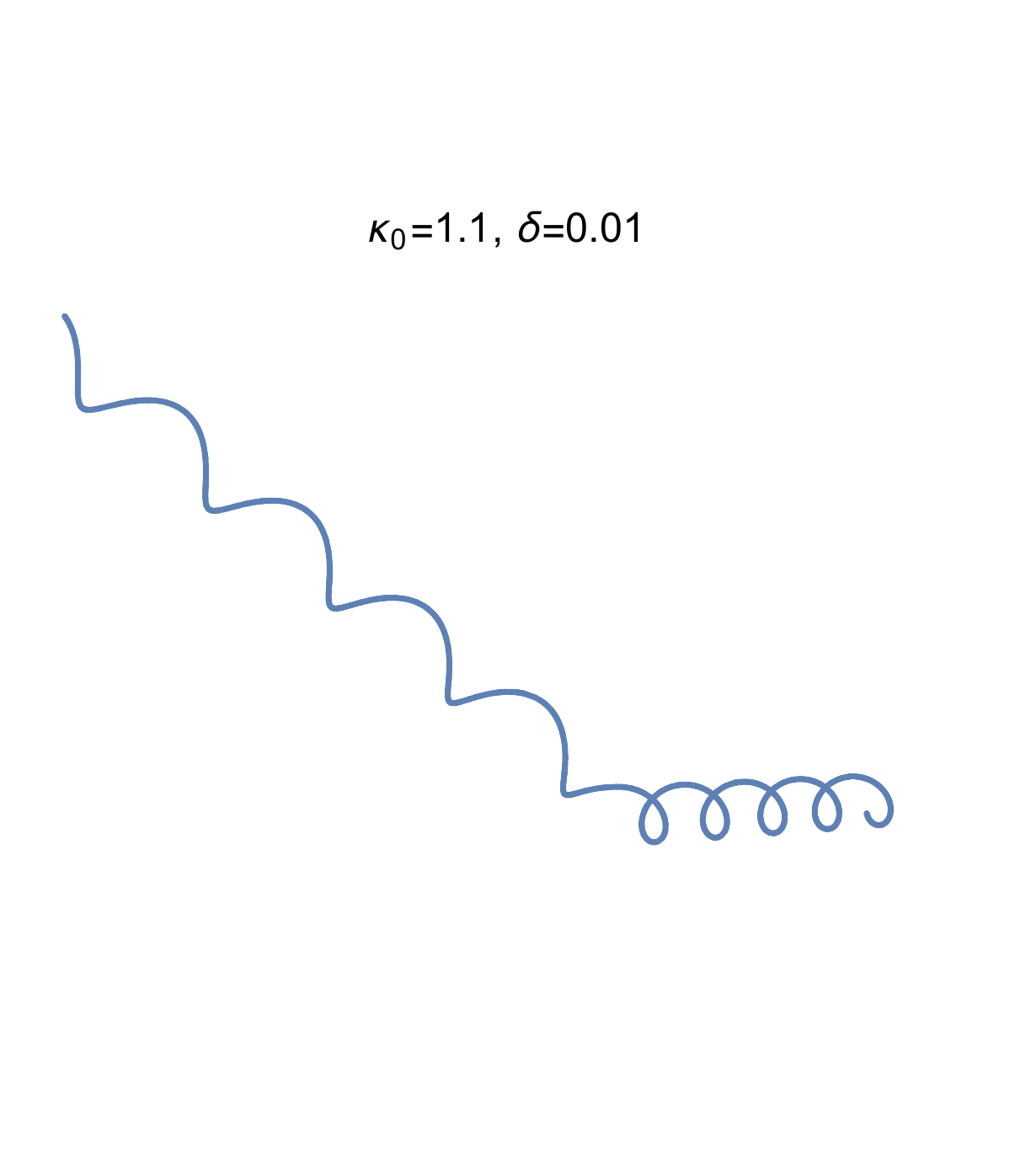}
\end{minipage}
\hfill{
\begin{minipage}{0.4\linewidth}
\centering
\includegraphics[width=0.9\linewidth,clip=true,trim=30pt 130pt 30pt 130pt]{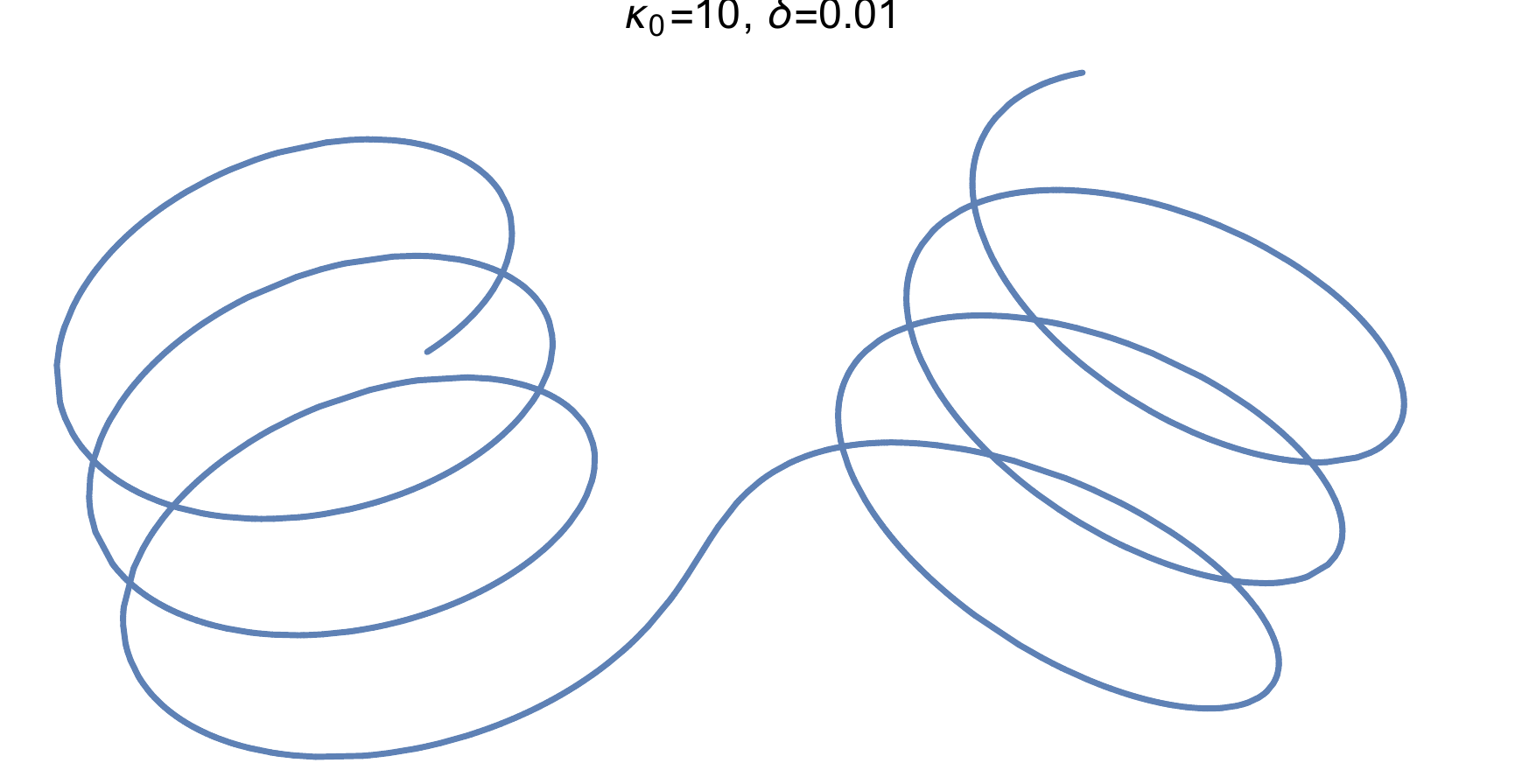}
\end{minipage}
}
\begin{minipage}{0.4\linewidth}
\centering
\includegraphics[width=0.9\linewidth,clip=true,trim=0pt 110pt 0pt 70pt]{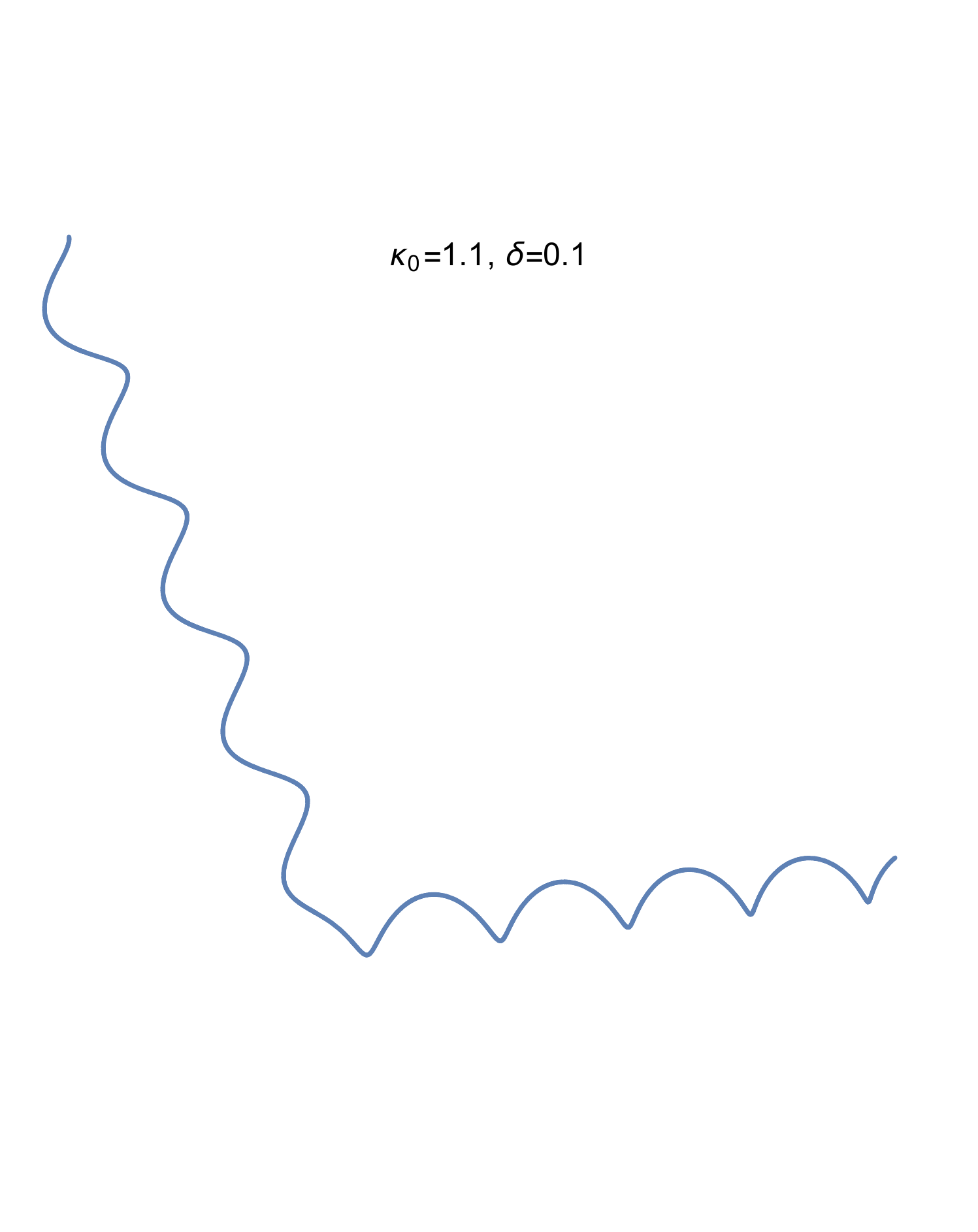}
\end{minipage}
\hfill{
\begin{minipage}{0.4\linewidth}
\centering
\includegraphics[width=0.9\linewidth,clip=true,trim=0pt 10pt 0pt 10pt]{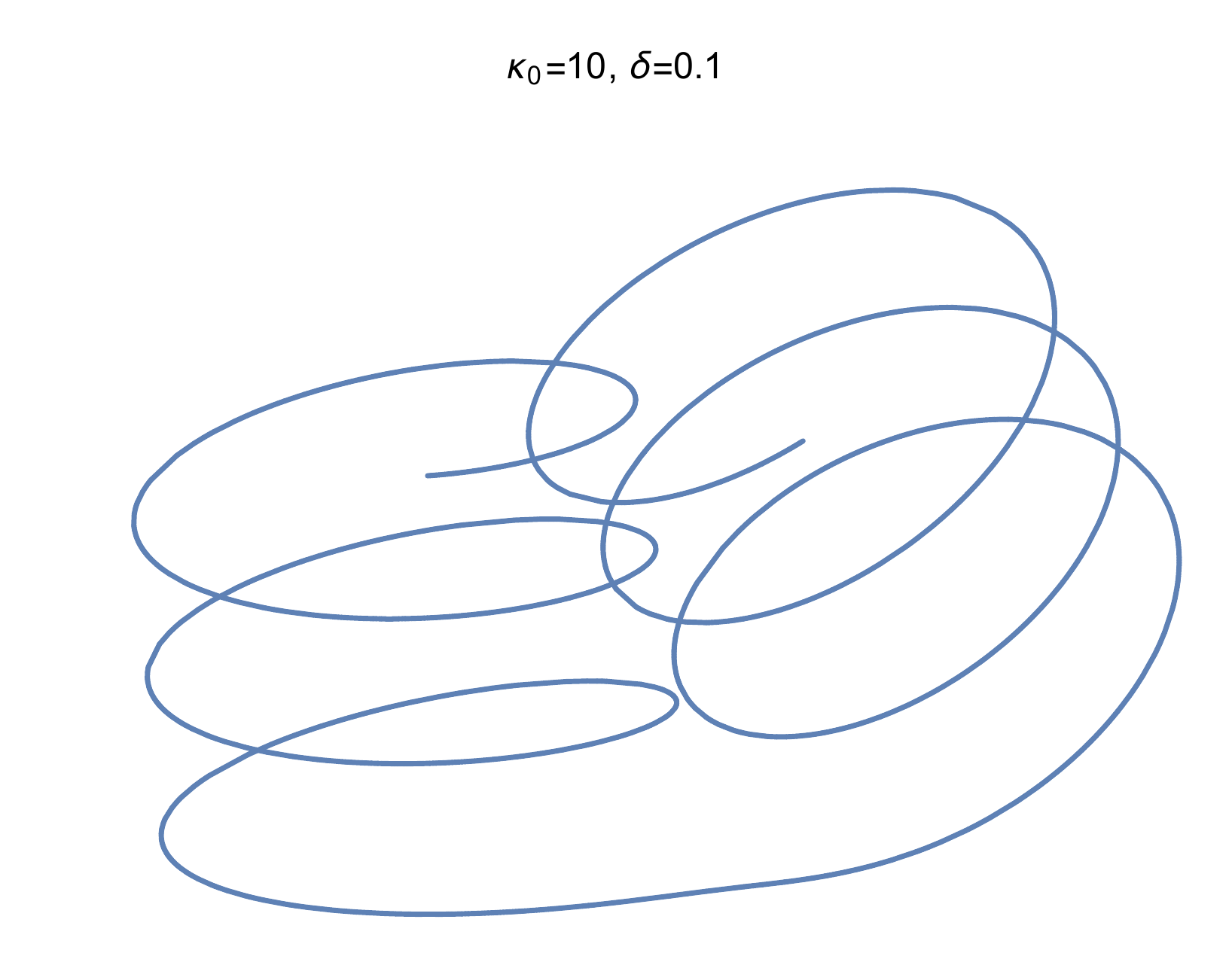}
\end{minipage}
}
\begin{minipage}{0.4\linewidth}
\centering
\includegraphics[width=0.9\linewidth,clip=true,trim=0pt 50pt 0pt 50pt]{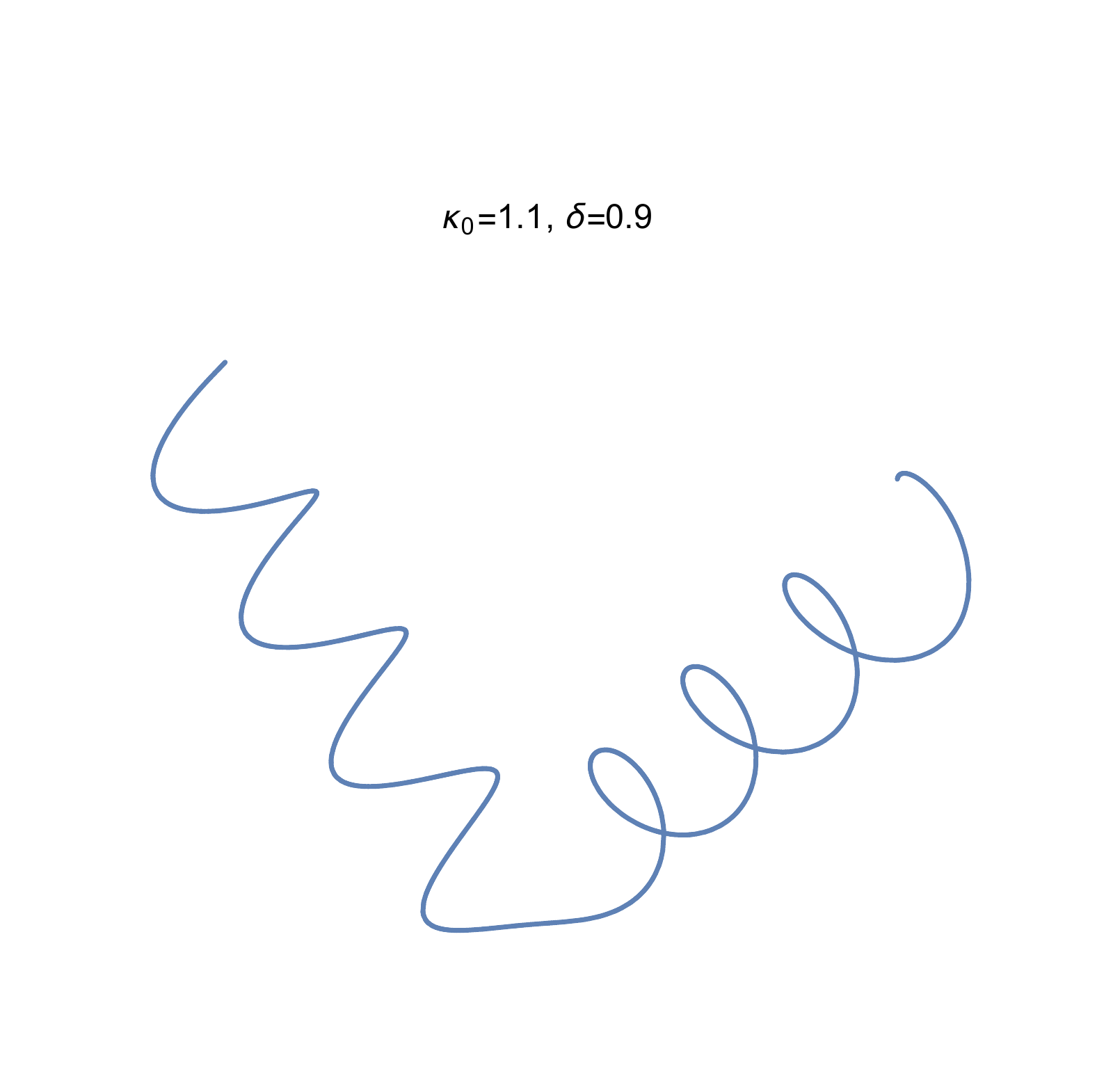}
\end{minipage}
\hfill{
\begin{minipage}{0.4\linewidth}
\centering
\includegraphics[width=0.9\linewidth,clip=true,trim=0pt 50pt 0pt 50pt]{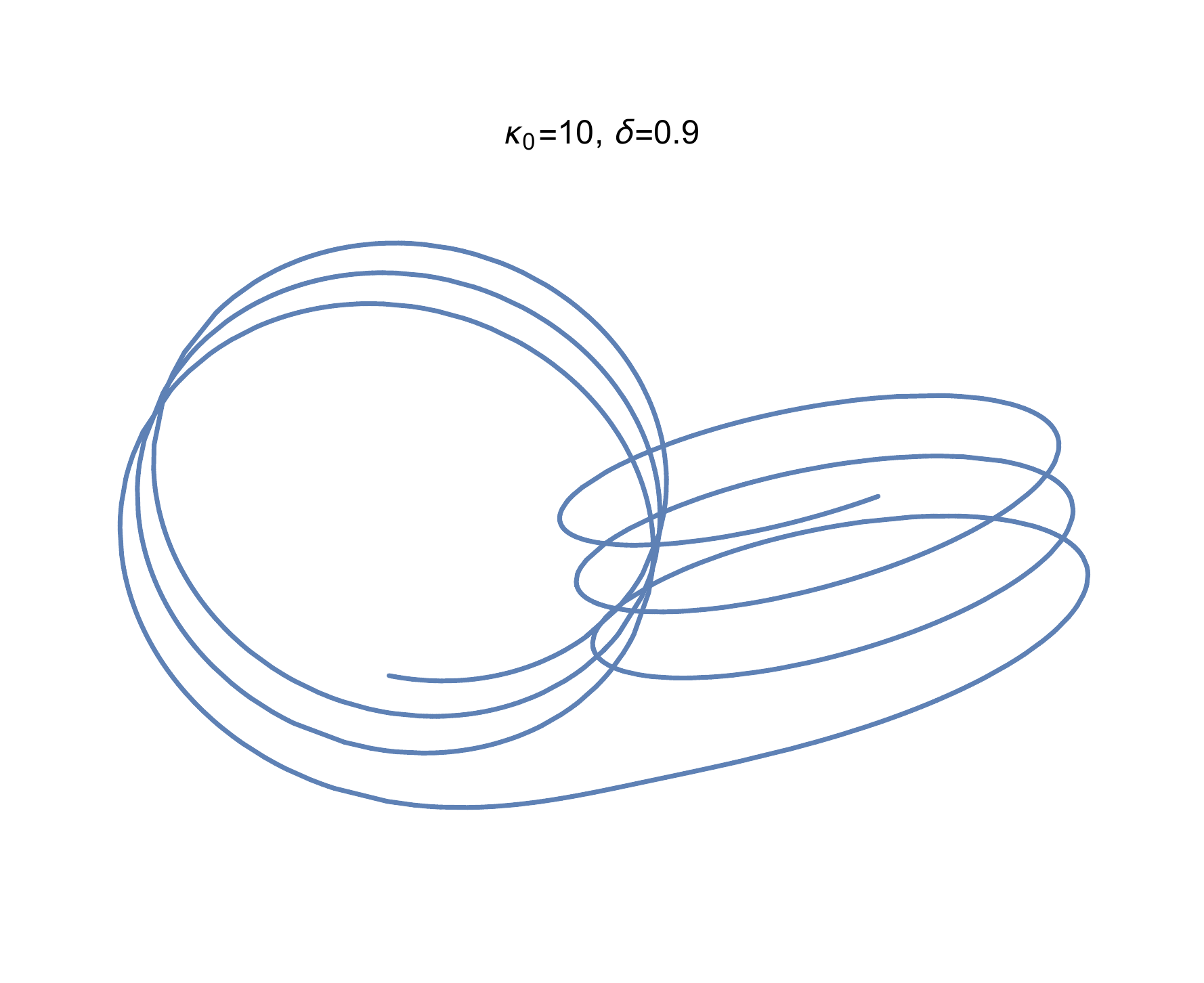}
\end{minipage}
}
\caption{Different kinks interpolating between two helical configurations. Different cases correspond to a different choice of the parameters in potential~(\ref{potGCdimfree}).}
\label{fig:kink}
\end{figure}

Apart from the ground state helices, the theory with potential~(\ref{regpot}) can have soliton-like solutions, which can be considered as excitations over the ground state. The solitons are the standard kinks of the double-well potential shown on figure~\ref{fig:regpotGC}. The soliton solution is expressed in terms of a quadrature
\be
\label{GenSol}
s-s_0 \ = \  \int\limits_{0}^\kappa \frac{dk}{\sqrt{\lambda}(\kappa_0^2-k^2)}\sqrt{\frac{k^2+\epsilon^2}{k^2+\kappa_1^2}}\,,
\ee
which can be reduced to elliptic functions. Some examples of curves with a kink in curvature and torsion defined by equation~(\ref{regtorsion}) is shown on figure~\ref{fig:kink}. Such curvature kinks in protein chains were extensively studied in references~\cite{Chernodub:2010xz,Molkenthin:2011,Hu:2011,Hu:2011wg,Krokhotin:2012}. From the point of view of the present discussion, those are canonical picture solitons, which we will discuss in the next section. 

To understand features of the curvature kinks in the geometry of curves one can reduce the number of parameters to two. We will choose one of the parameters to be $\delta=\epsilon/\kappa_1$, so that $0<\delta<1$. Consequently, it is convenient to measure the curvature in units of $\kappa_1$, $\kappa=\varkappa\kappa_1$. The second parameter is naturally the ``equilibrium" curvature $\kappa_0$ in units of $\kappa_1$, $\kappa_0=\varkappa_0\kappa_1$, $|\varkappa_0|>1$. Overall factor of potential~(\ref{regpot}) can be removed by rescaling the parameter $s$, so that the potential of the dimensionless problem becomes
\be
\label{potGCdimfree}
V(\varkappa) \ = \ \frac{(\varkappa^2-\varkappa_0^2)^2(\varkappa^2+1)}{2(\varkappa^2+\delta^2)}\,.
\ee
We have analyzed the geometry of kinks in several limiting cases. The results are shown on figure~\ref{fig:kink}. In particular, the figure intends to demonstrate the cases of $\delta\ll 1$, $\varkappa_0\simeq 1$ (top left), $\delta\ll 1$, $\varkappa_0\gg 1$ (top right), $\delta< 1$, $\varkappa_0\simeq 1$ (center left), $\delta< 1$, $\varkappa_0\gg 1$ (center right), $\delta\simeq 1$, $\varkappa_0\simeq 1$ (bottom left), $\delta\simeq 1$, $\varkappa_0\gg 1$ (bottom right). One can see that the size of the kink in comparison to the helical curvature scale set by $\kappa_0$ grows as $\varkappa_0$ is increased. Parameter $\delta$ controls the ``straightness" of the kink: as $\delta$ is increased one can observe longer straight segments at the position of the kink. The angle between the axes of the two asymptotic parts of the helix is a non-trivial function of the parameters. It would be interesting to see if there is any dynamical restrictions on the value of that angle.

Solutions illustrated by figure~\ref{fig:kink} can be compared with the helix-loop-helix \emph{structural motifs} in proteins. An example of a typical motif in the myoglobin molecule is shown on figure~\ref{fig:plotks} (left), where we used the PyMOL software~\cite{pymol} to visualize the relevant piece of the molecule. We will postpone the details of the comparison to section~\ref{sec:proteins}. In particular, we will see that the solitons in the canonical picture are more interesting from the point of view of the real proteins.

%%%%%%%%%%%%%%%%%%%%%%%%%%%%%%%%

\section{Field configurations in the canonical ensemble}
\label{sec:sols}

\subsection{Singular potential}

The case of the canonical ensemble can be considered similarly. The analysis is the same for any specific choice of $\Delta\eta$, since the difference is a mere shift of the potential energy in~(\ref{EffAction}). The corresponding term contributes to the energy of the ground state of the system, which we do not discuss in this paper.

The ``potential'' (terms, which do not contain derivatives) of the effective model is shown on figure~\ref{fig:potential} (right), where the parameters $m^2$ and $F$ are adjusted in such a way that the potential has a local minimum and a local maximum. Rewriting the potential in the form of equation~(\ref{potential}) with $0<\kappa_1<\kappa_0$, one obtains a theory with a local energy minimum at $\kappa=\kappa_0$ and the global one at $\kappa=0$. Parameter $\kappa_1$ is a simple zero of the potential.

As in the case of the grand canonical ensemble, there is a minimum energy solution $\kappa=\kappa_0$ of equations of motion~(\ref{eom}) corresponding to a constant-curvature, constant-torsion helix described in section~\ref{sec:helix}. However, in the canonical case, it is a metastable solution corresponding to a local minimum, as opposed to the stable one at $\kappa=0$. The stable solution is a straight line. The energy of the latter solution is unbounded from below. Regularizing this singularity leads to several interesting options that we discuss later.

Let us first look beyond the static homogeneous minimum energy solutions of the unbounded potential. In the potential of figure~\ref{fig:potential} (right) there exist ``dark soliton'' solutions, which correspond to a local deviation of the curvature field from the minimum energy configuration, rather than an interpolation between two minima, as in the case of a kink. To understand that such solutions exist, it is useful to think of problem~(\ref{eom}) as of a motion of a particle with ``coordinate'' $\kappa$ in ``time'' $s$. This motion occurs in the inverted potential $-V(\kappa)$. Imposing the kink boundary conditions, $\kappa(-\infty)=\kappa_0$, $\kappa'(s)=0$, the dark soliton solution corresponds to a bounce of the configuration off the inverted potential's wall at $\kappa=\kappa_1$ at time $s=0$. It is well known that such configurations are unstable and we will give a standard argument explaining this fact in the next section. Let us first construct the solution.

\begin{figure}[htb]
	\begin{minipage}{0.45\linewidth}
	\includegraphics[width=\linewidth]{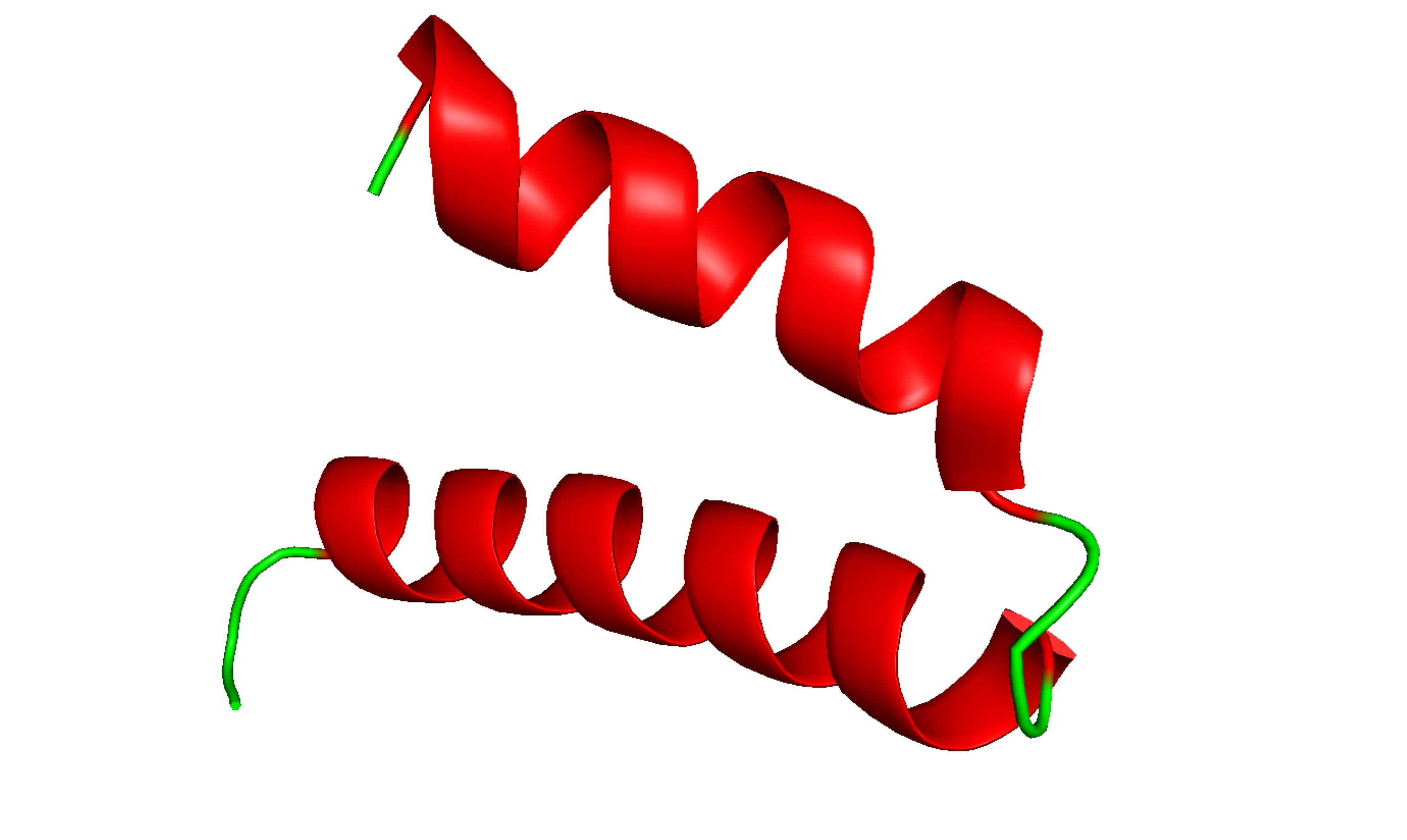}
	\end{minipage}
\hfill{
	\begin{minipage}{0.45\linewidth}
	\includegraphics[width=\linewidth]{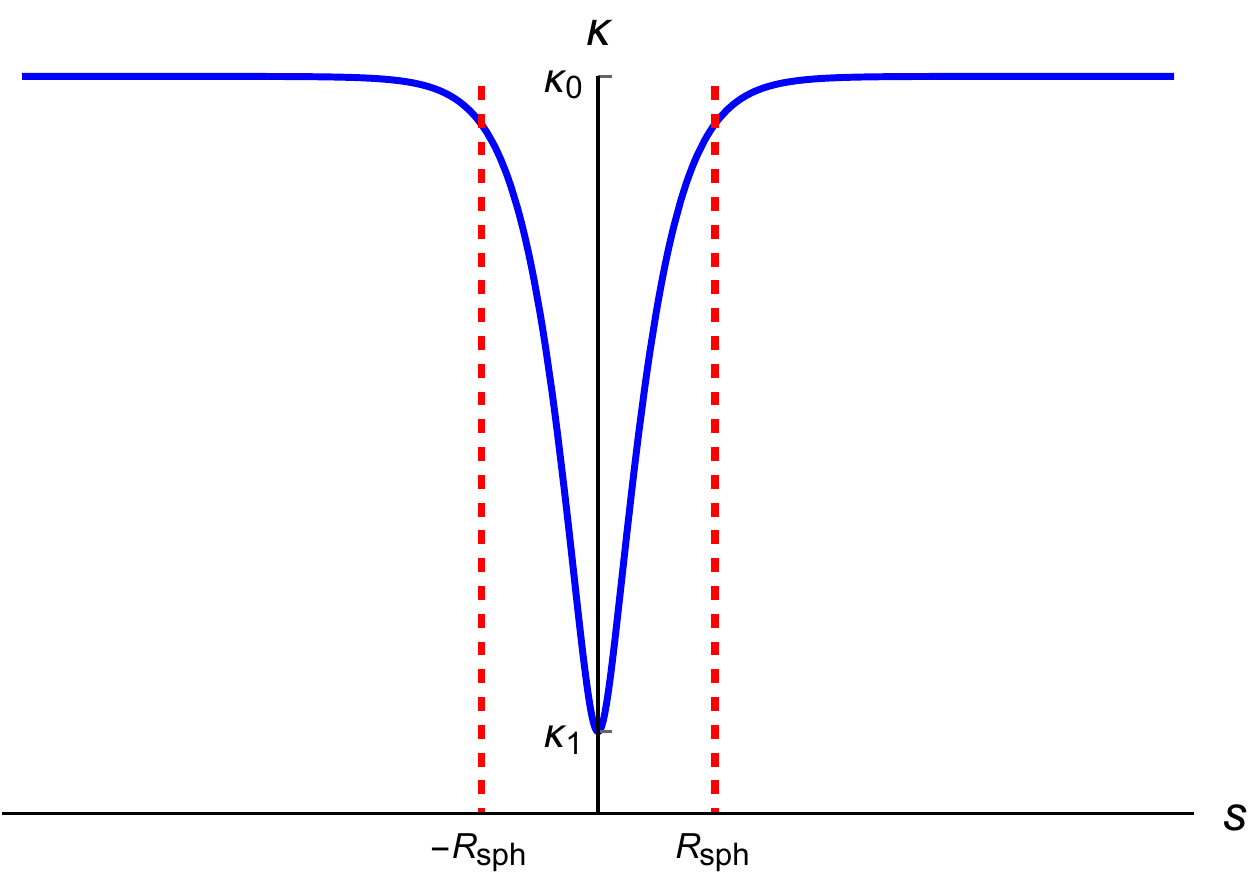}
	\end{minipage}
}
	\caption{(\emph{Left}) A Helix-Loop-Helix motif in the myoglobin molecule (PDB code \emph{1a6m}). The image is generated using the PyMOL software~\cite{pymol}. (\emph{Right}) Dark soliton (sphaleron) solution. The dashed lines show the characteristic radius $R_{\rm sph}$ of the soliton from equation~(\ref{SphaleronSize}).}
	\label{fig:plotks}
\end{figure}

With the above initial conditions, one needs to find a solution for a particle, which starts the motion at position $\kappa_0$ at time $s=-\infty$, with zero energy, then explores the minimum of the inverted potential, bounces back at time $s=0$ and position $\kappa=\kappa_1$ and, finally, returns to $\kappa_0$ at $s=\infty$. It is straightforward to find such a solution:
\be
\label{sphaleron}
\bar{\kappa}^2(s) \ = \ \kappa_1^2+(\kappa_0^2-\kappa_1^2)\tanh^2\left(\sqrt{\lambda}s\sqrt{\kappa_0^2-\kappa_1^2}\right).
\ee
In the limit $\kappa_1\to 0$ one recovers the ordinary kink solution, since that is the limit of the standard double well potential in equation~(\ref{potential}). Figure~\ref{fig:plotks}~(right) shows the curvature profile of the dark soliton.

From the form of the explicit solution in equation~\ref{sphaleron} one finds the characteristic radius of the dark soliton
\be
\label{SphaleronSize}
R_{\rm sph} \ \simeq \ \frac{1}{\sqrt{\lambda}\sqrt{\kappa_0^2-\kappa_1^2}}\,,
\ee
which is the size of the asymptotic region as $\kappa\to\kappa_0$. Figure~\ref{fig:plotks}~(right) shows this scale in comparison with the actual soliton.

Despite being unstable, these dark soliton solutions are interesting because they characterize the height of the potential wall separating the local (false) and the global (true) minima of the potential. Such solutions are known as ``sphalerons''~\cite{Klinkhamer:1984di} -- they can be used to estimate the rate of the decay of the false vacuum of a theory~\cite{Rubakov}. We will further discuss them in section~\ref{sec:stability}.

\subsection{Regularized potential}

\begin{figure}[htb]
\centering
\begin{minipage}{0.45\linewidth}
 \includegraphics[width=\linewidth]{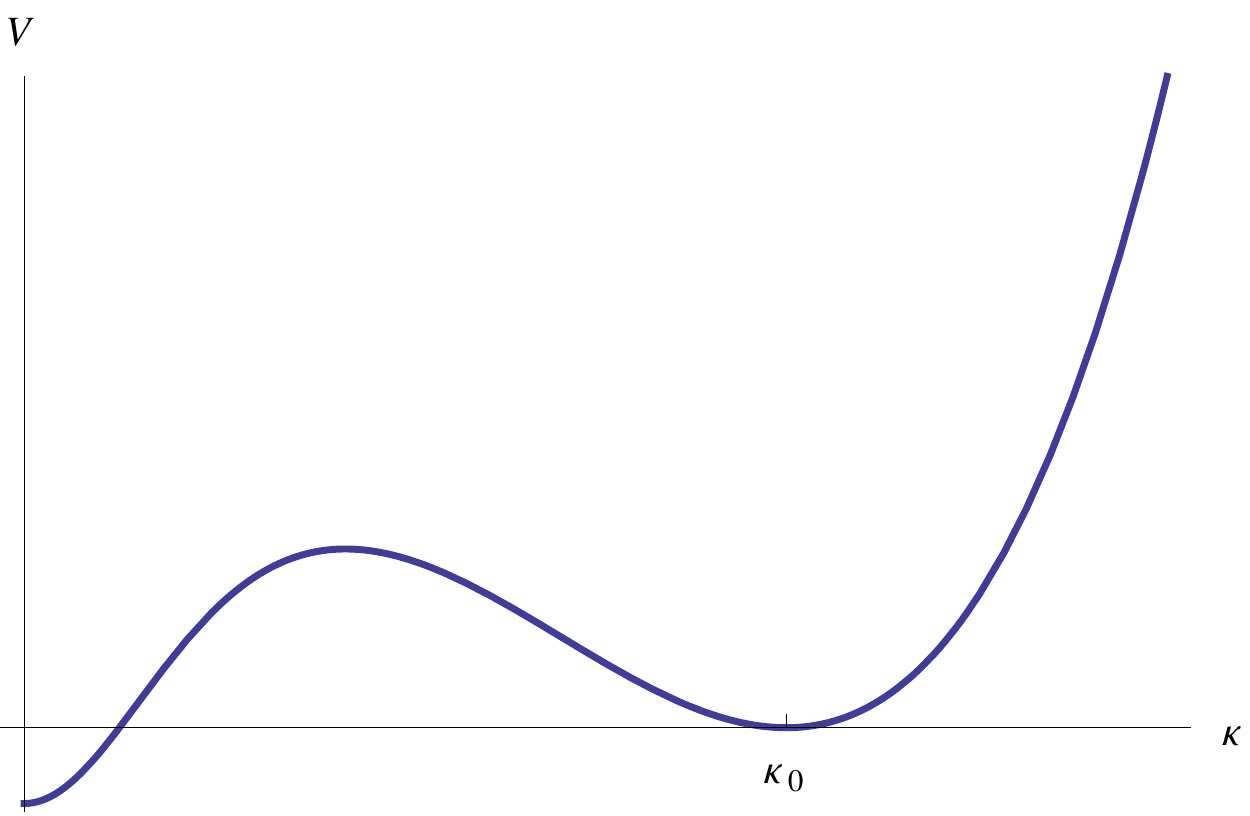}
\end{minipage}
\hfill{
\begin{minipage}{0.45\linewidth}
\includegraphics[width=\linewidth]{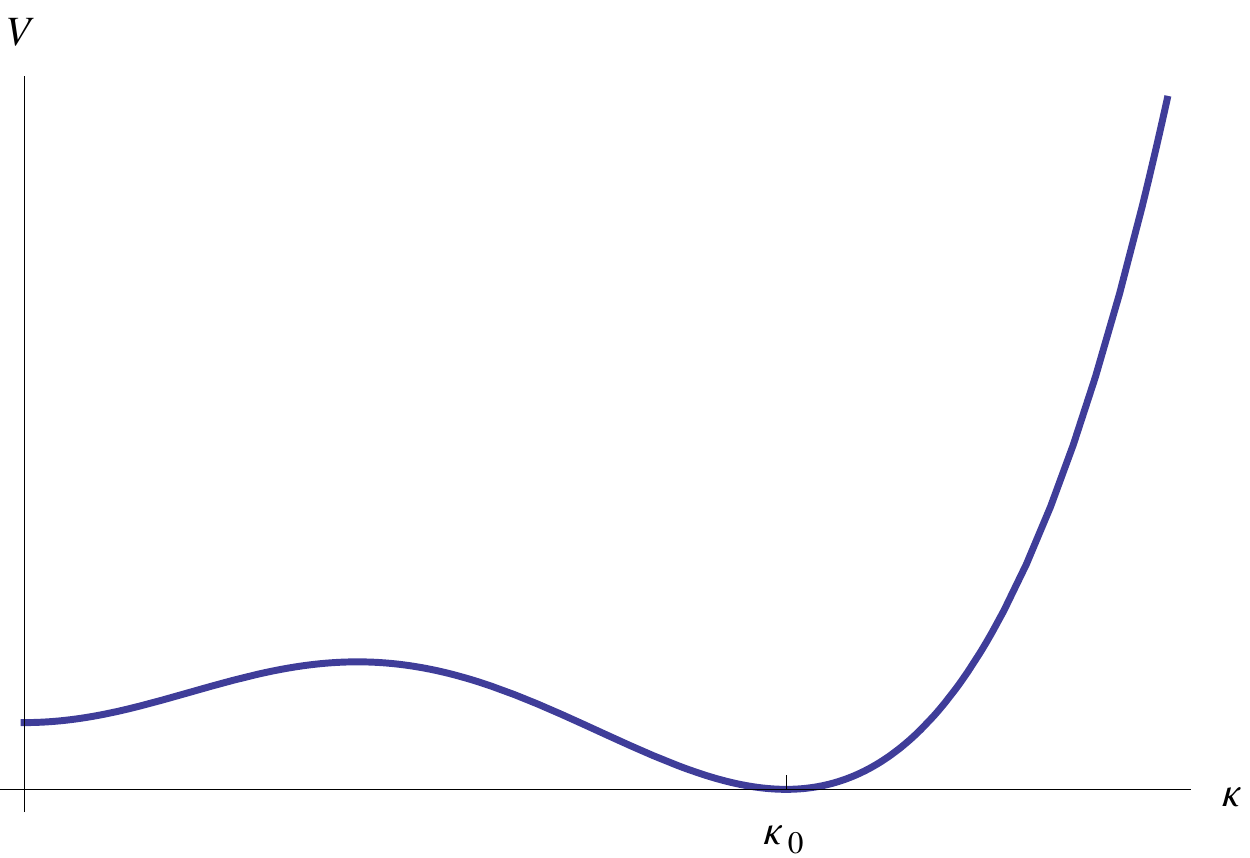}
\end{minipage}
}
\caption{(\emph{Left}) Regularized potential~(\ref{regpot2}). (\emph{Right}) The same potential with an imaginary value of the parameter $\kappa_1$, $|\kappa_1^2|< \kappa_0^2\epsilon^2/(\kappa_0^2+2\epsilon^2)$}
\label{fig:regpot}
\end{figure}

As in the discussion of the grand canonical case, we would like to regularize the potential to make the energy of the vacuum at $\kappa=0$ finite. This is achieved by introducing the parameter $\epsilon$ to obtain potential~(\ref{regpot}). There are two characteristic cases: $\kappa_1^2<0$ and $\kappa_1^2>0$, which are distinguished by whether potential~(\ref{regpot}) has or has not a zero at real $\kappa=\kappa_1$. Let us consider the first case. To keep all the parameters real, we use the sign of $\kappa_1^2$ as in equation~(\ref{potential}),
\be
\label{regpot2}
V(\kappa) \ = \ \frac{\lambda(\kappa^2-\kappa_0^2)^2(\kappa^2-\kappa_1^2)}{2(\kappa^2+\epsilon^2)}\,.
\ee
This potential has a finite-energy global minimum at $\kappa=0$, as in figure~\ref{fig:regpot} (left). The case of a small regulator, $\epsilon\ll \kappa_1$, is essentially no different from the case of the singular potential. Note that solution~(\ref{sphaleron}) does not explore the region $\kappa<\kappa_1$ and is weakly affected by a small cutoff. The only role of the cutoff, for $\kappa_1\gg \epsilon$, is to make the energy of the global minimum finite. If $\kappa_1$ is not large on the $\epsilon$ scale, the stability will depend on the ratio $\kappa_1/\epsilon$. In the limit $\kappa_1\to 0$, there are two degenerate minima and the sphaleron solution becomes stable.

The second class of models corresponds to $0\leq \kappa_1^2< \epsilon^2$ in terms of potential~(\ref{regpot}). A typical shape of such a potential is shown on figure~\ref{fig:regpot} (right). In particular, the minimum at $\kappa=0$ becomes the local minimum, while $\kappa=\kappa_0$ is the stable global minimum. In fact, the $\kappa=0$ minimum only exists if $\kappa_1^2< \kappa_0^2\epsilon^2/(\kappa_0^2+2\epsilon^2)$. Beyond that bound the potential looks more like in the case of the grand canonical ensemble. 

There are two types of solutions, beyond the constant ones, that can be discussed in the setup of the potential on figure~\ref{fig:regpot}~(right). First, the stable kink-like solution interpolating between $\kappa_0$ and $-\kappa_0$ and the unstable ``bright'' soliton solution associated with the local minimum $\kappa=0$. The bright soliton solution is a sphaleron.

\begin{figure}[htb]
 \begin{minipage}{0.5\linewidth}
  \includegraphics[width=\linewidth]{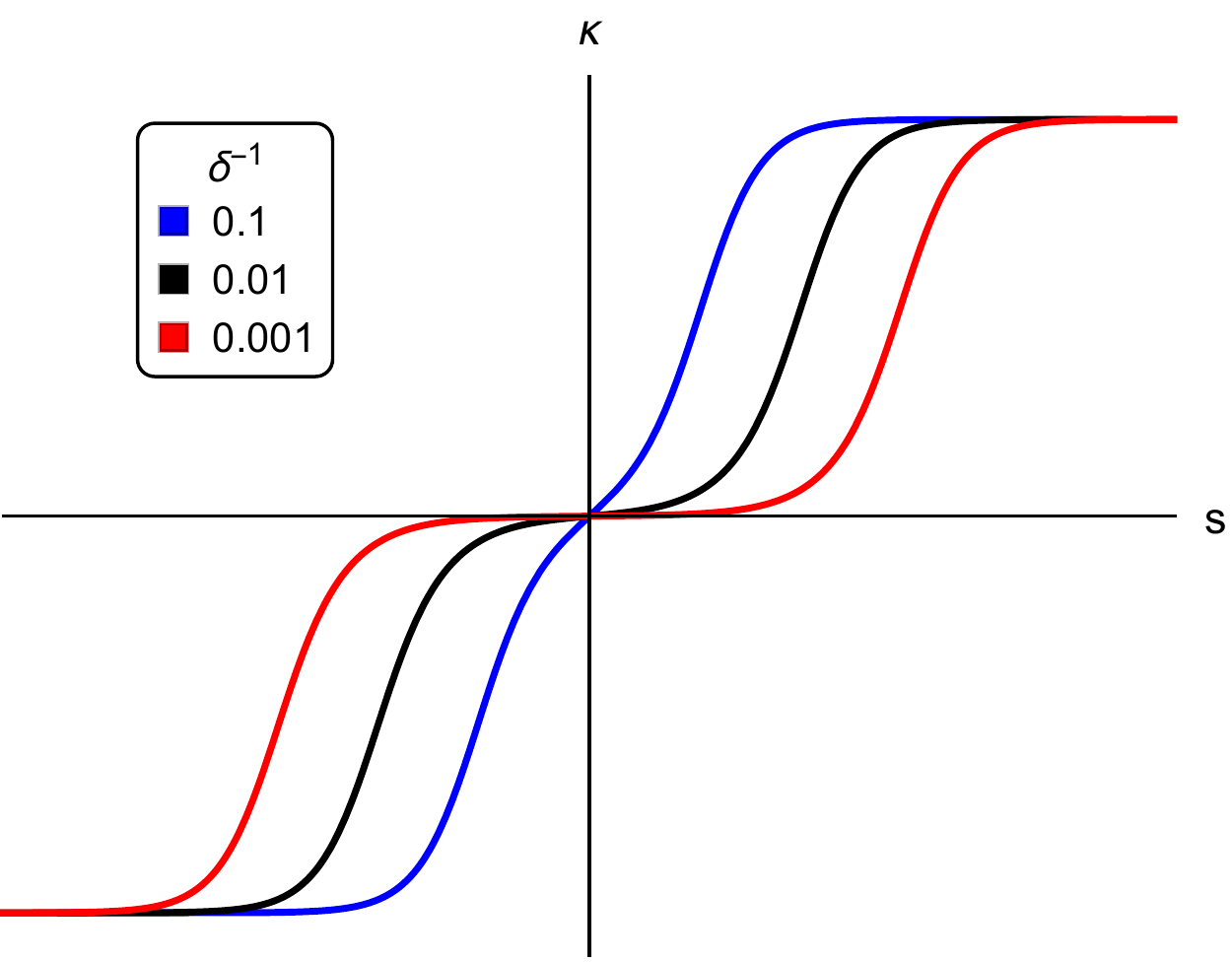}
  \end{minipage}
 \hfill{
\begin{minipage}{0.45\linewidth}
  \includegraphics[width=\linewidth,trim=50pt 200pt 50pt 200pt]{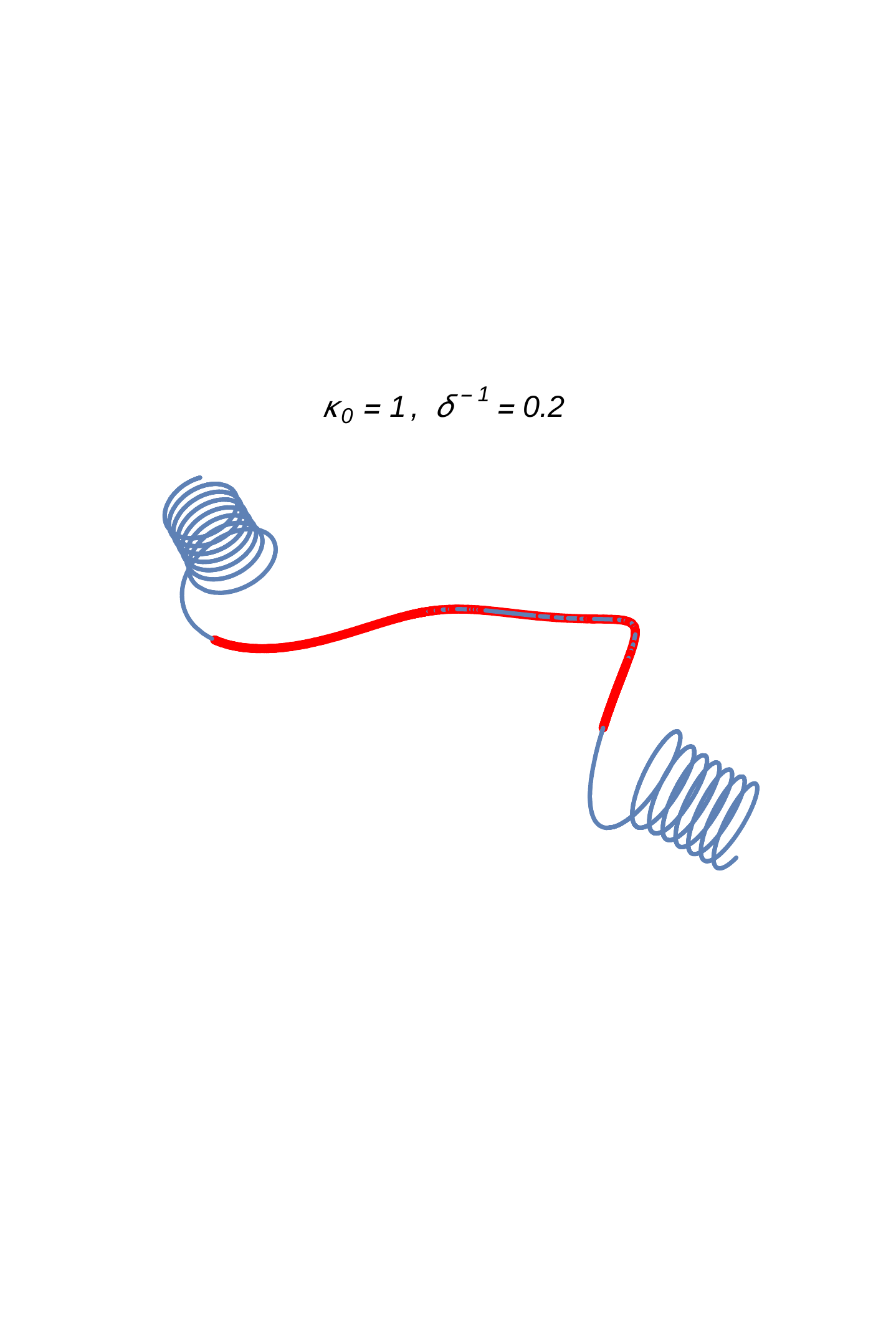}
\end{minipage}
 }
\caption{(\emph{Left}) A family of curvature profiles of the stable kink-like solution in the right potential of figure~\ref{fig:regpot} labeled by ratio $\delta=\epsilon/\kappa_1$. (\emph{Right}) The curve corresponding to a solution with $\delta=5$. Red color highlights the piece contributing approximately half of the total energy.}
\label{fig:RegSoliton}
\end{figure}

Regularized potential~(\ref{regpot}) can only be explicitly integrable in terms of elliptic functions, so we will mostly analyze the solutions numerically. On figure~\ref{fig:RegSoliton}~(left) we have shown the curvature profiles of the stable soliton for a family of potentials labeled by the parameter $\delta=\epsilon/\kappa_1$.\footnote{Similar potentials and solitons appeared in previous literature~\cite{Christ,Bazeia}. However, the potential of our model is non-local, even though originating from a local theory.} We remind that in the grand canonical case $0\leq\delta\leq 1$, while here $\delta\geq 1$. In the limit $\delta=1$, one recovers the standard double well potential and the regular kink. For $\delta\gg 1$, an intermediate step is developed in the center of the kink. In the precise limit $\delta=\infty$, the kink splits in two half-kinks. An example of the curve corresponding to the kink solution is shown on figure~\ref{fig:RegSoliton}~(right). The local minimum corresponds to a long, quasilinear segment inserted between two helices. Similarly to the grand canonical case, the size of the straight segment connecting the two helices is growing as $\delta$ increases. The curve on figure~\ref{fig:RegSoliton}~(right) is an analytical continuation of the solitons on figure~\ref{fig:helices} to large $\delta>1$.

\begin{figure}[htb]
\begin{minipage}{0.45\linewidth}
 \includegraphics[width=\linewidth]{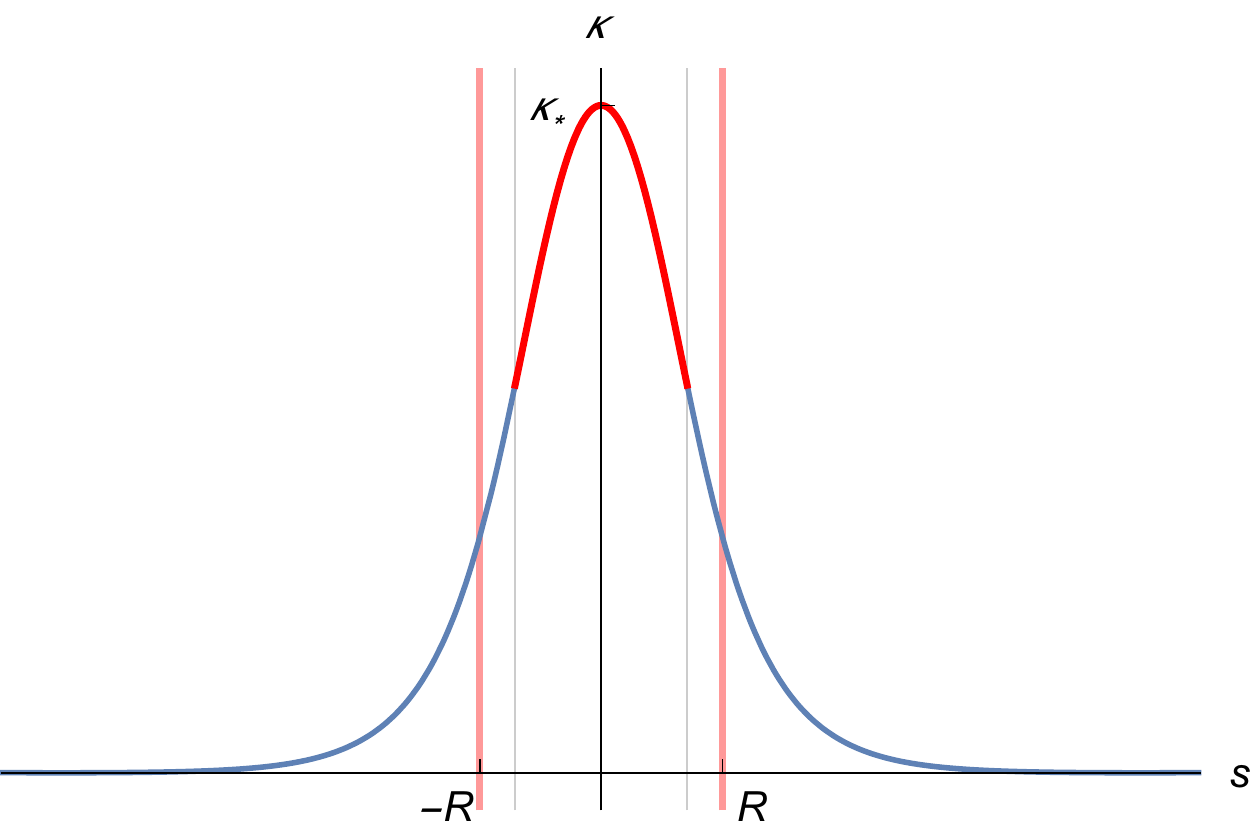}
 % plotbs.pdf: 0x0 pixel, 0dpi, 0.00x0.00 cm, bb=
\end{minipage}
\hfill{
\begin{minipage}{0.45\linewidth}
 \includegraphics[width=\linewidth,trim=0pt 50pt 0pt 70pt]{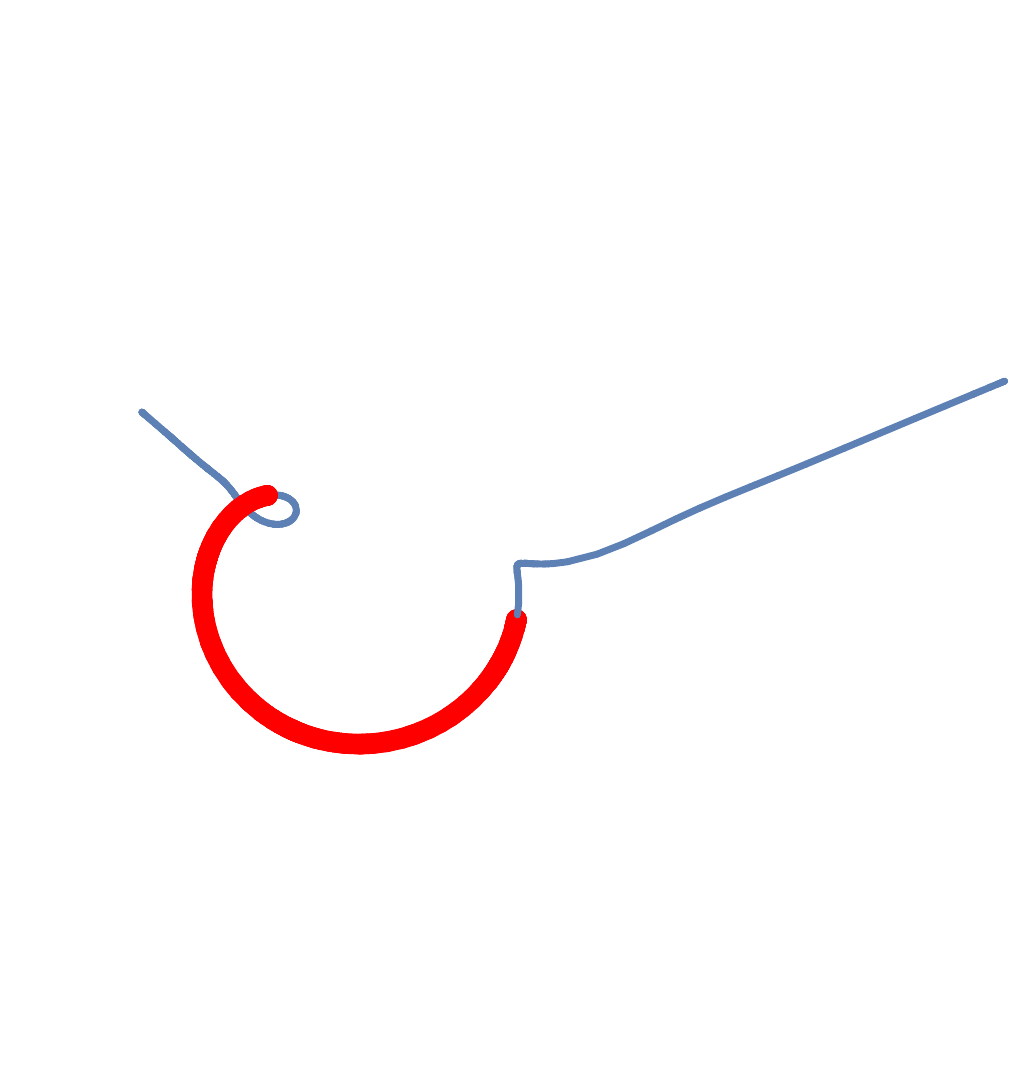}
 % plotbs.pdf: 0x0 pixel, 0dpi, 0.00x0.00 cm, bb=
\end{minipage}
}
\caption{(\emph{Left}) Curvature profile of the unstable bright soliton in the right potential of figure~\ref{fig:regpot}. Vertical lines indicate the estimate of the soliton size by equation~(\ref{SphaleronSize2}) (\emph{Right}) The curve corresponding to the bright soliton. Red color highlights the piece contributing approximately half of the total energy.}
\label{fig:Regsphaleron}
\end{figure}

On figure~\ref{fig:Regsphaleron} we show the curvature profile of the unstable bright soliton (sphaleron) solution. Asymptotically the soliton interpolates to the zero-curvature vacuum, reaching a maximal value $\kappa=\kappa_\ast$ in the center. One can make an estimate of the size of the asymptotic region, similar to equation~(\ref{SphaleronSize}) by expanding equation~(\ref{GenSol}) close to $\kappa=0$,
\be
\label{SphaleronSize2}
R_{\rm sph} \ \simeq \ \frac{\epsilon^2}{\sqrt{\lambda}\kappa_0\sqrt{\kappa_0^2(\epsilon^2-\kappa_1^2)-2\epsilon^2\kappa_1^2}}\,.
\ee
 
Summarizing this section, we have discussed different solution of the original and regularized models in the canonical ensemble approach. In this approach the models have two competing minima corresponding to helices and straight lines. There is a series of stable and unstable soliton solutions that interpolate between the minima, depending on their relative depth. The stable solutions are kinks, which connect two helical regions via a lower curvature loop. In the limit, in which the two vacua are degenerate, these kinks become large with a long straight piece inserted in the middle. The unstable sphalerons are also interesting solutions, which characterize the stability of the metastable minima of the model. In the next section we will derive basic estimates for the decay probabilities of the unstable minima.

\section{Stability of false vacua}
\label{sec:stability}

\subsection{Local minimum at $\kappa\neq 0$}
\label{sec:StabKneq0}

The kink configuration of the phi-four model is classically stable. Its stability is guaranteed by the fact that this solution interpolates between two physically distinct vacua. It is said to be topologically protected, since the two vacuum configurations cannot be smoothly deformed into each other: in the infinite volume case, one needs to pay infinite energy in order to move the field configuration from one vacuum to another vacuum, or to make a soliton configuration.

If the vacua are not degenerate, that is there is a true vacuum and a false vacuum, then one cannot have a solution that interpolates between different vacua. Instead one can have bounce-like solutions that start and end in the same metastable vacuum. Let us first discuss the example of solution~(\ref{sphaleron}) in potential~(\ref{potential}). This solution starts in the local vacuum of figure~\ref{fig:potential} (right), rolls over the hill, reaching the position $\kappa_1$, bouncing and returning to the original minimum. Such a solution is only marginally stable -- small perturbation would drive it to either the local or the global minimum. It is a type of well-known sphaleron solutions~\cite{Klinkhamer:1984di}.\footnote{For a review on sphalerons in field theory see for example~\cite{Rubakov}.}

To see that the solution is indeed unstable, we should first introduce the time dependence. Now, instead of the energy functional, we consider the following effective action
\be
\label{TDepAction}
S \ = \ \int dt \int ds\ \frac{1}{2}\left(\dot{\kappa}^2-{\kappa'}^2 + m^2\kappa^2 - \lambda\kappa^4 + \frac{F^2}{\kappa^2}\right).
\ee
The fact that this action is essentially Lorentz-invariant is a mere coincidence -- we have just added the simplest kinematic term (the same happens if one considers a continuous limit of a chain of oscillators, the speed of sound plays the role of the speed of light). Stability is encoded in the spectrum of small time-dependent perturbations
\be
\kappa \ = \ \bar{\kappa} + \delta\kappa(s){\rm e}^{i\omega t}\,,
\ee
over $\bar\kappa$, which here represents solution~(\ref{sphaleron}). If the linearized equation
\be
\label{Schroedinger}
-\delta\kappa'' - m^2\delta\kappa + 6\lambda\bar\kappa^2\delta\kappa - \frac{3F^2}{\bar\kappa^4}\,\delta\kappa \ = \ w^2\delta\kappa\,
\ee
has imaginary eigenmodes, the solution $\bar\kappa$ is unstable. This is a Sturm-Liouville problem with boundary conditions
\be
\delta\kappa\big|_{s=\pm\infty} \ = \ 0\,.
\ee

First, one notices that there is a zero-mode solution to this problem. Indeed, from the translational invariance of energy functional~(\ref{EffAction}), in the limit of infinite curve, $\delta\kappa=\bar\kappa'(s)$ is the zero mode of equation~(\ref{Schroedinger}) satisfying boundary conditions. Second, the zero-mode cannot be lowest-energy eigenmode of the Sturm-Liouville problem, since $\kappa'(s)$ has a zero (at the turning point of the bounce solution). There exists another eigenmode with $\omega^2<0$.

The sphaleron solution sits at the top of the potential well in the space of field configurations, which separates the true and the false vacua of the model. On the contrary, a kink of the phi-four theory is metastable -- it sits in a local minimum, separated by an infinite energy barrier. Note that the derivative of the kink solution is non-zero everywhere inside the infinite interval, so the zero mode is the lowest eigenmode. 

Sphaleron is a useful notion in field theory, since it helps to calculate the probability of the thermal phase transition. Besides, sphalerons also appear as the Euclidean solution in the quantum tunneling problem, though, in comparison to the thermal transition problem, sphaleron is the Euclidean solution of a lower-dimensional problem. From the phase transition point of view, sphaleron is the critical bubble of the stable phase, inside the metastable one. Hence, one can estimate the size of such bubble and the temperature of the phase transition.

It is convenient to work with the dimensionless version of the problem:
\be
\sigma \ = \ \sqrt{\lambda}\kappa_0 s\,, \qquad \xi \ = \ \frac{\kappa_1}{\kappa_0}\,,
\ee
where $\sigma$ is the dimensionless coordinate and $\xi$ is the ratio of the two parameters of the potential. In this case, energy~(\ref{EffAction}) of sphaleron solution~(\ref{sphaleron}) can be cast in the form
\be
\label{dimless}
E_{\rm sph} \ = \ \sqrt{\lambda}\kappa_0^3D(\xi)\,,
\ee
where $D(\xi)$ is a fixed dimensionless function of $\xi$. In the case of solution~(\ref{sphaleron}) this function can be computed explicitly:
\be
\label{D}
D(\xi) \ = \ \int\limits_{-\infty}^{\infty} \frac{(k^2(\sigma)-1)^2(k^2(\sigma)-\xi^2)}{k^2(\sigma)}\, d\sigma\,,
\ee
where dimensionless function $k(\sigma)$ is equal to
\be
k^2(\sigma) \ = \ \xi^2 + (1-\xi^2)\tanh^2\left(\sqrt{1-\xi^2}\sigma\right)\,.
\ee
\begin{figure}[htb]
 \begin{minipage}{0.45\linewidth}
 \includegraphics[width=\linewidth]{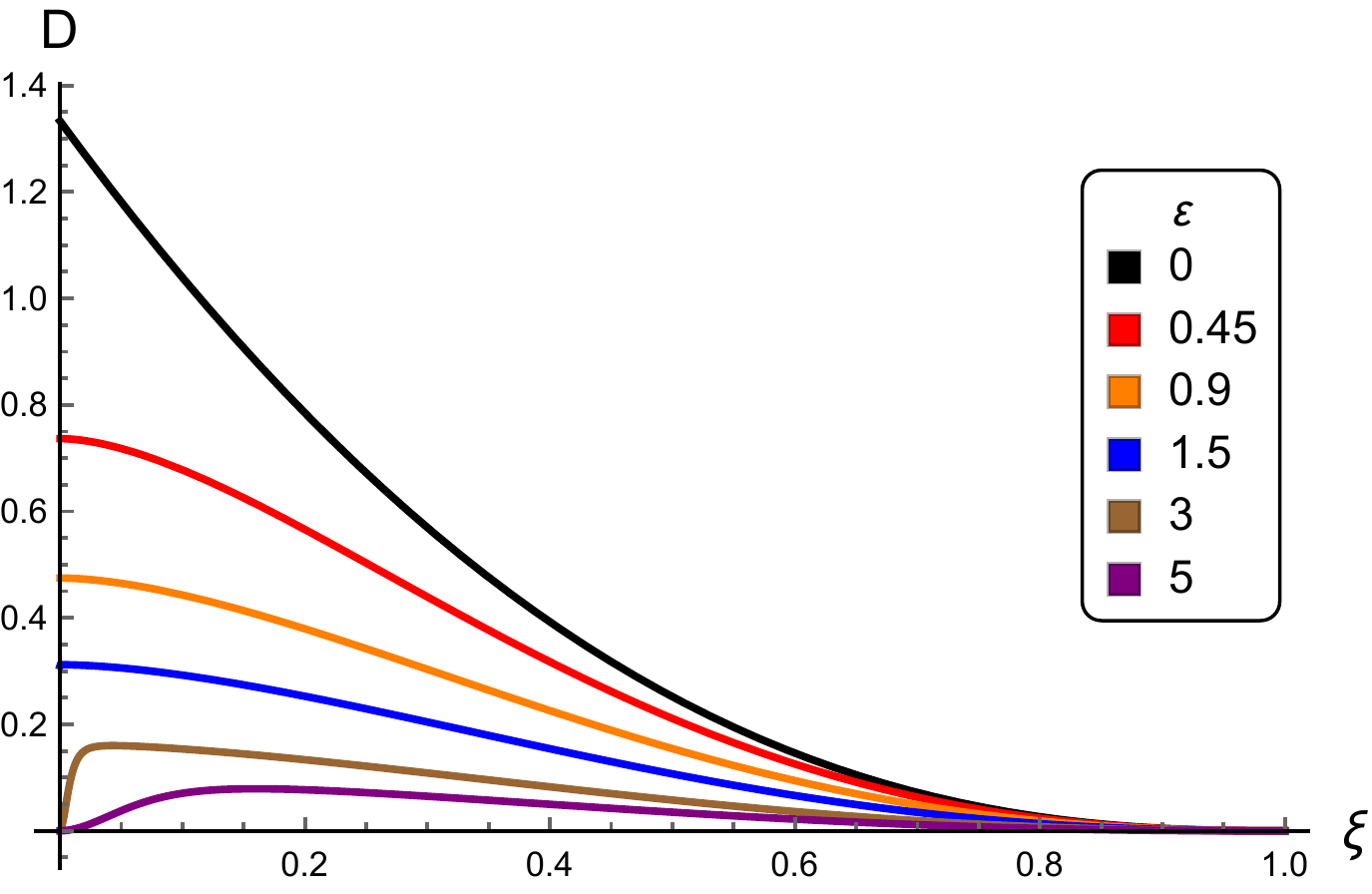}
 \end{minipage}
 \hfill{
 \begin{minipage}{0.45\linewidth}
 \includegraphics[width=\linewidth]{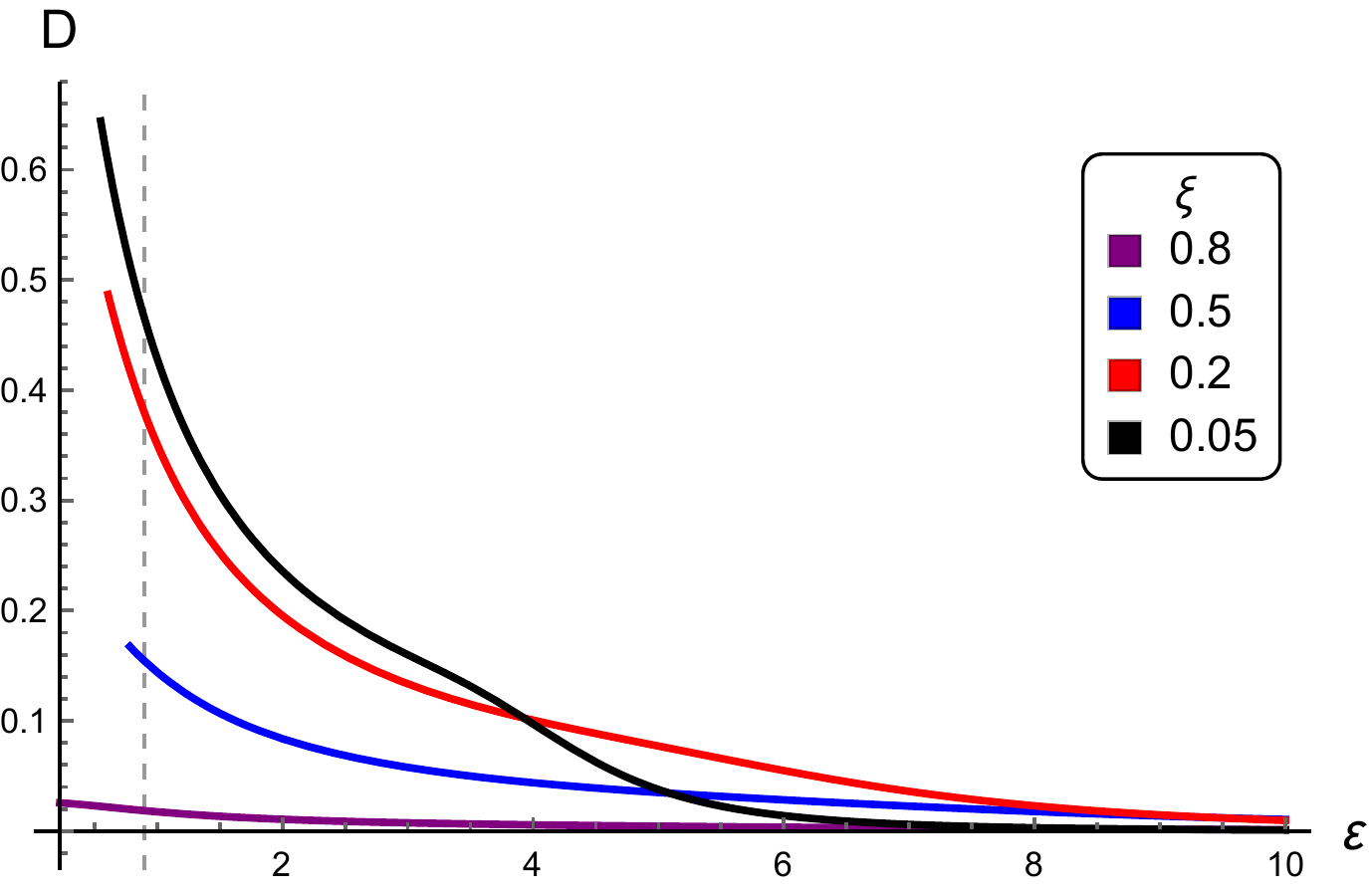}
 \end{minipage}
 }
 \caption{(\emph{Left}) Plots of the dimensionless energy $D(\xi,\varepsilon)$ of the sphaleron. Color of the lines corresponds to variation of $\varepsilon$ between $\varepsilon=0$ (black) and $\varepsilon=5$ (magenta). (\emph{Right}) Plots of $D(\xi,\varepsilon)$ as a function of $\varepsilon$ for different values of $\xi$ between $\zeta=0.05$ (black) and $\zeta=0.8$ (magenta). Vertical line indicates the phenomenological value $\varepsilon\simeq 0.9$.}
 \label{fig:dxi}
\end{figure}

Consequently,
\be
\label{dxi}
D(\xi) \ = \ \frac{2}{3}\,\sqrt{1-\xi ^2}(\xi ^2+2) -{2}\xi \, {\rm arctan}\left(\frac{\sqrt{1-\xi ^2}}{\xi }\right).
\ee
The plot of this function is the black curve on figure~\ref{fig:dxi} (left). In the limit $\xi\to 0$ one reproduces the energy of the phi-four kink $E=4\sqrt{\lambda}\kappa_0^3/3$. Energy~(\ref{dimless}) is thus our estimate for the phase transition temperature
\be
\label{temphase}
T\ \sim \ E_{\rm sph}\,.
\ee
This phase transition is analogous to the so-called \emph{helix-coil transition}, in which the alpha helices of the protein ``melt'' and the backbone assumes a random coil structure~\cite{Charkrabartty:1996,Doig:2008}. In section~\ref{sec:proteins} we will make phenomenological estimates for the parameters of the model and the helix-coil transition temperature.

Now let us discuss the quantum stability of the metastable vacuum solution $\kappa=\kappa_0$.\footnote{Usually quantum effects are assumed to be weak due to the size of the protein molecules. We will see that in our model they are not necessarily negligible, but turn out to be so in the phenomenologically most viable scenario.} The metastable solution can tunnel through the energy barrier given by the $E_{\rm sph}$. The probability of such a process is given in the leading semiclassical approximation by the Euclidean action computed on the Euclidean bounce solution:
\be
\label{amptun}
\Gamma \  \sim \ {\rm e}^{-\frac{1}{\hbar}\,2S_{\rm E}}\,.
\ee

To compute the Euclidean action one needs to solve time-dependent equations of motion of~(\ref{TDepAction}) in the Euclidean time $t=i\tau$:
\be
-\kappa_{\tau\tau} - \kappa_{ss} - \omega^2\kappa + 2\lambda\kappa^3 + \frac{F^2}{\kappa^3} \ = \ 0\,.
\ee
One is interested in spherically symmetric solution, so it is convenient to introduce the radial coordinate
\be
r^2 \ = \ \tau^2 + s^2\,,
\ee
and assume $\kappa=\kappa(r)$. Hence the equation, which needs to be solved is
\be
\label{InstEq}
\frac{d^2}{dr^2}\kappa + \frac{1}{r}\frac{d}{dr} \kappa + \omega^2\kappa - 2\lambda\kappa^3 - \frac{F^2}{\kappa^3} \ = \ 0\,.
\ee
The Euclidean bounce solution satisfies this equation with boundary conditions
\be
\kappa(r\to\infty) = \kappa_0\,, \qquad \text{and} \qquad \frac{d}{dr}\kappa(r=0)\ = \ 0\,.
\ee
This equation is not integrable, so we will present the results of the numerical analysis.

For the numerical study one casts the equations in the dimensionless form. It is more convenient to use the parametrization in terms of $\kappa_0$ and $\kappa_1$:
\be
\label{EucEoM}
\frac{d^2}{dr^2}\kappa + \frac{1}{r}\frac{d}{dr} \kappa - \frac{dV}{d\kappa}  \ = \ 0\,,
\ee
where the potential is given by equation~(\ref{potential}).

Again, we can measure the radius in units of $1/\sqrt{\lambda}\kappa_0$, introduce dimensionless $\xi=\kappa_1/\kappa_0$ and make the curvature function dimensionless,
\be
\kappa(r) \ = \ \kappa_0 k(\rho)\,, \qquad \rho \ =\ {\sqrt{\lambda}\kappa_0}r\,.
\ee
As before, this allows us to express the energy, or Euclidean action, computed on the solution as a combination
\be
\label{SEuclid}
S_{\rm E} \ = \ 2\pi \kappa_0^2 G(\xi)\,,
\ee
where factor $2\pi$ comes from the integration over the angular directions in the space $(\tau,s)$ and the integral of the action should be taken with an appropriate measure
\be
\int\limits_0^\infty r\,dr \ = \ \frac{1}{\lambda\kappa_0^2}\int\limits_0^\infty \rho\,d\rho \,.
\ee
In the dimensionless presentation equation~(\ref{EucEoM}) reads
\be
\label{DlessInstEq}
k''(\rho) + \frac{1}{\rho}k'(\rho) + (2+\xi^2)k(\rho) - 2k^3(\rho) - \frac{\xi^2}{k^3(\rho)} \ = \ 0\,.
\ee
We need to find a solution, which satisfies the boundary conditions
\be
\label{bconditions2}
k(\rho\to\infty) \ = \ 1\,, \qquad k'(\rho=0)\ = \ 0\,.
\ee

\begin{figure}[htb]
\begin{minipage}{0.45\linewidth}
  \includegraphics[width=\linewidth]{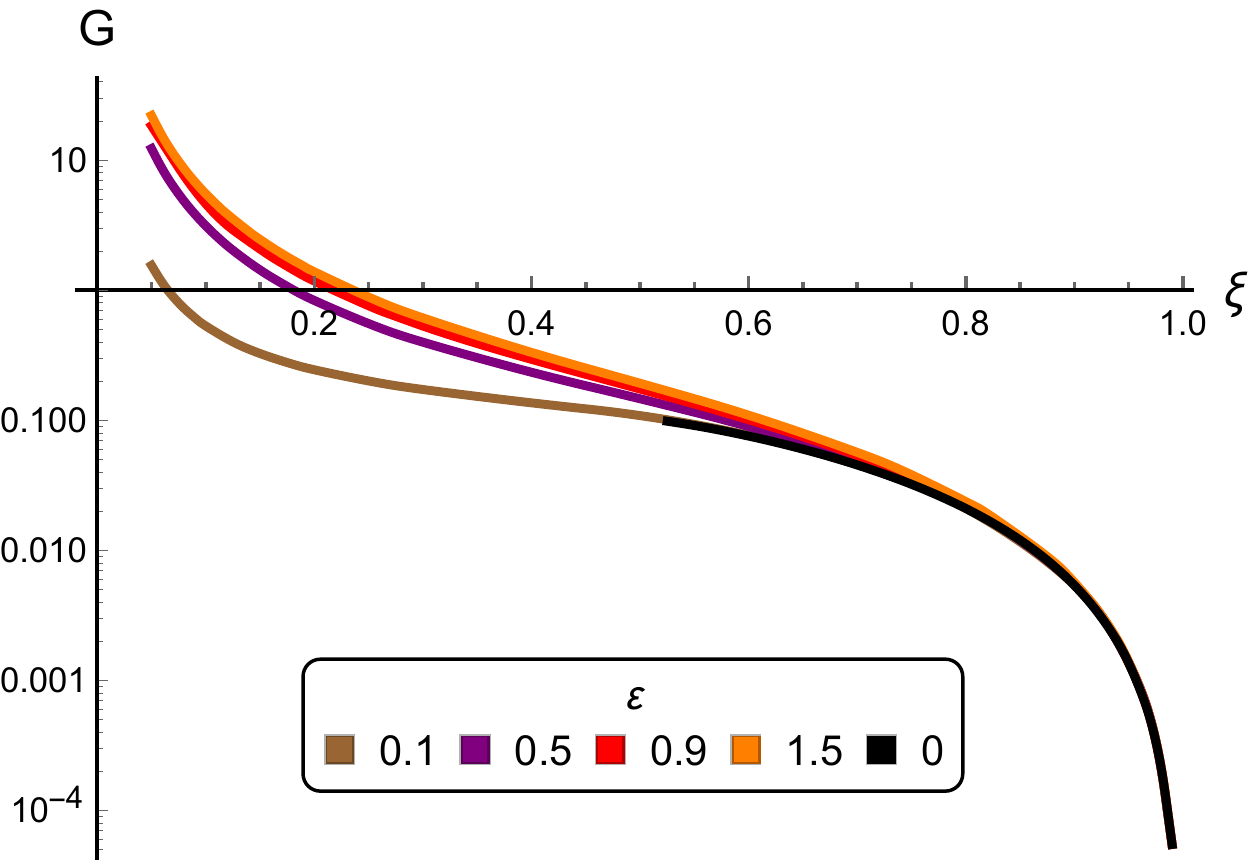}
\end{minipage}
\hfill{
\begin{minipage}{0.45\linewidth}
\includegraphics[width=\linewidth]{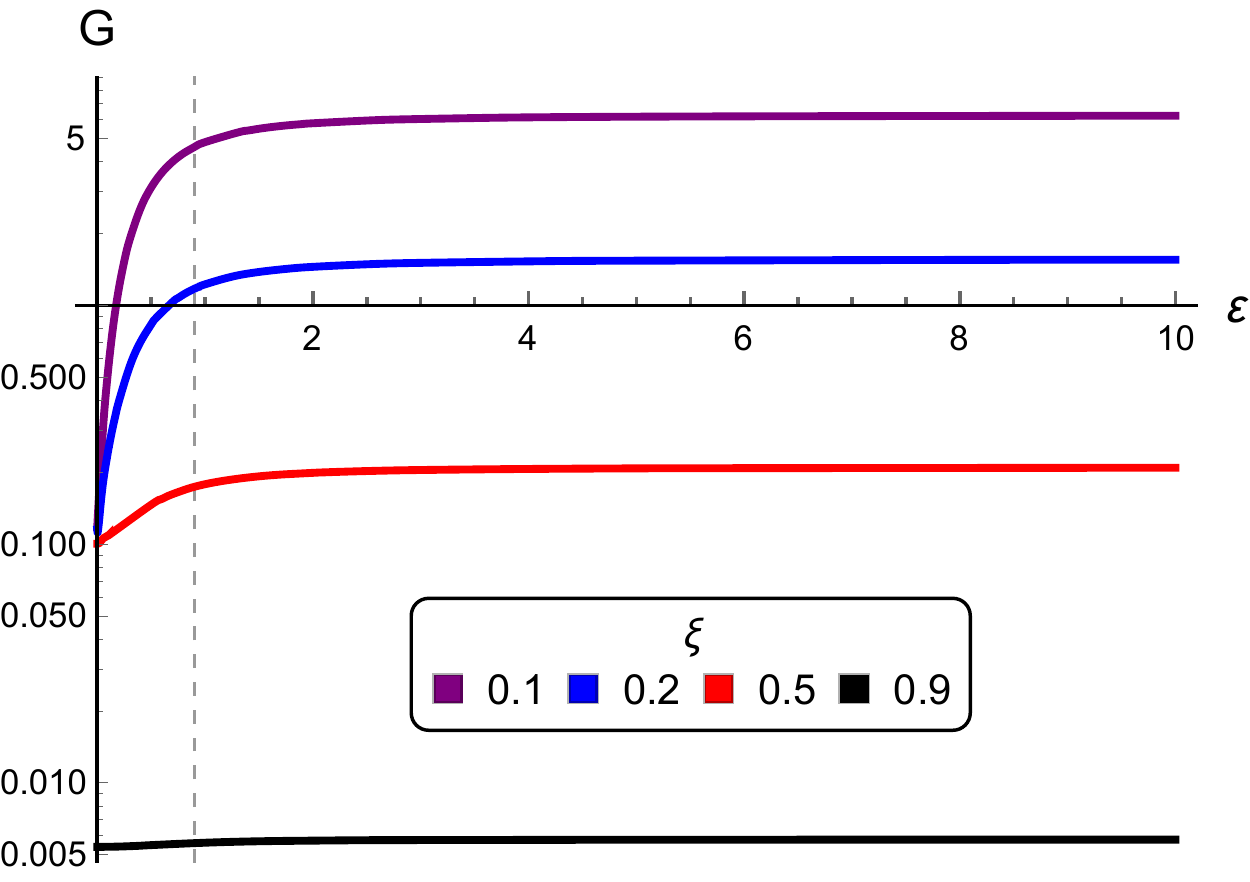}
\end{minipage}
}
 \caption{(\emph{Left}) Log plots of the dimensionless Euclidean action $G(\xi,\varepsilon)$ on the sphaleron configuration. Color of the lines corresponds to variation of $\varepsilon$ between $\varepsilon=0$ (black) and $\varepsilon=1.5$ (yellow). (\emph{Right}) Log plots of $G(\xi,\varepsilon)$ as a function of $\varepsilon$ for different values of $\xi$ between $\zeta=0.1$ (magenta) and $\zeta=0.9$ (black). Vertical line indicates the phenomenological value $\varepsilon\simeq 0.9$.}
 \label{fig:F2}
\end{figure}

The existence of such a solution can be easily understood (see for example~\cite{Rubakov}). Equation~(\ref{DlessInstEq}) describes classical motion of a particle in the inverted potential $\tilde{V}=-V(\kappa)$, where $V(\kappa)$ is given by equation~(\ref{potential}), see also figure~\ref{fig:potential} (right). We need to find a ``bounce'' solution that starts at $\kappa_0$ at Euclidean time $\tau=-\infty$, bounces from the potential wall for $\kappa<\kappa_0$ at $\tau=0$ and returns to $\kappa_0$ at $\tau=+\infty$. Note however, that the equation has a dissipation term proportional to the first derivative. Hence the particle will loose energy on its way. Consequently, we should find a solution which corresponds to a descent from the wall in the left part of the inverted potential starting somewhere above the zero energy level to make sure that, after some energy is lost to dissipation, the particle still makes it to the top of the barrier at $\kappa_0$. Adapting accordingly the numerical experiment we find dimensionless $G(\xi)$ illustrated by the black curve on figure~\ref{fig:F2}~(left). The numerical results indicate that no solution exists for values $\xi<0.52$. In terms of the classical analogy, the dissipation makes it impossible to climb the hill at $\kappa_0$ for any initial potential energy. This should be interpreted as quantum stability of the vacuum $\kappa=\kappa_0$ for $\xi<0.52$. 

We can also discuss how the stability is affected if the potential is regularized, as in equation~(\ref{regpot2}). Let us introduce a dimensionless quantity $\varepsilon=\epsilon/\kappa_0$. When $\varepsilon\leq \xi$, the stability analysis is weakly affected by the regularization effects, since neither sphaleron, nor solution to the Euclidean problem probe the region of $\kappa\sim\epsilon$. Numerical analysis of the sphaleron solutions in the regularized potential demonstrate that larger values of $\epsilon$ make the local minimum at $\kappa_0$ less stable classically. For fixed $\xi$, the right plots of figure~\ref{fig:dxi} show monotonic decrease of the dimensionless sphaleron energy $D(\xi,\varepsilon)$ as a function of $\varepsilon$. One can also see from the left plots of the same figure that for fixed $\varepsilon>0.5$, function $D(\xi,\varepsilon)$ develops a ridge of optimal values of $\xi$ (maxima of function $D(\xi)$).

The probability of quantum tunneling in the regularized potential, computed by Euclidean action~(\ref{SEuclid}), can be expressed in terms of a dimensionless function $G(\xi,\varepsilon)$. In this case one uses equation~(\ref{EucEoM}) with potential~(\ref{regpot2}). The numerical solutions for $G(\xi,\varepsilon)$, shown on figure~\ref{fig:F2}, reveal that for large $\varepsilon$ the probability of quantum phase transition becomes $\varepsilon$-independent. 

The large $\varepsilon$ limit can be understood as follows. For $\epsilon\gg\kappa_0$ the potential becomes
\be
\label{eps2inf}
V \ \to \ \frac{\lambda}{2\epsilon^2}(\kappa^2-\kappa_0^2)^2(\kappa^2-\kappa_1^2)\,.
\ee
Hence $\epsilon$ essentially rescales the parameter $\lambda$. Note that $\lambda$ appears in expression~(\ref{dimless}) for the sphaleron energy, but not in equation~(\ref{SEuclid}). So one may expect the following general behavior for $D$ and $G$, 
\be
\label{DGeps2inf}
D(\varepsilon)\  \to \ {\rm const}/\varepsilon\,, \qquad G(\varepsilon) \ \to \ {\rm const}\,, \qquad \text{if} \qquad \varepsilon\ \to \infty\,.
\ee

In summary, in this section we derived theoretical estimates for the probabilities of the classical and quantum helix-coil transitions of the metastable configurations $\kappa=\kappa_0$ ($\alpha$-helices) of the model discussed in section~\ref{sec:model} with boundary condition $\Delta\eta \ = \ 0$. The probability of the phase transition driven by thermal fluctuations is set by the height of the energy barrier effectively computed by the sphaleron solution: the height of this barrier is
\be
\label{SphEn2}
E_{\rm sph} \ = \ \sqrt{\lambda}\kappa_0^3D(\kappa_1/\kappa_0,\epsilon/\kappa_0)\,,
\ee
where the behavior of function $D(\xi,\varepsilon)$, in general computed numerically, is demonstrated on the plots of figure~(\ref{fig:dxi}). The leading probability of the quantum tunneling through the same barrier is computed by the exponential of the Euclidean bounce solution
\be
\label{EucA2}
S_{\rm E} \ = \ 2\pi\kappa_0^2 G(\kappa_1/\kappa_0,\epsilon/\kappa_0)\,.
\ee
The dimensionless function $G$ is also determined numerically. The plots of this function for different values of the parameters is shown on figure~\ref{fig:F2}. One feature of the behavior that we observe numerically, is the absence of quantum tunneling for $\epsilon=0$ and $\kappa_1<0.52\kappa_0$.

%%%%%%%%%%%%%%%%%%%%%%%%%%%%%%%%%%%%%%%%%%%%%%%%%%%%%%%%%%%%%%%%%%%%%%%%%

\subsection{Global minimum at $\kappa\neq 0$}

In this section we repeat the analysis of the previous subsection in the regime of the right potential on figure~\ref{fig:regpot}. This regime is defined by the values of the parameters subject to constraints $0\leq \kappa_1^2\leq \kappa_0^2\epsilon^2/(\kappa_0^2+2\epsilon^2)$ and $\epsilon^2\geq \kappa_1^2$ in the potential specified by equation~(\ref{regpot}), so that the potential has a global minimum at $\kappa=\kappa_0$ and a local one at $\kappa=0$. Let us discuss the stability of the local minimum.

\begin{figure}[htb]
\begin{minipage}{0.45\linewidth}
 \includegraphics[width=\linewidth]{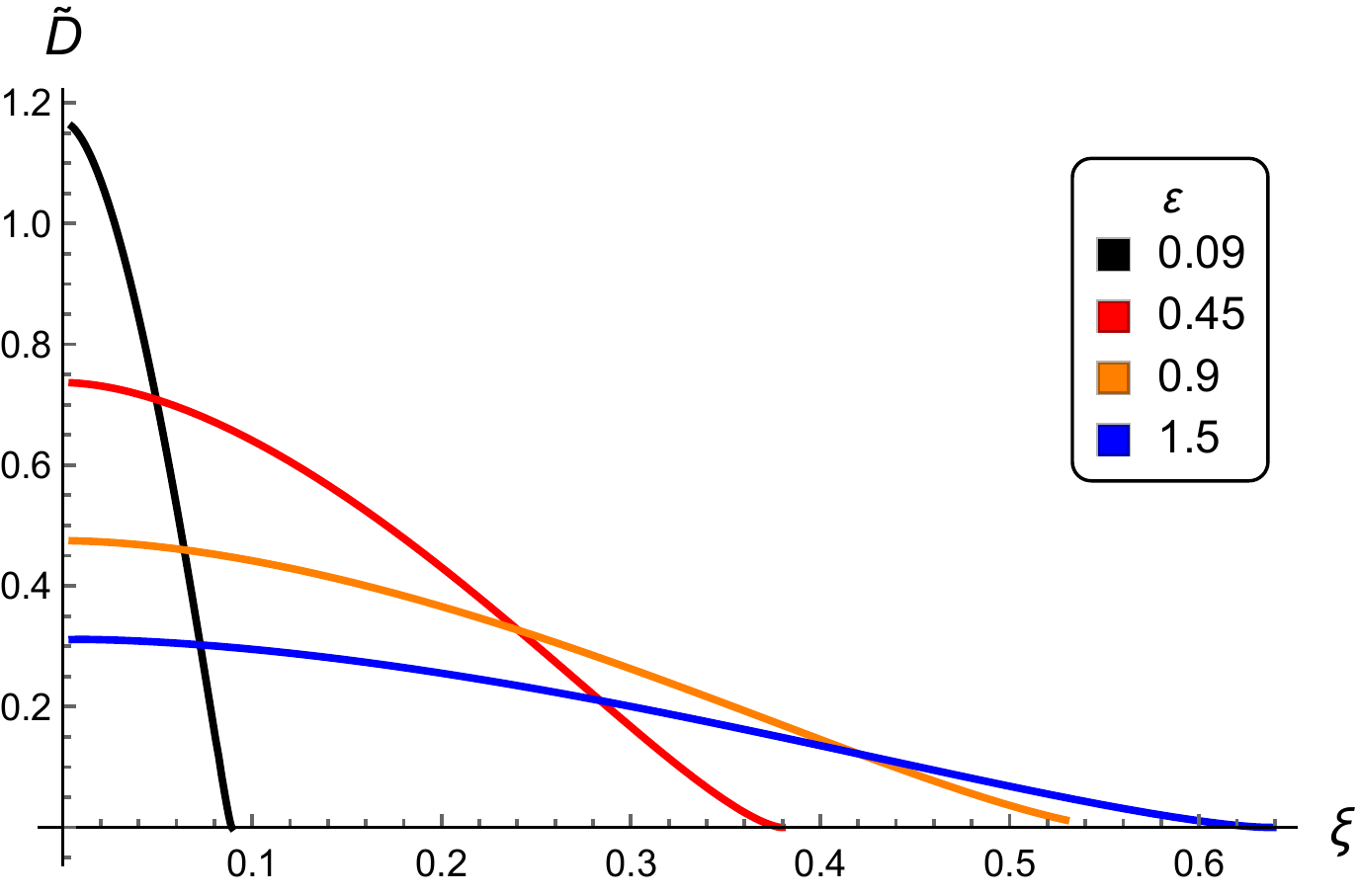}
\end{minipage}
\hfill{
\begin{minipage}{0.45\linewidth}
 \includegraphics[width=\linewidth]{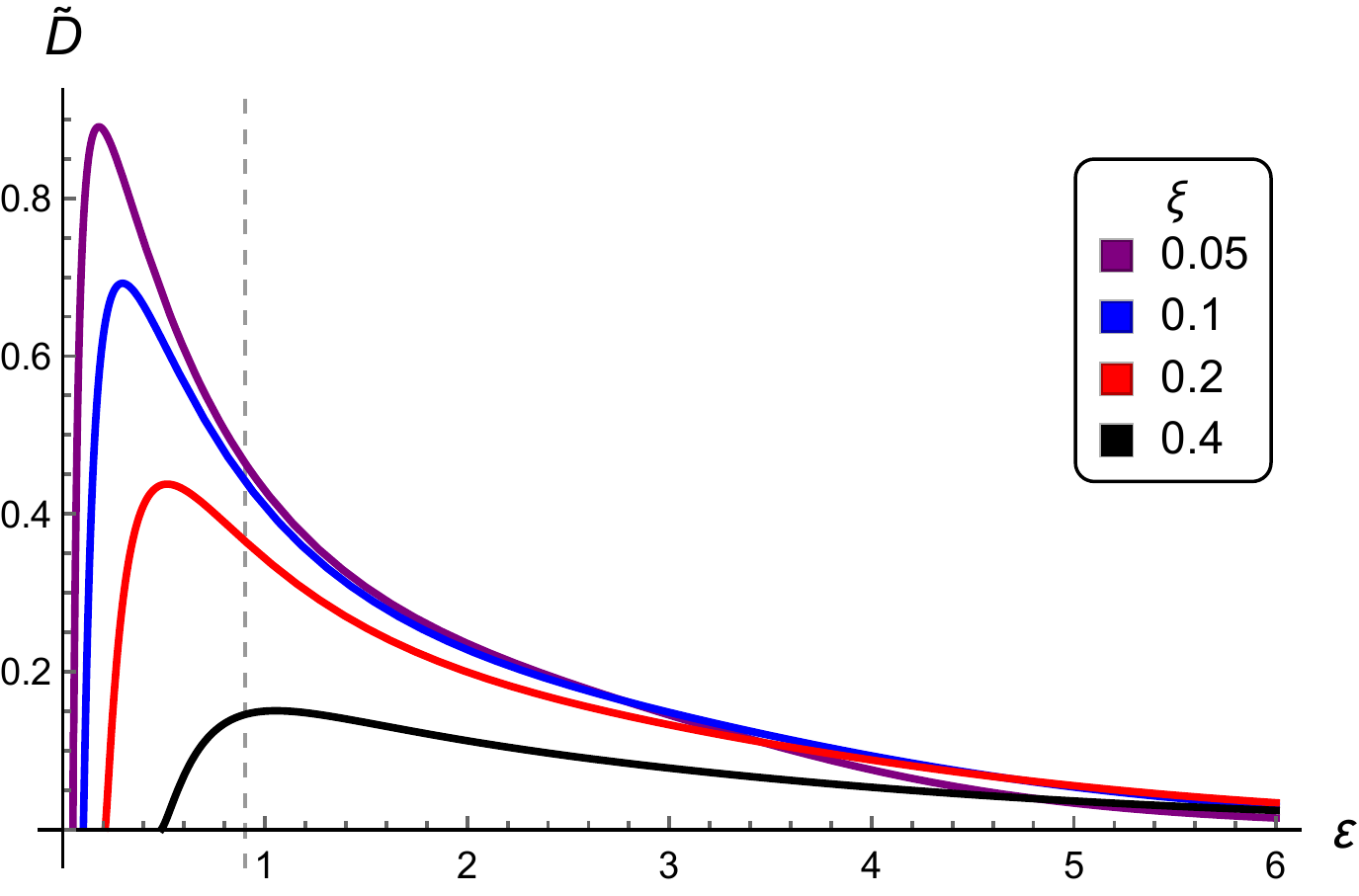}
\end{minipage}
}
\caption{(\emph{Left}) The dimensionless energy $\tilde{D}(\xi,\varepsilon)$ of the sphaleron in the right potential of figure~\ref{fig:regpot} plotted as a function of $\xi$ for the indicated values of $\varepsilon$. (\emph{Right}) The same dimensionless energy plotted as a function of $\varepsilon$ for the indicated values of $\xi$. Vertical line indicates the phenomenological value $\varepsilon\simeq 0.9$.}
\label{fig:D3}
\end{figure}

The classical stability requires computation of the energy on the sphaleron solution, which in this case corresponds to a solution of equation
\be
\frac{d\kappa}{ds} \ = \ -\sqrt{2V(\kappa)-2V(0)}\,,
\ee
with $\kappa(\pm\infty)=0$ and $\kappa(0)=\tilde\kappa$, where $\tilde\kappa$ is a root of equation
\be
V(\tilde\kappa) - V(0) \ = \ 0\,,
\ee
in terms of potential~(\ref{regpot}). The root we need satisfies condition $V'(\tilde\kappa)<0$. Expressing the energy in terms of a dimensionless function of dimensionless parameters,
\be
\label{SphEn3}
E_{\rm sph} \ = \ \sqrt{\lambda}\kappa_0^3\tilde{D}(\xi,\varepsilon)\,,
\ee
its dependence on the parameters $\xi=\kappa_1/\kappa_0$ and $\varepsilon=\epsilon/\kappa_0$ can be determined numerically (figure~\ref{fig:D3}). Since the minimum at $\kappa=0$ disappears for $\epsilon\to\xi+0$ (right plot), or when $\xi^2\to \varepsilon^2/(1+2\varepsilon^2)-0$ (left plot), the probability of the decay goes to one in this limit. The minimum is also unstable in the limit $\epsilon\to\infty$, when it becomes very shallow. This is the same behavior, as in equations~(\ref{DGeps2inf}) of the previous section.

There is an optimal value of $\epsilon$ at which $\tilde{D}(\xi,\epsilon)$ has a maximum and the metastable state enjoys maximum stability. In other words, an appropriate regularization makes the minimum at $\kappa=0$ more stable. The absolute maximum of stability is reached at $\xi=0$, at which the two vacua are degenerate and the sphaleron is a stable kink interpolating between them.

\begin{figure}[htb]
\begin{minipage}{0.45\linewidth}
 \includegraphics[width=\linewidth]{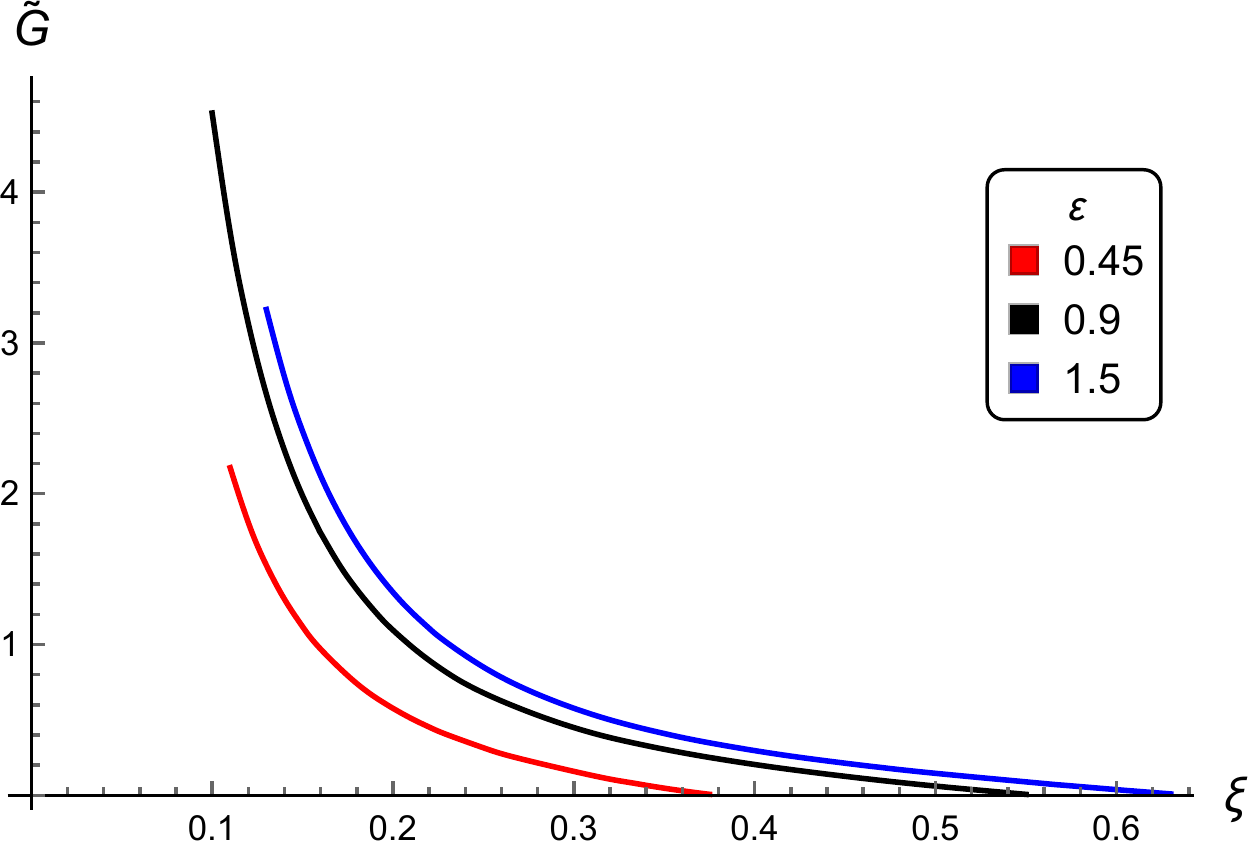}
\end{minipage}
\hfill{
\begin{minipage}{0.45\linewidth}
 \includegraphics[width=\linewidth]{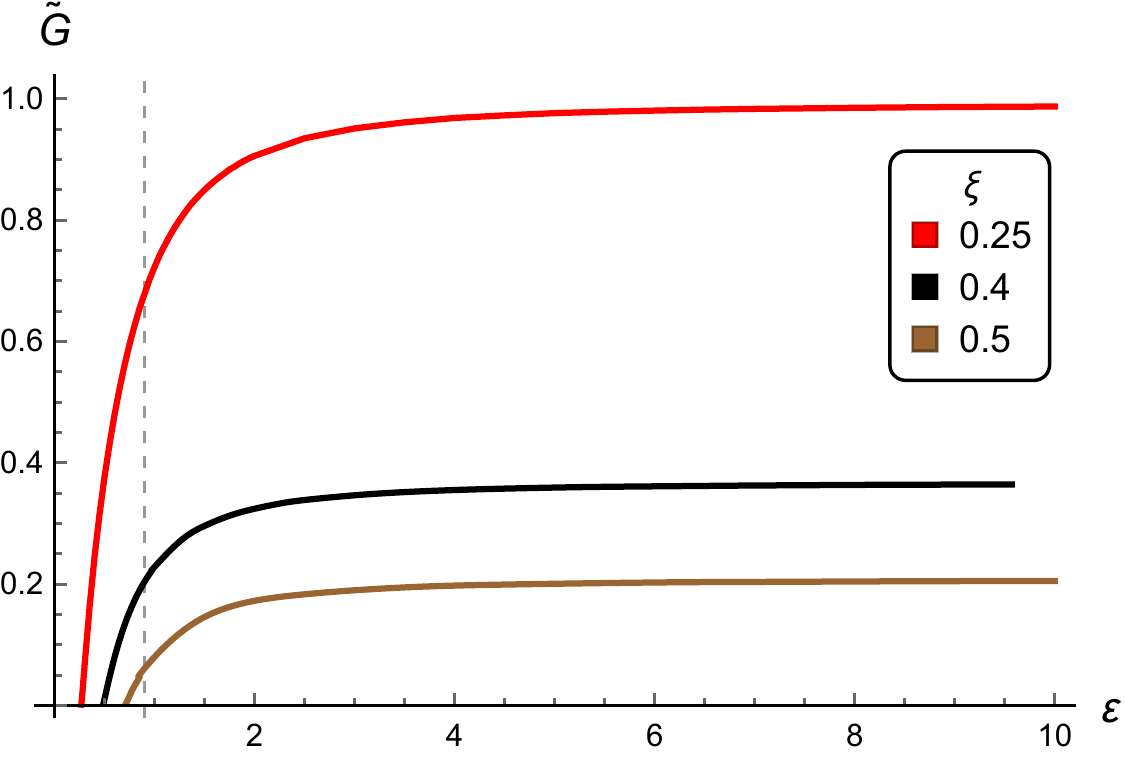}
\end{minipage}
}
\caption{Plots of the dimensionless Euclidean action $\tilde{G}(\xi,\varepsilon)$ defining the tunneling probability of the metastable state in the right potential of figure~\ref{fig:regpot}. Vertical line indicates the phenomenological value $\varepsilon\simeq 0.9$. For $\varepsilon<0.1$ the numerical analysis becomes subtle and demands very high accuracy.}
\label{fig:F3}
\end{figure}

The quantum decay of the metastable vacuum is controlled by the Euclidean action computed on the bounce solution of equations of motion~(\ref{EucEoM}) with boundary conditions
\be
\kappa(r\to\infty) = 0\,, \qquad \text{and} \qquad \frac{d}{dr}\kappa(r=0)\ = \ 0\,.
\ee
As in the previous subsection, the Euclidean action can be expressed in terms of a dimensionless function of parameters $\xi$ and $\varepsilon$,
\be
\label{EucA3}
S_{\rm E} \ = \ 2\pi\kappa_0^2 \tilde{G}(\xi,\varepsilon)\,.
\ee
Function $\tilde{G}$ can be determined numerically. Its behavior is illustrated by figure~\ref{fig:F3}. As expected, the probability of quantum tunneling becomes unity, when $\epsilon\to \xi+0$ or  $\xi^2\to \varepsilon^2/(1+2\varepsilon^2)-0$, because the unstable vacuum disappears in this limit. Again, in contrast with the classical case, in the limit $\epsilon\to\infty$, the quantum tunneling probability saturates at a constant rate. As explained around equation~(\ref{DGeps2inf}), this is expected, because $\lambda$ scales out from equation~(\ref{EucA3}).

We note that our stability analysis was based on the translational invariance of the system, that is it assumed infinite and smooth curves. In the case of finite curves and, even more so, discrete curves (polygons), the translational symmetry is broken. This potentially stabilizes the weakly unstable solutions. Hence one can expect all of them to be present in a more natural setup under some favorable conditions. We leave the detailed study of the translational symmetry breaking effects for a future work.

%%%%%%%%%%%%%%%%%%%%%%%%%%%%%%%%%%%%%%%%%%%%%%%%%%%%%%%%%%%%%%%%%%%%%%%%%

\section{From curves to proteins}
\label{sec:proteins}

In this section we would like to put the features of the ACSH model discussed above in the context of real protein molecules. We can both check the compatibility of the model with the real proteins as well as derive numerical estimates for the quantities studied theoretically. The first step is to quantify the minimum energy solutions (helices) described in sections~\ref{sec:GC} and~\ref{sec:sols}.

\subsection{Fitting the proteins}

\begin{figure}[htb]
	\begin{minipage}{0.45\linewidth}
		\includegraphics[width=\linewidth]{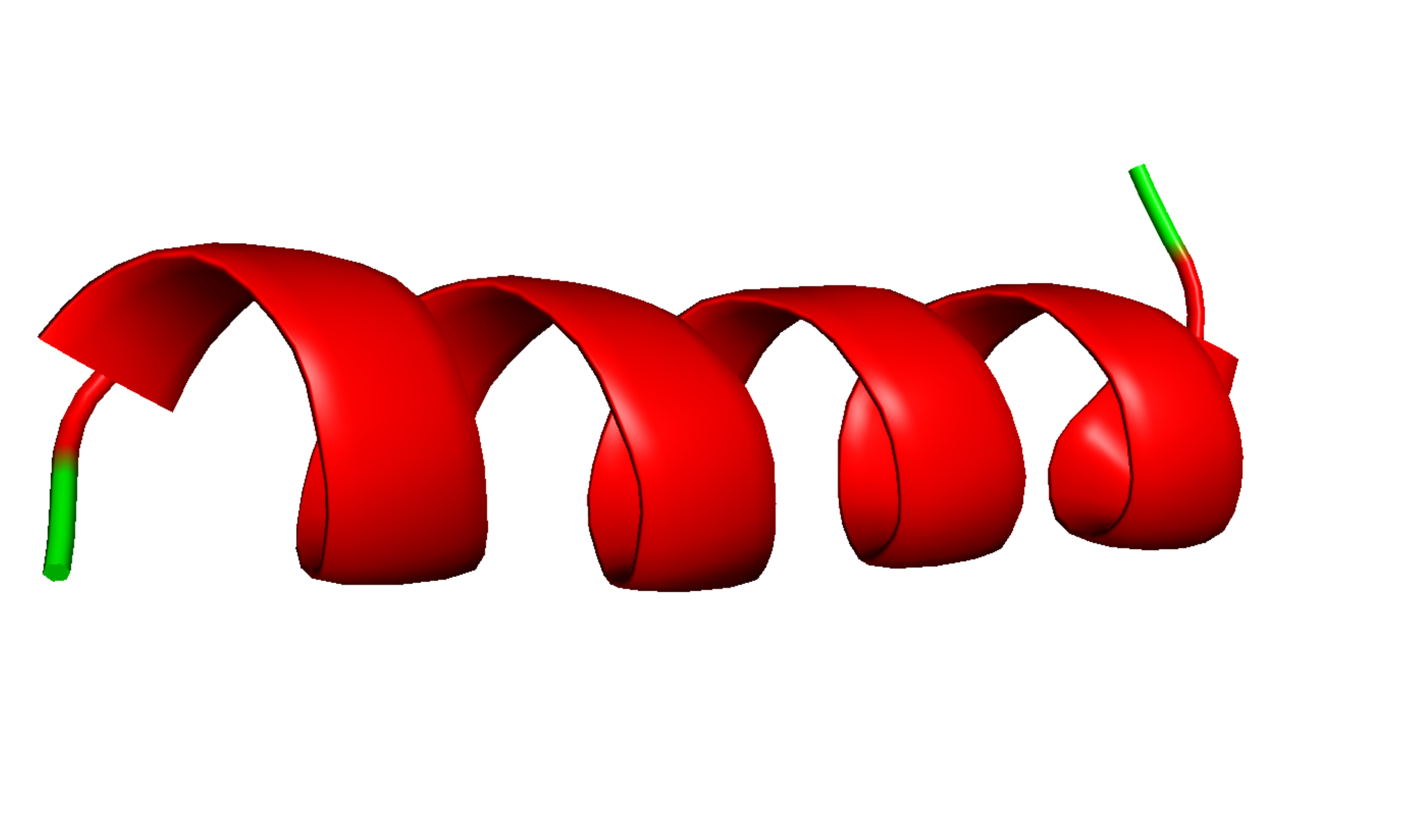}
	\end{minipage}
	\hfill{
		\begin{minipage}{0.45\linewidth}
			\includegraphics[width=\linewidth]{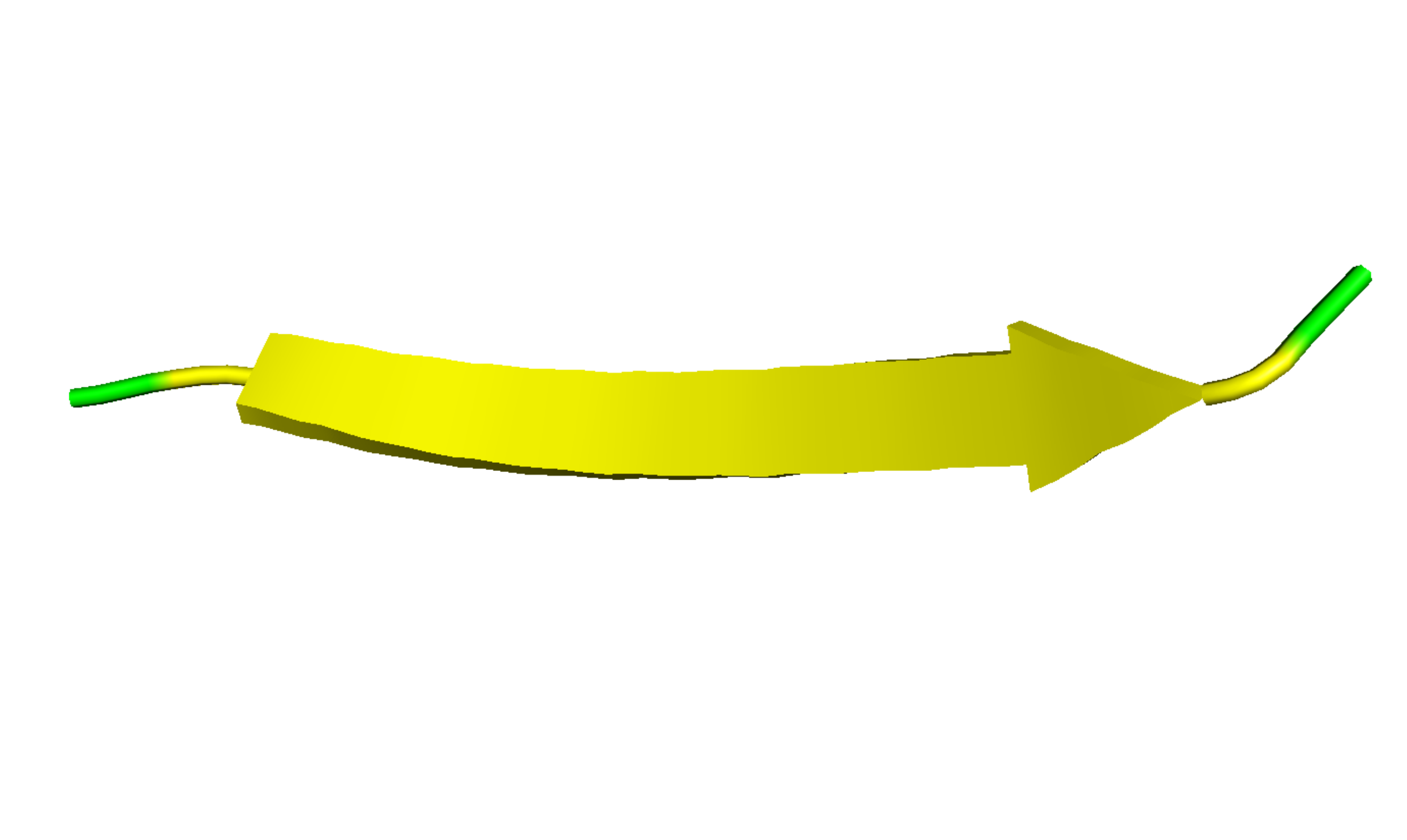}
		\end{minipage}
	}
	\caption{An alpha helix of the myoglobin molecule \emph{1am6} (left) and a beta strand of bucandin \emph{1f94} (right). Images generated using the PyMOL software~\cite{pymol}.}
	\label{fig:helix}
\end{figure}

On the ``secondary'' structure level protein is roughly a sequence of alpha helices and beta strands (see figure~\ref{fig:helix}) connected by structural motifs (loops and turns). While alpha helices have a clear helical structure, it is not so obvious in the case of beta strands. However, as figure~\ref{fig:helix} also suggests, beta strands are commonly viewed as ribbons with some amount of torsion. The orientation of the ribbon in the standard visualization typically follows the orientation of the peptide planes of the amino acids, so that the ribbon is a ``framed version" of the protein backbone curve, corresponding to the peptide framing~\cite{Melnikov:2019len}.  Here we would like to propose to view the beta strands, and more precisely, their backbone curves as pieces of helices~(\ref{xhelix})-(\ref{zhelix}) in the regime $\gamma \gg 1$ (what we called beta-helices on figure~\ref{fig:helices}).  The compatibility of this assumption can be tested  by fitting parameters $\kappa_{0}$ and $\tau_{0}$ in equations like~(\ref{torsion}), or its regularized form~(\ref{regtorsion}).

\begin{figure}[htb]
	\begin{minipage}{0.45\linewidth}
		\includegraphics[width=\linewidth]{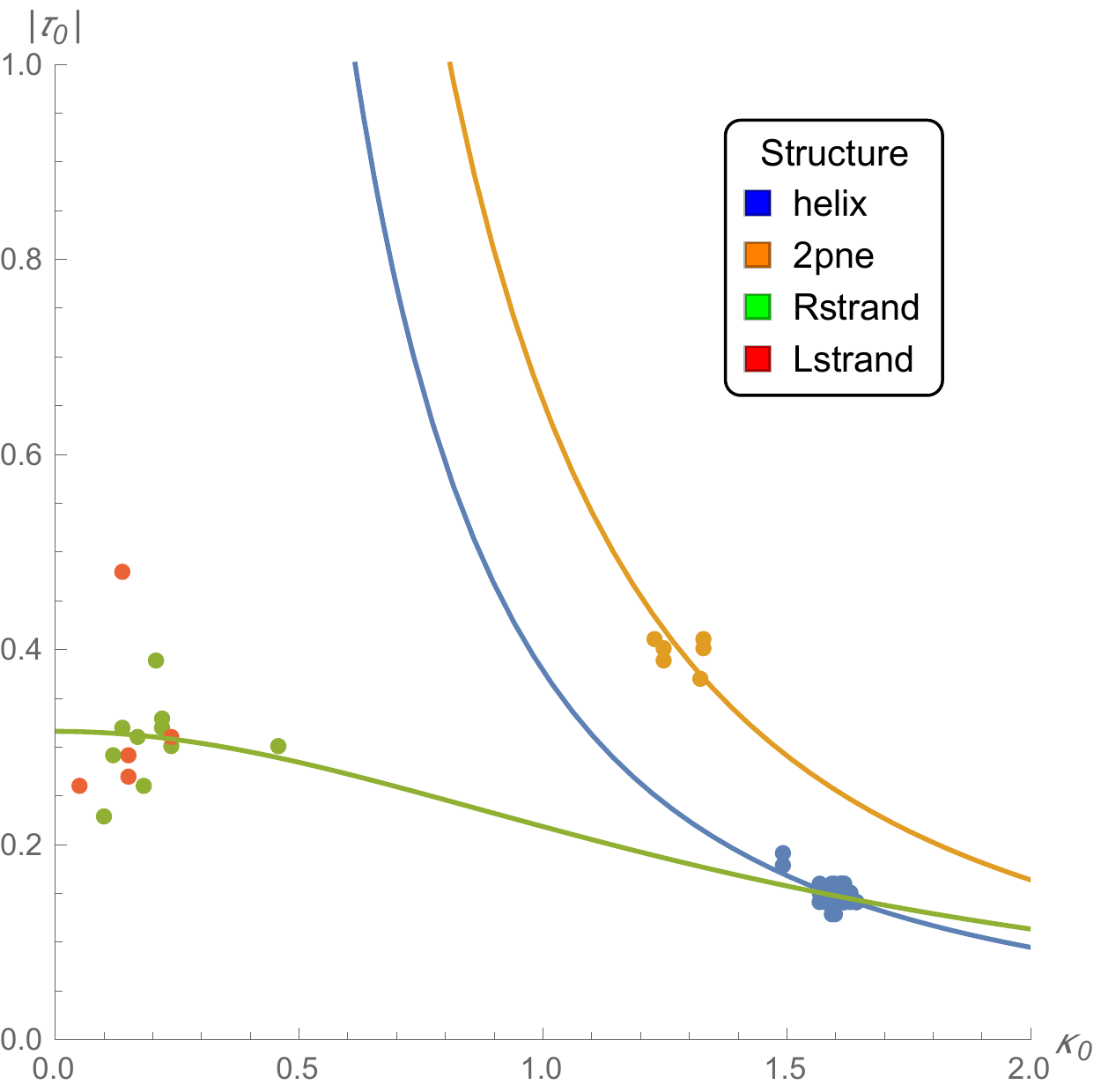}
	\end{minipage}
	\hfill{
		\begin{minipage}{0.45\linewidth}
			\includegraphics[width=\linewidth,trim=0pt 0pt 0pt 0pt]{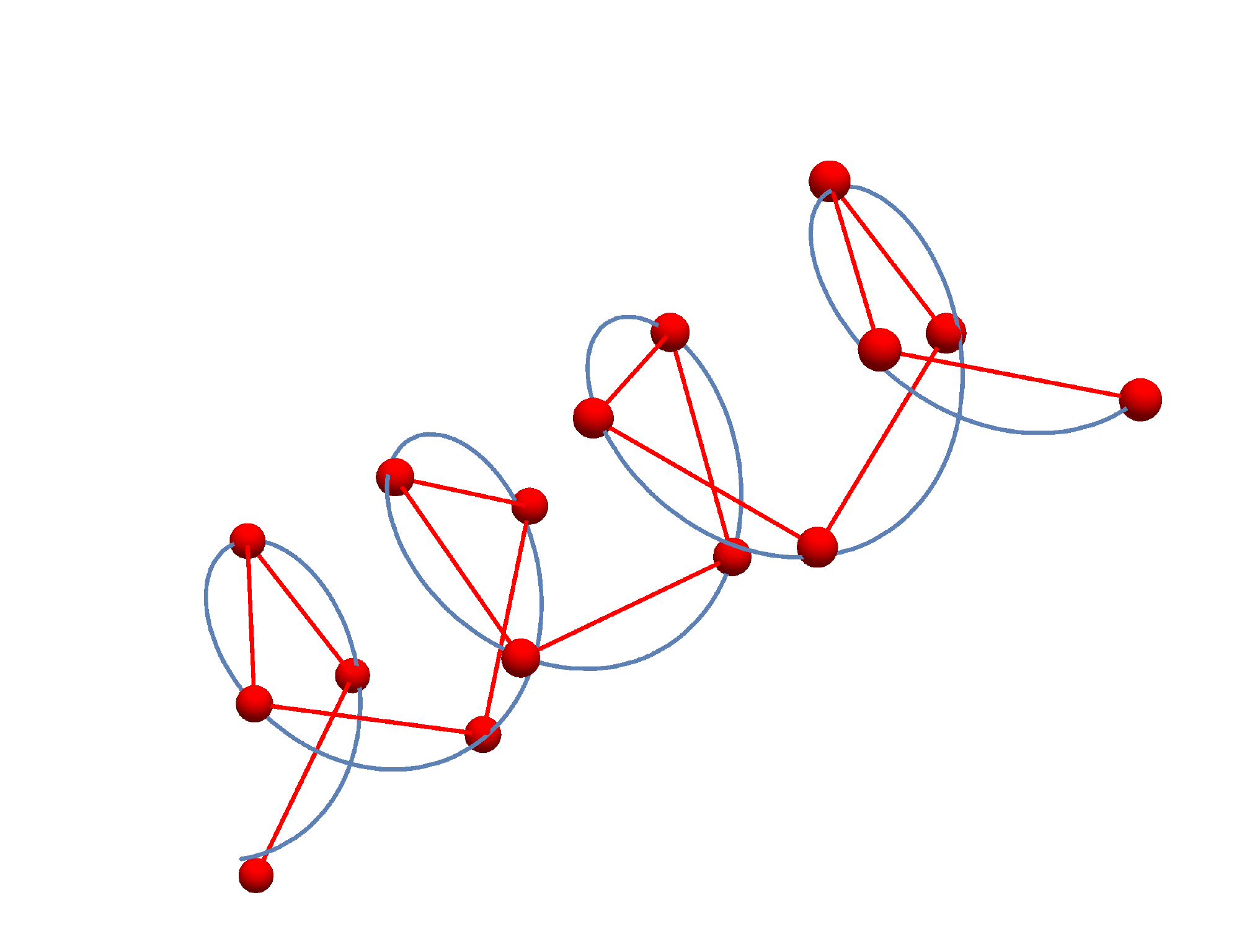}
		\end{minipage}
	}
	\caption{(\emph{Left}) Statistics of the pairs $(\kappa_0,|\tau_0|)$ of the alpha helices (blue) and beta strands (red and green) from tables~\ref{tab:params} and~\ref{tab:params1} respectively. Green and red points correspond to positive and negative torsion strands respectively. Orange points correspond to the case of the \emph{2pne} protein, whose helical structure is half way between alpha helices and beta strands. The blue line is a fit of relation~(\ref{torsion}) considering only helices. The light green line is a fit of relation~(\ref{regtorsion}), with beta strand points included. (\emph{Right}) A 3D fit of an alpha helix of the zinc protease \emph{1c7k}.}
	\label{fig:ktcorrelation}
\end{figure}

Using the Protein Data Bank (PDB) data~\cite{pdb} we analyzed $60$ alpha helices in 19 protein molecules and $21$ beta strands in $9$ proteins, for which we extracted estimates of values of $\kappa_0$ and $\tau_0$. The results are summarized in tables~\ref{tab:params} and~\ref{tab:params1}, and plotted on figure~\ref{fig:ktcorrelation}. Let us give some details of how the data were obtained.

For the fits we used proteins, whose structure was measured using diffraction data with the resolution better than 1.0{\AA} and passed additional consistency checks~\cite{Hinsen:2013}. From the list of such proteins we randomly selected a number of examples, which visually appeared to have the most regular helical and strand structures. In the case of the alpha helices we were selecting those which did not have apparent irregularities, such as helices ``broken" in the middle (that should rather be viewed as several helices connected together), or short helices with one or two loops. Our definition of the helices did not fully match the identification of the PDB files, also if a helix had a ``loose end'' (an apparently non-helical piece). Similarly, in the case of the beta strands, we selected those strands that could be thought as of a straight but twisted ribbon.

Given a set of $C\alpha$ carbon atoms, organized in the form of a selected helix, we fit their coordinates with an equation of a helix in a generic parameterization
\be
\label{genparam}
x(t) \ = \ x_0 + A\cos(\omega t + \varphi)\,, \qquad y(t) \ = \ y_0 + A\sin(\omega t + \varphi)\,, \qquad z(t) \ = \ z_0 + v t\,,
\ee
with the two additional fit parameters given by the orientation angles of the axis of the helix in the coordinate frame of the PDB file. Note that equations~(\ref{xhelix})-(\ref{zhelix}) above correspond to a helix in the special parameterization ${x(s),y(s),z(s)}$, in terms of the arclength parameter $s$. In general parametrization~(\ref{genparam}), parameters $A$, $\omega$ and $v$ determine $\kappa_0$ and $\tau_0$ through
\be
\kappa_0 \ = \ \frac{A\omega^2}{\sqrt{v^2+\omega^2A^2}}\,, \qquad \tau_0 \ = \ -\, \frac{v\omega}{v^2+\omega^2A^2}\,.
\ee
An example of the result of a fit of an alpha helix is shown on figure~\ref{fig:ktcorrelation} (right). Fit values of $\kappa_0$ and $\tau_0$ are summarized in table~\ref{tab:params}, where we also compute the value of the parameter $F$ assuming relation~(\ref{torsion}). The alpha helix values of the curvature and torsion appear as blue dots on figure~\ref{fig:ktcorrelation} (left).

\begin{table}[H]
	\begin{subtable}{.55\textwidth}
		
	\begin{tabular}{||c|c|c|c|c||}
			\hline 
			PDB code & helix & $\kappa_{0}$ & $\tau_{0}$ & $F$ \\ 
			\hline
			1a6m & $4$ & $1.61$ & $0.16$ & $0.40$ \\
		\hline 
		1a6m & $6$ & $1.64$ & $0.14$ & $0.39$ \\ 
		\hline 
		1a6m & $8$ & $1.61$ & $0.15$ & $0.39$ \\
		\hline 
		1i1w & $3$ & $1.59$ & $0.14$ & $0.34$ \\
		\hline 
		1i1w & $5$ & $1.59$ & $0.14$ & $0.36$ \\ 
		\hline 
		1i1w & $7$ & $1.57$ & $0.15$ & $0.36$ \\ 
		\hline  
		1i1w & $9$ & $1.59$ & $0.13$ & $0.35$ \\ 
		\hline 
		1i1w & $10$ & $1.60$ & $0.14$ & $0.36$ \\ 
		\hline
		1i1w & $12$ & $1.59$ & $0.14$ & $0.36$ \\
		\hline 
		1i1w & $15$ & $1.62$ & $0.14$ & $0.37$ \\
		\hline
		1c7k & $2$ & $1.60$ & $0.15$ & $0.37$ \\ 
		\hline 
		1c7k & $4$ & $1.63$ & $0.15$ & $0.39$ \\ 
		\hline 
		1c7k & $6$ & $1.61$ & $0.15$ & $0.38$ \\
		\hline
		1exr & $1$ & $1.63$ & $0.14$ & $0.38$ \\ 
		\hline 
		1exr & $5$ & $1.59$ & $0.15$ & $0.37$ \\ 
		\hline 
		1g66 & $2$ & $1.57$ & $0.16$ & $0.39$ \\ 
		\hline 
		1g66 & $4$ & $1.58$ & $0.14$ & $0.36$ \\ 
		\hline 
		1g66 & $5$ & $1.60$ & $0.16$ & $0.40$ \\ 
		\hline 
		1iee & $1$ & $1.60$ & $0.14$ & $0.36$ \\ 
		\hline 
		1iee & $2$ & $1.58$ & $0.15$ & $0.37$ \\ 
		\hline
		1iee & $4$ & $1.62$ & $0.14$ & $0.39$ \\
		\hline
		1cex & $5$ & $1.64$ & $0.14$ & $0.38$ \\ 
		\hline
		1cex & $6$ & $1.60$ & $0.15$ & $0.38$ \\ 
		\hline 
		1nwz & $3$ & $1.61$ & $0.15$ & $0.39$ \\ 
		\hline
		5o41 & $1$ & $1.63$ & $0.15$ & $0.41$ \\ 
		\hline 
		5o41 & $2$ & $1.59$ & $0.14$ & $0.37$ \\ 
		\hline
		5o41 & $5$ & $1.62$ & $0.15$ & $0.40$ \\ 
		\hline
		5o41 & $7$ & $1.49$ & $0.19$ & $0.43$ \\ 
		\hline
		5o41 & $8$ & $1.49$ & $0.18$ & $0.41$ \\ 
		\hline
		2ddx & $1$ & $1.58$ & $0.14$ & $0.36$ \\ 
		\hline 
		
		\end{tabular}
		
	\end{subtable}
	\vspace{0.1cm}  
	\begin{subtable}{.5\textwidth}
		
		\begin{tabular}{||c|c|c|c|c||}
			\hline 
			PDB code & helix & $\kappa_{0}$ & $\tau_{0}$ & $F$ \\ 
			\hline
		2ddx & $4$ & $1.60$ & $0.15$ & $0.37$ \\ 
		\hline		
		2ddx & $6$ & $1.59$ & $0.15$ & $0.37$ \\ 
		\hline
		2ddx & $10$ & $1.61$ & $0.14$ & $0.37$ \\ 
		\hline
		2ddx & $12$ & $1.64$ & $0.14$ & $0.37$ \\ 
		\hline
		2ddx & $15$ & $1.58$ & $0.14$ & $0.36$ \\ 
		\hline
		3ne0 & $1$ & $1.63$ & $0.15$ & $0.40$ \\ 
		\hline
		3ne0 & $3$ & $1.61$ & $0.16$ & $0.41$ \\ 
		\hline
		3a02 & $1$ & $1.62$ & $0.14$ & $0.38$ \\ 
		\hline
		3a02 & $2$ & $1.62$ & $0.14$ & $0.36$ \\ 
		\hline
		3a02 & $3$ & $1.60$ & $0.15$ & $0.38$ \\ 
		\hline 
		2p5k & $1$ & $1.64$ & $0.14$ & $0.38$ \\ 
		\hline 
		2p5k & $2$ & $1.61$ & $0.15$ & $0.40$ \\
		\hline 
		2p5k & $3$ & $1.61$ & $0.15$ & $0.39$ \\ 
		\hline
		2r31 & $6$ & $1.61$ & $0.14$ & $0.36$ \\ 
		\hline
		2r31 & $13$ & $1.62$ & $0.16$ & $0.42$ \\ 
		\hline
		2rbk & $1$ & $1.61$ & $0.16$ & $0.40$ \\ 
		\hline
		2rbk & $5$ & $1.61$ & $0.14$ & $0.38$ \\ 
		\hline
		1j0p & $4$ &  $1.59$ & $0.16$ & $0.39$ \\ 
		\hline
		2vxn & $1$ & $1.57$ & $0.14$ & $0.36$ \\ 
		\hline
		2vxn & $3$ & $1.60$ & $0.13$ & $0.33$ \\ 
		\hline
		2vxn & $6$ & $1.59$ & $0.14$ & $0.36$ \\ 
		\hline
		2vxn & $8$ & $1.58$ & $0.14$ & $0.36$ \\ 
		\hline
		2wfi & $3$ & $1.62$ & $0.16$ & $0.41$ \\ 
		\hline
		3bwh & $2$ & $1.62$ & $0.15$ & $0.39$ \\ 
		\hline
		3bwh & $5$ & $1.59$ & $0.14$ & $0.37$ \\ 
		\hline
		3bwh & $7$ & $1.62$ & $0.14$ & $0.37$ \\ 
		\hline
		3bwh & $8$ & $1.59$ & $0.15$ & $0.38$ \\ 
		\hline
		3bwh & $10$ & $1.61$ & $0.15$ & $0.39$ \\ 
		\hline
		3bwh & $11$ & $1.64$ & $0.14$ & $0.38$ \\ 
		\hline
		3bwh & $12$ & $1.58$ & $0.14$ & $0.35$ \\ 
		\hline
		\end{tabular}
		
	\end{subtable}
	\caption{The values of parameters $\kappa_{0}$ and $\tau_{0}$ for a selected set of helices in a set of proteins. The first column contains the PDB codes of the proteins. The second column contains the identification numbers of the helices from the pdb files. The last column corresponds to the flux parameter $F$ computed from equation~(\ref{torsion}) with $\kappa=\kappa_0$ and $\tau=\tau_0$. All the values are given in the inverse {\AA} units.}
	\label{tab:params}
\end{table}

Beta strands are harder to analyze since the positions of $C\alpha$ atoms do not obviously resemble a helix, but rather a zigzag. As already mentioned we would like to think of the zigzag as of a  twisted ribbon, as the one shown on figure~\ref{fig:helix} (right). To measure the curvature and torsion of the ribbon we fit the positions of the midpoints between the consecutive pairs of $C\alpha$. In other words, the zigzag structure implies that the angles between two consecutive peptide planes is small, or rather that the second plane is rotated with respect to the first by an angle close to $180^\circ$. Instead of looking at the peptide planes one can look at the rotation of ``hyper'' planes formed by three consecutive $C\alpha$ atoms, which we characterize by the chain of midpoints. 

\begin{figure}[htb]
	\begin{minipage}{0.45\linewidth}
		\includegraphics[width=\linewidth,trim=200pt 300pt 200pt 350pt]{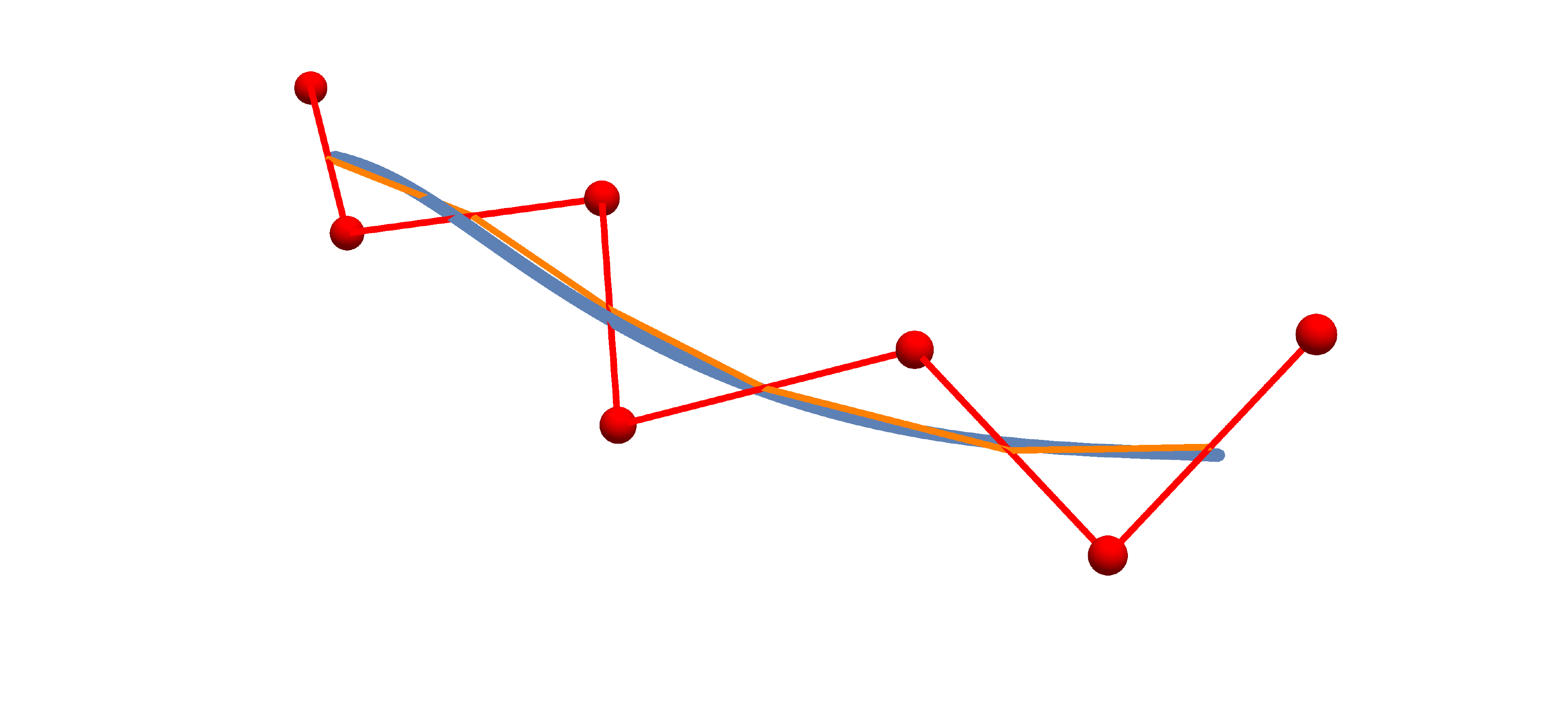}
	\end{minipage}
	\hfill{
		\begin{minipage}{0.45\linewidth}
			\includegraphics[width=\linewidth,trim=0pt 100pt 0pt 30pt]{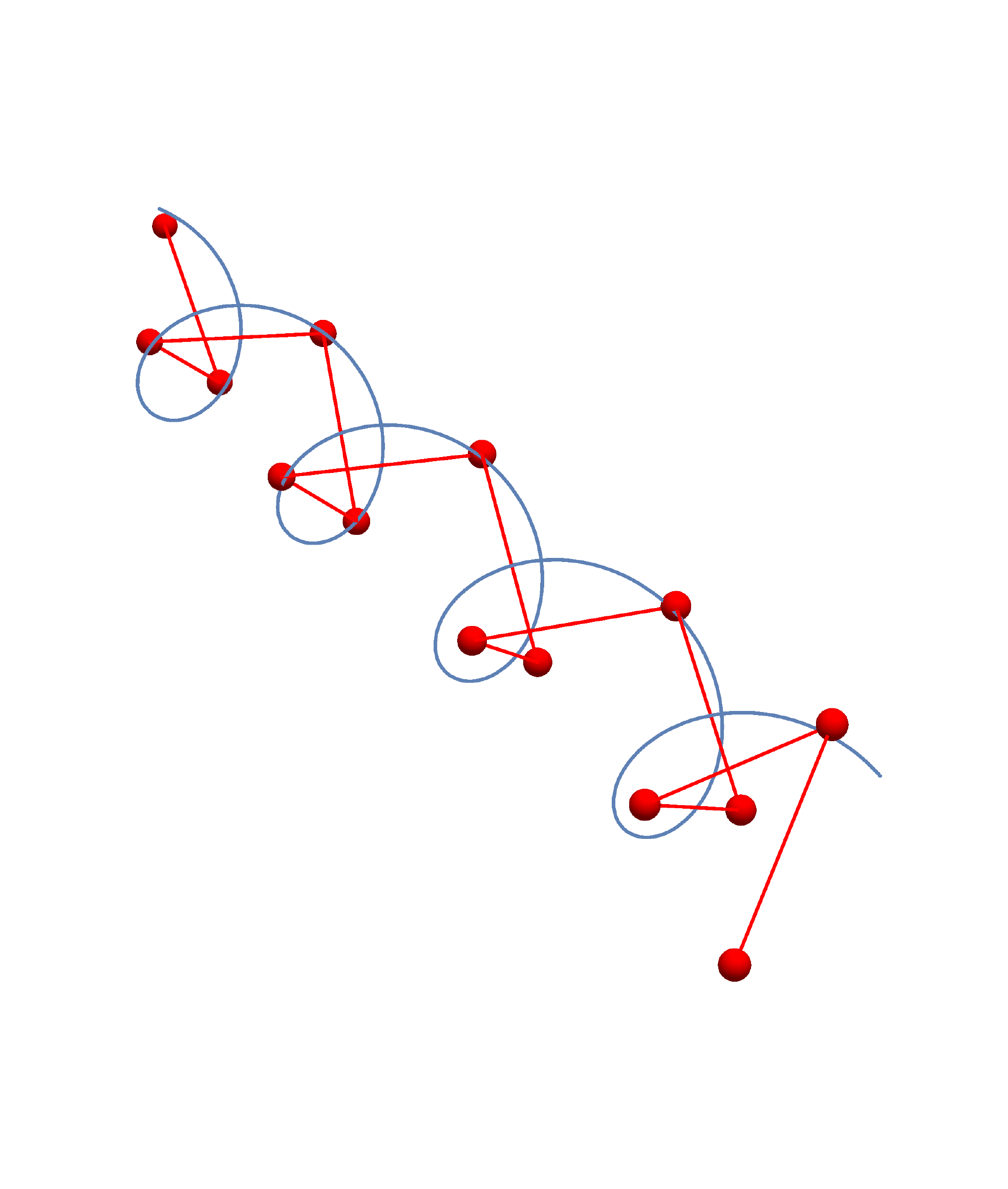}
		\end{minipage}
	}
	\caption{Examples of 3D fits of beta strands in bucandin \emph{1f94} (left) and in the snow flea antifreeze protein \emph{2pne} (right). Red dots show positions of the $C\alpha$ atoms, while blue curves are the fits by equations (\ref{genparam}). We note that the helices of the \emph{2pne} protein are classified as beta strands. In the case of the \emph{1f94} strand the midpoints between $C\alpha$ atoms, here connected by an orange polygon, are fitted.}
	\label{fig:fitplot}
\end{figure}

One can also find that many beta strands are not really ribbons of constant torsion (curvature). Such ribbons, from the point of view of the present studies, should be viewed as several ribbons with a constant torsion joined together. In the fits we only selected ribbons (or their pieces) that are suitable to be considered as twisted but straight ribbons, which could be approximated by beta helices. The results are shown in table~(\ref{tab:params1}) as green (positive torsion) and red (negative torsion) dots on figure~\ref{fig:ktcorrelation} (left). An example of a beta strand fit is shown on figure~\ref{fig:fitplot} (left).

An interesting example of beta strands, which are half way between alpha helices and beta strands, are found in the protein \emph{2pne}. Its strands are rather good, albeit stretched, helices, which can be fit with equations~(\ref{genparam}), in the same way as alpha helices. An example of such a fit is shown on figure~\ref{fig:fitplot} (right). The values of $\kappa_0$ and $\tau_0$ are for different strands in the \emph{2pne} molecules are contained in table~\ref{tab:params1} and plotted on figure~\ref{fig:ktcorrelation} (left) as orange points. 

\begin{table}[htb]
	
	\begin{center}
		\begin{tabular}{||c|c|c|c|c||}
			\hline 
			PDB code & strand & $\kappa_{0}$ & $\tau_{0}$ & $F$ \\ 
			\hline
		2pne	 & $1$ & $1.25$ & $-0.40$ & $-0.62$ \\ 
		\hline
		2pne	 & $2$ & $1.33$ & $-0.41$ & $-0.72$ \\ 
		\hline
		2pne	 & $3$ & $1.25$ & $-0.39$ & $-0.62$ \\ 
		\hline
		2pne	 & $4$ & $1.33$ & $-0.40$ & $-0.70$ \\ 
		\hline
		2pne	 & $5$ & $1.23$ & $-0.41$ & $-0.62$ \\ 
		\hline
		2pne 	& $6$ & $1.32$ & $-0.37$ & $-0.65$ \\
		\hline
		1f94 & 2A & $0.46$ & $0.30$ & $0.063$ \\
		\hline
		1f94  & 2B  & $0.17$ & $0.31$ & $0.0093$ \\
		\hline
		1f94 & 3B & $0.14$ & $-0.48$ & $-0.0100$ \\
		\hline
		2pnd & 3A & $0.051$ & $-0.26$ & $-0.00067$ \\
		\hline
		2ids & 2B & $0.15$ & $-0.29$ & $-0.0068$ \\
		\hline
		3ll2 & 1D & $0.24$ & $-0.31$ & $-0.019$ \\
		\hline 
		3ll2 & 4D & $0.22$ & $0.33$ & $0.016$ \\
		\hline	
		2e4t & 4B & $0.18$ & $0.26$ & $0.0083$ \\
		\hline	
		2e4t & 3B & $0.21$ & $0.39$ & $0.016$ \\
		\hline	
		2e4t & 3C & $0.14$ & $0.32$ & $0.0063$ \\
		\hline
		2fou &  6B & $0.15$ & $-0.27$ & $-0.0059$ \\
		\hline
		2fou &  8B & $0.24$ & $0.30$ & $0.017$ \\
		\hline
		3wvm & 3A & $0.12$ & $0.29$ & $0.0038$ \\
		\hline
		3wvm &  7A & $0.10$ & $0.23$ & $0.0025$ \\
		\hline
		1x8q &  5B & $0.22$ & $0.32$ & $0.015$ \\
		\hline
			\end{tabular}
	\end{center}
	\caption{The values of curvature $\kappa_{0}$, torsion $\tau_{0}$ and flux parameter $F$ for a set of beta strands selected to satisfy regularity conditions. As before, the first column contains the PDB codes, while the second column identifies the strands in the pdb file. All the values are given in the inverse {\AA} units.
	}
	\label{tab:params1}
\end{table}

If one ignores the special case of the \emph{2pne} protein, figure~\ref{fig:ktcorrelation} (left) supports the following picture. There are two regions populated by common types of alpha helices (blue points) and beta strands (green and red points). This structure is clearly just another presentation of the Ramachandran plot. However, from the point of view of the effective model, the figure contains some additional information about the parameters. One can see that the points can be fit with a relation of the form~(\ref{regtorsion}) assuming some universal mean values for parameters $F$ and $\epsilon$. Data on figure~\ref{fig:ktcorrelation} indeed support the claim that relation~(\ref{torsion}) is regularized in real proteins bounding the value of torsion for small curvature. The fits indicate to the following universal values,
\be
\label{FEpsEstimate}
F\simeq 0.70~\text{\AA}^{-1}\,, \qquad \epsilon\simeq 1.5~\text{\AA}^{-1}\,.
\ee
In other words, we can say that at the level of geometry the regularized ACSH model appears to be in the same universality class with the full microscopic theory of real protein molecules. At large scales the values of the parameters, which in general depend the chemical composition and beyond-nearest neighbor interactions, appear to be distributed around the above universal values. In the meantime, protein \emph{2pne} belongs to a different universality class.

We note that in our preliminary analysis  we have not performed a proper study of errors of our results. It is clear that for alpha helices and \emph{2pne} the errors are reasonably small. It is less obvious in the case of beta strands, but as the tables and plot~\ref{fig:ktcorrelation} (left) demonstrate, the values obtained for different strands are compatible. We leave a more systematic analysis for a future work.

%%%%%%%%%%%%%%%%%%%%%%%%%%%%%%%%

\subsection{Decay rates} 
\label{sec:lambda}

In this section we would like to estimate the parameters which characterize the decay rates of the unstable minima of the two-minimum potentials in the canonical ensemble approach. Consequently, we would like to estimate a characteristic temperature of the transition and compare the probabilities of the classical and quantum phase transitions. 

The classical stability is controlled by the classical energy of the sphaleron solutions given by equations~(\ref{dimless}), (\ref{SphEn2}) and~(\ref{SphEn3}) for different ranges of the model parameters. The leading order quantum stability is determined by the value of the Euclidean action in equations~(\ref{SEuclid}), (\ref{EucA2}) and~(\ref{EucA3}). These expressions depend on the curvature parameter $\kappa_0$ (which by convention is the curvature of the minimum $\kappa\neq 0$), on coupling parameter $\lambda$ as well as on $\kappa_1$, which controls the simple zero of the potential, and on the regulator $\epsilon$. 

From analyzing the statistics of the secondary structure of different molecules we were able to extract some ``universal" values of $\kappa_0$, $F$ and $\epsilon$ for the ACSH model. We will assume that $\epsilon$ and $F$ take the values from equation~\ref{FEpsEstimate}. Values of $\kappa_0$ can be taken from table~\ref{tab:params}. The mean value for the helices is $\kappa_0\simeq 1.60$~\AA$^{-1}$. In this section we will get estimates of $\kappa_1$ and $\lambda$. Note that the latter two parameters are not independent. They are related through equations
\be
\label{Frel1}
 F^2 \ =\ {\lambda}(\kappa_0^2+\epsilon^2)^2(\epsilon^2-\kappa_1^2)\,,
\ee
for potential~(\ref{regpot}) and relation 
\be
\label{Frel2}
 F^2 \ =\ {\lambda}(\kappa_0^2+\epsilon^2)^2(\kappa_1^2+\epsilon^2)\,,
\ee
for potential~(\ref{regpot2}). 

\paragraph{Local minimum at $\kappa\neq 0$.} We first consider the case of potential~(\ref{regpot2}) of figure~\ref{fig:regpot} (left), with local minimum at $\kappa\neq 0$.  In order to determine the values of model parameters $\lambda$ and $\kappa_{1}$ in such a potential, one should either actually measure the sphaleron, that is the critical bubble of the true vacuum inside the false one, or measure the transition temperature, which gives some information about the height of the potential barrier separating the two vacua. We have no such data available. Instead, we would like to make a prediction of the transition temperature.

We will assume, in this case, that the size of a sphaleron (or a critical bubble) is of the typical size of the loops connecting two helical pieces of a protein. One motivation for this is to think that finite size effects stabilize the sphaleron and the regions of smaller curvature, which characterize the loops, \emph{e.g.} figure~\ref{fig:plotks} (left), are the metastable sphalerons. More generally, this assumption will give a bound on the values of the parameters.

The radius of the sphaleron, in potential~(\ref{regpot2}) is controlled by combination
\be
\label{Rsph1}
R_{\rm sph} \ \simeq \ \frac{1}{\sqrt{\lambda}\kappa_0}\sqrt{\frac{1+\varepsilon^2}{1-\xi^2}}\,,
\ee
where, as before, $\varepsilon=\epsilon/\kappa_0$ and $\xi=\kappa_1/\kappa_0$. More precisely, equation~(\ref{Rsph1}) determines the size of the asymptotic region, where $\kappa\simeq \kappa_0[1-\exp(-|s|/R_{\rm sph})]$. In case of potential~(\ref{regpot2}) $2R_{\rm sph}$ is a good approximation for the size of the entire sphaleron.   

Since, in this model, $0\leq \xi < 1$, there is a lower bound on the sphaleron size. When $\kappa_1=\xi=0$, the two vacua of potential~(\ref{regpot2}) become degenerate, and the size of the sphaleron is at least
\be
\label{Rmin}
R_{\rm sph} \geq R_{\rm min} \ = \  \frac{\epsilon(\epsilon^2+\kappa_0^2)^{3/2}}{F\kappa_0^2}\,.
\ee
Smaller values of $R_{\rm sph}$ are only compatible with the other form of the potential, equation~(\ref{potential}), for which $\kappa_0$ is the global minimum. We note that in the degenerate potential with $\kappa_1=0$ the sphaleron becomes a stable kink, interpolating between the two degenerate minima. This  kink, is characterized by an asymptotic region of size~(\ref{Rmin}) close to $\kappa=\kappa_0$ and by another asymptotic region of size 
\be
\hat{R} \ \simeq \ \frac{\epsilon}{\sqrt{\lambda}\kappa_0^2}\,,
\ee
next to $\kappa=0$.  Using the above values of parameters the minimum value is $R_{\rm min}\simeq 9$~{\AA} and the corresponding maximum value of $\lambda=\lambda_{\rm max}\simeq 0.01$. The value of the second asymptotic region for $\kappa_1\to 0$ is $\hat{R}\simeq 6$~\AA. In general, we expect the sphalerons to be at least twice as large as $R_{\rm min}$. However, one can find examples among actual proteins with shorter loops. Thus, protein \emph{5o41} has a very short loop $L_{\rm loop}\simeq 11$~\AA, incompatible with the model potential~(\ref{regpot2}). 

For longer loops we give our estimates for $\kappa_1$ and $\lambda$ in table~\ref{tab:params2}. We consider loops in proteins \emph{1a6m}, \emph{2p5k}, \emph{1exr}, \emph{3a02} and \emph{5o41}. The size of the loop $L_{\rm loop}$ is calculated by summing  distances between neighboring $C\alpha$ atoms in the loop. Assuming that $\kappa_1$ is not much smaller than $\kappa_0$, so that $L_{\rm loop}\simeq 2R_{\rm sph}$ we evaluate $\kappa_1$ and $\lambda$ from equations~(\ref{Rsph1}) and~(\ref{Frel2}).

\begin{table}[htb]
	
	\begin{center}
		\begin{tabular}{||c|c|c|c|c|c|c|c|c|c||}
			\hline 
			PDB & helix 1 & helix 2 & $L_{\rm loop}$, \AA &  $\kappa_{1}$, \AA$^{-1}$  & $\lambda$ & $E_{\rm sph}$, \AA$^{-1}$ & $S_{\rm E}$ & $T$, K & $Br$\\ 
			\hline
			1a6m & $4$ & $5$ & $22.8$    & $0.78$ & $0.0074$ & 0.056 & 3.0 & 170 & 230\\
			\hline 
			2p5k & $1$ & $2$ & $19.0$   & $0.42$ & $0.0087$ & 0.124 & 11.2 & 370 & $10^9$\\
			\hline
			1exr & $4$ & $5$ & $27.7$  & $1.02$ & $0.0064$  & 0.024 & 1.31 & 70 & 10 \\ 
			\hline 
			3a02 & $1$ & $2$ & $19.0$  & $0.42$ & $0.0087$ & 0.124 & 11.2 & 370 & $10^9$ \\ 
			\hline 
			1g66 & $4$ & $5$ & $43.0$  & $1.33$ & $0.0052$ & 0.0036 & 0.260 & 10 & 2\\
			\hline
		\end{tabular}
	\end{center}
	\caption{The estimates of the values of $\kappa_1$ and $\lambda$ in case of potential~(\ref{regpot2}). We assume that $L_{\rm loop}\simeq 2R_{\rm sph}$ and use values~(\ref{FEpsEstimate}) and $\kappa_0\simeq 1.6$~\AA$^{-1}$. The obtained values are used to estimate energy $E_{\rm sph}$ of the sphaleron of potential~(\ref{regpot2}), the associated transition temperature $T$ and the Euclidean action $S_E$ controlling the probability of the quantum decay of the false vacuum. $Br$ computes the ratio of classical and quantum probabilities using equation~(\ref{Br}).
	}
	\label{tab:params2}
\end{table}

Although this choice of the potential is not quite compatible with the observed loop length, we can ignore short loops and estimate the probabilities of quantum and classical phase transitions. In table~\ref{tab:params2} we used the estimated values of $\kappa_1$ and $\lambda$ to find the sphaleron energy $E_{\rm sph}$ and the Euclidean action~$S_{\rm E}$. We can use the former as an estimate of the classical transition temperature $T\simeq E_{\rm sph}$ and the latter to estimate the ratio $Br$ of the classical and quantum transition probabilities at room temperature,
\be
\label{Br}
Br \ = \ \frac{\exp(-E_{\rm sph}/T_{\rm room})}{\exp(-2S_{\rm E}/\hbar)} \,.
\ee
Note that $E_{\rm sph}$ is computed in units of inverse angstroms and we need a correct conversion factor to get the value in more appropriate energy units. The secondary structures, such as alpha helices and beta strands are formed due to hydrogen bonds, whose characteristic energy is $5-6$~kcal/mol for an isolated bond and $0.5-1.5$~kcal/mol for a protein in solution~\cite{Sheu:2003}. To get sensible values for the transition temperature we  assume that 
\be
\label{conversion}
\Lambda \ = \ 1~\text{\AA}^{-1} \ \simeq \ 6\, \frac{\text{kcal}}{\text{mol}}\frac{1}{k_BN_A} \ \simeq \ 3000~\text{K}.
\ee
This conversion factor gives the estimates of $T$ and $Br$ in table~\ref{tab:params2}. With the exception of the longest loop, the quantum phase transition is consistently suppressed.

\paragraph{Local minimum at $\kappa=0$.}

A similar analysis can be performed in the case of potential~(\ref{regpot}) shown on figure~\ref{fig:regpot} (right). We would like to see whether it is more compatible with the observed loop length. We remind that the present choice implies the following bounds on the parameters: $0\leq \kappa_1^2\leq \epsilon^2\kappa_0^2/(\kappa_0^2+2\epsilon^2)$ and $\epsilon^2\geq\kappa_1^2$. This ensures that there is a global minimum at $\kappa=\kappa_0$ and the local one at $\kappa=0$. 

As we discuss in section~\ref{sec:sols} this potential has stable kink solutions, which are more natural to associate with the loops connecting alpha helices. The kinks are solutions interpolating between minima $\kappa_0$ and $-\kappa_0$, so they have a scale, similar to $R_{\rm sph}$ above, which characterize the asymptotic behavior close to the minima,
\be
\label{Rkink}
R_{\rm kink} \ \simeq \ \frac{1}{\sqrt{\lambda}\kappa_0}\sqrt{\frac{1+\varepsilon^2}{1+\xi^2}}\,.
\ee

As in the previous case, the set of equations is supplemented with relation~(\ref{Frel1}), allowing to determine $\kappa_1$ and $\lambda$. Since potential~(\ref{regpot}) is an analytic continuation of the previous potential~(\ref{regpot2}) to imaginary values of $\kappa_1$ and $F$, the scale $R_{\rm min}$ is now the maximal possible value for $R_{\rm kink}$, $R_{\rm kink}\leq R_{\min}$. On the other hand, one can also find longer loops among real proteins, violating $L_{\rm loop}\leq 2R_{\rm min}$. This is not a problem, however, because the size of the kinks is also controlled by other scales, as can be seen from figure~\ref{fig:RegSoliton}. There is a second asymptotic region, close to $\kappa\sim 0$, which appears when $\kappa_1$ is sufficiently small, and the two minima are nearly degenerate. We can also make estimates of the size of that region.

In the regime $\kappa_1=0$ and $\epsilon\gg \kappa_0$ the soliton is simply the phi-four kink. Let us assume $0\simeq\kappa_1\ll \kappa_0\simeq \epsilon$. In this case, there is a new kink in the center of the standard kink. The size of the new kink is composed of the approximately linear core, $\kappa\ll \kappa_1$, and the crossover region, $\kappa_1\ll \kappa\ll \kappa_0$. The core radius can be approximated by
\be
R_{\rm core}\ \simeq \ \int\limits_{0}^{\kappa_1} \frac{dk}{\sqrt{\lambda}\kappa_0^2}\frac{\epsilon}{\sqrt{k^2+\kappa_1^2}} \simeq \frac{\epsilon}{\sqrt{\lambda}\kappa_0^2}\simeq  \frac{\epsilon^2(\kappa_0^2+\epsilon^2)}{F\kappa_0^2}\,.
\ee 
where the upper limit of integration is chosen to be $\kappa_1$, where the approximation breaks down. Interestingly this upper bound is independent from $\kappa_1$. The numerical value of the core size is $R_{\rm core}\simeq 6.0~\text{\AA}$.

The crossover length can be estimated as
\be
R_{\rm cross}\ \simeq \ \int\limits_{\kappa_1}^{\kappa_0} \frac{dk}{\sqrt{\lambda}\kappa_0^2}\frac{\epsilon}{k} = \frac{\epsilon}{\sqrt{\lambda}\kappa_0^2}\log\frac{\kappa_0}{\kappa_1}\,.
\ee
This scale diverges for $\kappa_1\to 0$, but rather slowly. This logarithmic scaling can also be seen on figure~\ref{fig:RegSoliton}.

\begin{figure}[htb]
\begin{minipage}{0.45\linewidth}
\includegraphics[width=\linewidth]{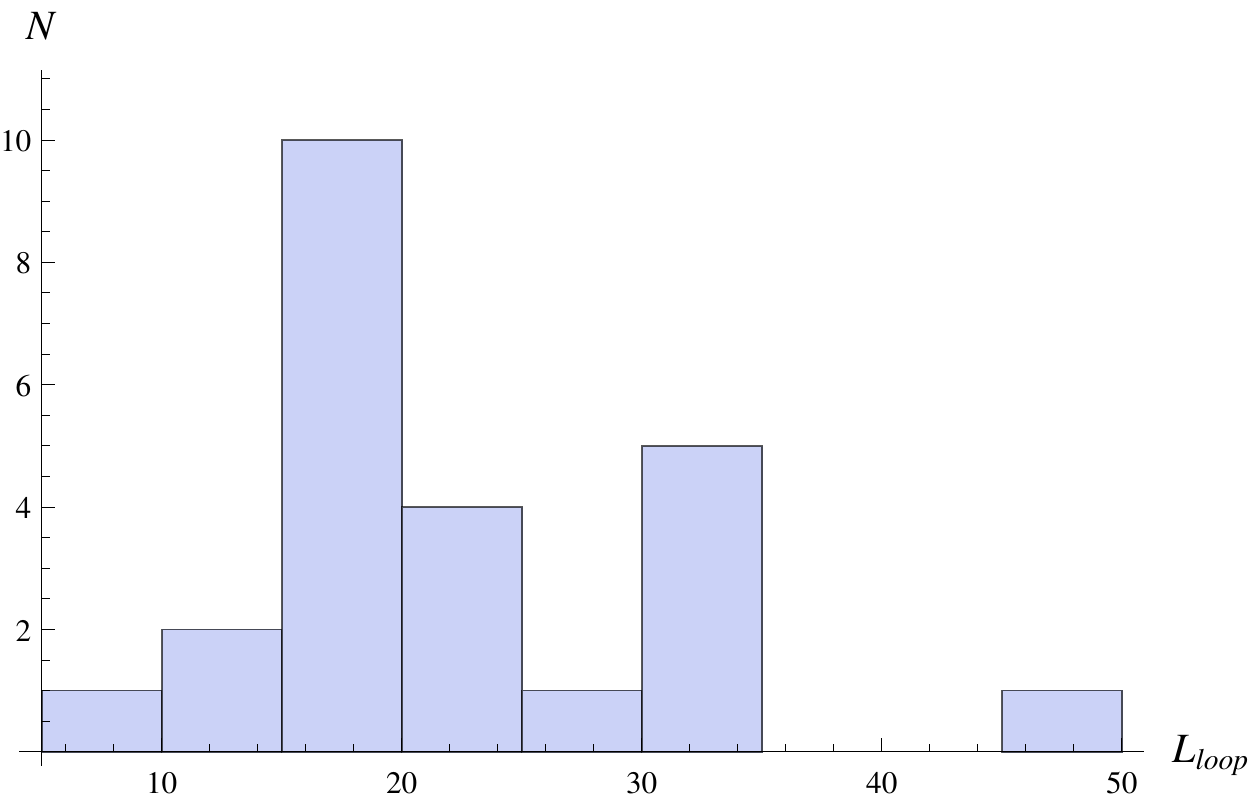}
\end{minipage}
\hfill{
\begin{minipage}{0.45\linewidth}
\includegraphics[width=\linewidth, trim=0pt 100pt 0pt 100pt]{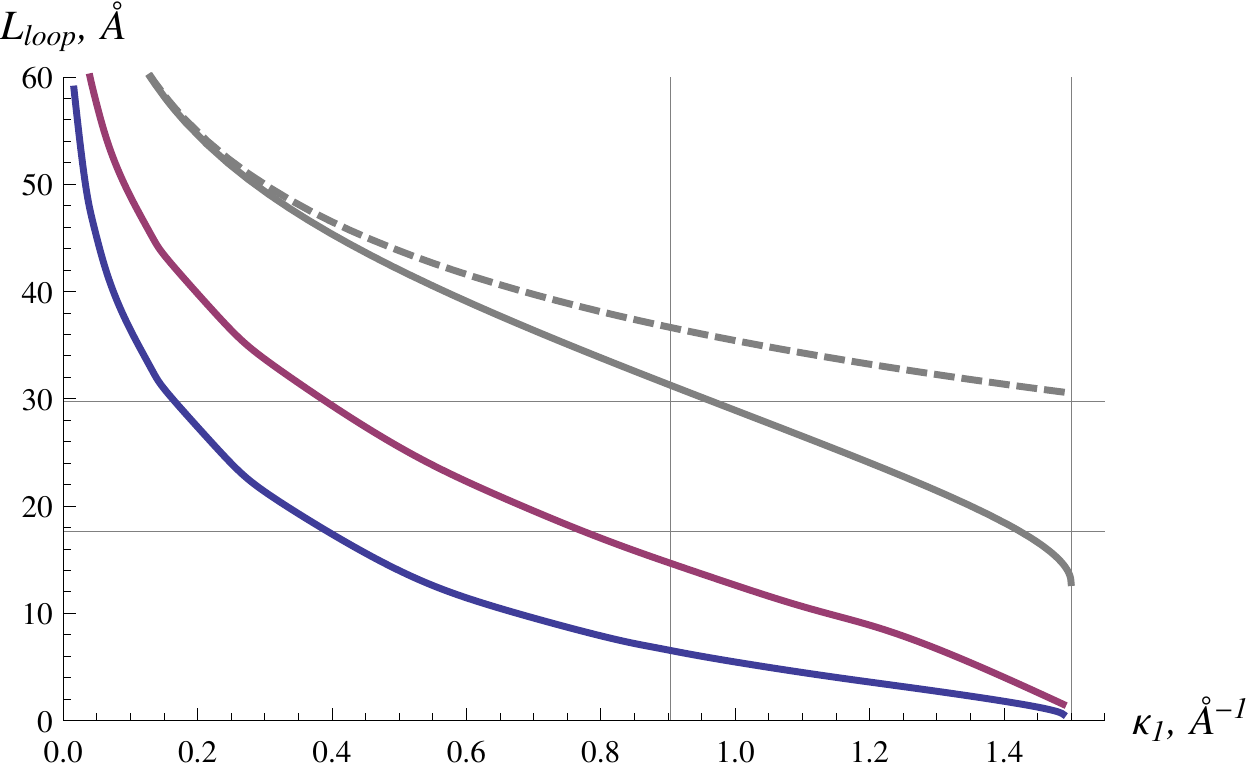}
\end{minipage}
}
 \caption{(\emph{Left}) Distribution of loop lengths $L_{\rm loop}\leq 50$~{\AA} for a selected list of about hundred proteins. (\emph{Right}) Dependence of the size of the loop connecting two helical vacua on the value of $\kappa_1$. The size is determined numerically as the locus of 50\% (blue) or 90\% (magenta) of the total energy of the kink. The gray lines show the approximate upper bound estimated by equation~(\ref{Rmax}) (dashed line) and a corrected expression, which is cut at $\kappa_1=\epsilon$. The first vertical line separates the cases of potential with a single minimum (right) and double minimum (left). The second vertical line shows the position of the point $\kappa_1=\epsilon$. Horizontal lines show the values of $2R_{\rm min}$ and constant part of $(\ref{Rmax})$.}
 \label{fig:RHelDistrib}
\end{figure}

Hence we can estimate the length of the loop in this model as
\be
\label{Rmax}
L_{\rm loop} = 2R_{\min} + 2R_{\rm core} + 2R_{\rm cross}= \frac{2\epsilon^2(\kappa_0^2+\epsilon^2)}{F\kappa_0^2}\left(1+\sqrt{1+\varepsilon^{-2}} +\log\frac{\kappa_0}{\kappa_1}\right)\simeq 30 + 12\log\frac{\kappa_0}{\kappa_1}~\text{\AA} .
\ee
This expression can be viewed as an upper bound for the length, which is reasonably consistent with the observations. On figure~\ref{fig:RHelDistrib} (left) we show the distribution of the lengths of the loop-like structures connecting alpha helices, which are shorter than approximately 50~{\AA}. For 24 identified structures, satisfying this condition, we observe a peak around $L_{\rm loop}\simeq 20~\text{\AA}$. Figure~\ref{fig:RHelDistrib} (right) shows the results of the numerical studies of the loop length for given values of the parameter $\kappa_1$. The blue curve shows the length of the kink segment bearing a half of its energy, while the magenta curve corresponds to the 90\% energy segment. Gray dashed curve is the upper bound~(\ref{Rmax}) and its version (solid curve) improved by exact relation~(\ref{Rkink}), rather than simply $R_{\min}$ in that equation. The latter modification  cuts the function at $\kappa_1\to\epsilon$, since there are no kinks beyond that limit.

From the data on the loop sizes we can estimate the values of parameters $\kappa_1$ and $\lambda$. In table~\ref{tab:params4} we calculate those values for a few representative loop sizes. It is interesting to note that while $\kappa_1$ and the loop length can vary by a factor $3-4$, the variation in $\lambda$  is rather small. Its value is close to the value at $\kappa_1=0$,  
\be
\lambda \ \simeq \ \frac{F^2}{\epsilon^2(\kappa_0^2+\epsilon^2)^2} \ \simeq \ 0.01\,.
\ee
With the calculated values we go ahead and estimate the temperature of the transition destroying the unstable minimum $\kappa=0$ (table~\ref{tab:params4}). For assumed conversion factor~(\ref{conversion}) we get quite reasonable results for the transition temperatures. In particular, these temperatures are higher than those obtained in the case of the stable minimum at $\kappa=0$ (table~\ref{tab:params2}). More importantly, we observe that the transition temperature decreases as the characteristic length of the loop increases, or equivalently, parameter $\kappa_1$ decreases. This means that varying the latter parameter, one can make straight metastable structures less stable. With a specific choice of the conversion factor, we obtained that for the loop length of order $20$~\AA (maximum on the left part of figure~\ref{fig:RHelDistrib}), the straight structures (that is our approximation of the beta strands) become disfavored close to room or body temperature. Also, this version of the model predicts that quantum transitions are strongly suppressed.

\begin{table}[htb]
	
	\begin{center}
		\begin{tabular}{||c|c|c|c|c|c|c|c|c||}
			\hline 
			PDB code & helix 1 & helix 2 & $L_{\rm loop}$, \AA  & $\kappa_{1}$, \AA$^{-1}$ & $\lambda$ & $E_{\rm sph}$, \AA$^{-1}$ & $S_{\rm E}$ & $T$, K\\ 
			\hline
			1a6m & $4$ & $5$ & $22.8$  & 0.275 & 9.7$\times 10^{-3}$ & 0.154 & 25 & 462\\
			\hline 
			2p5k & $1$ & $2$ & $19.0$   & 0.355  & 10.0$\times 10^{-3}$ & 0.138 & 15 & 410 \\
			\hline
			1exr & $4$ & $5$ & $27.7$  & 0.193 &  9.6 $\times 10^{-3}$ & 0.167 & 51 & 500\\ 
			\hline 
			3a02 & $1$ & $2$ & $19.0$  & 0.355 & 10.0 $\times 10^{-3}$ & 0.138 & 15 & 410\\ 
			\hline 
			1g66 & $4$ & $5$ & $43.0$  & 0.059 & 9.4 $\times 10^{-3}$ & 0.181 & $>$88 & 540\\
			\hline 
			5o41 & $1$ & $2$ & $11.0$  & 0.623 & 11.3 $\times 10^{-3}$ & 0.070 & 3.8 & 210\\
			\hline 
		\end{tabular}
\end{center}
	\caption{The estimated values of parameters $\kappa_1$ and $\lambda$ in the model with potential~(\ref{regpot}) based on a few measured lengths of the loops connecting alpha helices and numerical estimates assuming 50\% of the energy. The values are used to obtain the estimates for $E_{\rm sph}$ and $S_{\rm E}$. For the latter, the numerical analysis is complicated for small $\kappa_1$, so we only get an lower bound estimate in case of \emph{1g66}.}
	\label{tab:params4}
\end{table}

One can also discuss the existence of longer loops, violating the upper bound~(\ref{Rmax}). Such loops, should not be considered as kinks of this model, but rather pieces of metastable configurations with $\kappa=0$. If one associates them to beta strands, then it is possible that  strands are connected with each other by sphalerons, stabilized by finite size effects. The characteristic size of a sphaleron is controlled by equation~(\ref{SphaleronSize2}), whose minimum value is $R_{\rm sph}\simeq 12$~\AA, when $\kappa_1=0$.  Such sphalerons (shown on figure~\ref{fig:Regsphaleron}) are curved motifs connecting low-curvature pieces (straight lines). One common motif of this type in proteins is called beta-hairpin.

\section{Discussion and outlook}
\label{sec:conclusions}

In this paper we described the one-dimensional Abelian Chern-Simons-Higgs model as a possible basic phenomenological model of the secondary structure of proteins. In general, this model provides an effective Hamiltonian for a 3D curve parametrized by curvature $\kappa$ and torsion $\tau$. We concluded that the minimal version of that model with possible phenomenological interest must contain at least four parameters in the effective Hamiltonian. The Chern-Simons term plays an important role, introducing chiral helical configurations. Due to its presence there is another hidden parameter in the theory, related to the choice of boundary conditions for the gauge field (torsion) at the endpoints of the protein molecule. The choice of the boundary conditions may significantly affect the properties of the model.

We first considered Neumann boundary conditions, which correspond to treating the theory in the grand canonical ensemble. The role of the chemical potential in this case is played by the helicity (flux) parameter $F$. To avoid the local divergence of the potential energy a regulator $\epsilon$ has been introduced through a Proca mass term for the gauge field. The grand canonical picture appears less appealing. Its basic phenomenology consists of helical curves, possibly connected by kinks -- regions of small and vanishing curvature interpolating between two helical pieces. The kinks have the interpretation of the structural motifs (loops) connecting basic secondary structures (helices) in proteins.

A more interesting case is that of the Dirichlet boundary conditions, or the canonical ensemble. The potential energy in this case may have two minima: one at zero curvature~$\kappa=0$ and the second one at some $\kappa=\kappa_0$. A straightforward interpretation of the two minima is in terms of the two loci of bond angles on the Ramachandran plots. The one with non-zero curvature corresponds to alpha helix configurations in proteins. The zero-curvature one corresponds to beta strands, because beta strands look more like nearly flat ribbons with a small amount of torsion. However there are some important details. 

First, which of the two minima is local, and which is global, depends on the values of parameters $F$, $\epsilon$, $\kappa_0$ and a relative scale parameter $\lambda$. One possibility is that the global minimum is at $\kappa=0$. Then, as it stands, the model predicts that finite-curvature helices are metastable, while zero-curvature beta strands are the fundamental states of the proteins. The theory also possesses unstable classical configurations, which look like small, but non-vanishing, curvature kinks inserted between pairs of alpha helices. These are the sphalerons describing the critical bubbles in the classical phase transition from the metastable vacuum (helices) to the true vacuum of the theory (strands).

A more appealing scenario is the global minimum at $\kappa=\kappa_0\neq 0$. Such a model would have stable ``alpha helices'' and metastable ``beta strands''. Besides it would have stable loops (kinks) interpolating between pairs of alpha helices and unstable ``hairpins'' interpolating between the beta strands. In the new picture the hairpins would be critical bubbles separating the true (helix) and false (strand) vacua.

An interesting question to answer is whether the unstable sphaleron solutions can become metastable in a more realistic model. Indeed, the instability of sphalerons is tightly connected to translational symmetry. In real proteins the translational symmetry is broken to a discrete subgroup by the discreteness of molecules. It is broken even further by the finiteness of their length. Breaking of translational symmetry would make the zero mode massive, so the question is what happens with the lowest (instability) mode. This question will be addressed elsewhere.

After the qualitative analysis we have also made a quantitative analysis to estimate the phenomenological values of the parameters. We first notice that ACSH model predicts a very special type of helical solutions. Namely, the curvature and torsion of possible helices are not independent. In the regularized theory they are related through equation~(\ref{regtorsion}). We used PDB data for a set of regular helices in a set of selected proteins to check, whether in general there is a constraint on their curvature and torsion. The result is shown on figure~\ref{fig:ktcorrelation}.

It is hard to confirm the relation if only alpha helices are considered, because their distribution is localized around values $\kappa_0\simeq 1.6$ and $\tau_0\simeq 0.15$ in the inverse Angstrom units. However, beta strands can also be fit in this picture as low curvature, high torsion structures, as suggested by comparison of figures~\ref{fig:helices} (right) and~\ref{fig:helix} (right). Indeed, our estimates shows that curvature and torsion of the beta strand ribbons are distributed over a relatively small region in the $\kappa$-$\tau$ diagram~\ref{fig:ktcorrelation}. The loci of the alpha helices and beta strands are compatible with equation~(\ref{regtorsion}) and confirm the necessity of regularization parameter $\epsilon$. It is then possible to extract universal values of parameters $F\simeq 0.70$~\AA$^{-1}$ and $\epsilon\simeq 1.5$~\AA$^{-1}$.

We also find an example with a non-universal behavior. It is provided by protein~\emph{2pne}. This protein contains stretched helices which are classified as beta strands. Unlike the alpha helices studied in our analysis, the ``beta helices" of \emph{2pne} have left chirality. In terms of effective theories and renormalization group, this protein belongs to a different ``universality class".

Having fixed parameters $\kappa_0$, $F$ and $\epsilon$, there is a remaining parameter of the model, the relative scale  (or coupling constant) $\lambda$. This parameter, or more precisely $\sqrt{\lambda}$, shows how much the size of a soliton (a kink, or a sphaleron) is different from the scale set by $\kappa_0$. We assumed that the soliton size is of the order of a characteristic small loop connecting two alpha helices. By looking at different proteins we observed a variation of sizes of loops, which implies an ensemble of values of $\lambda$. It turns out that $\lambda$ itself is not a very convenient parameter, because its variation is rather small, compared to its average, which is another universal value in our model,  approximately $\lambda\simeq 0.01$. 

It is more convenient to discuss the properties of the model in terms of a related ``protein modulus" $\kappa_1$ (see \emph{e.g.}  definition~\ref{regpot}), which is a known function of $\lambda$ and the remaining parameters. In fact, varying $\kappa_1$ allows to cover all the discussed physical situations in the ACSH model. When $\kappa_1\geq\epsilon>0$, one is in the grand canonical ensemble, which we ruled less interesting, since it only has one energy minimum. When $\kappa_1$ is imaginary, one is in the situation of the global minimum at $\kappa=0$. This case does not have stable soliton solutions. Hoping that the unstable solitons might stabilize in a more realistic setup, we still find that this case sets a minimum size on the loops connecting helices, which is violated by the observations. So this situation can also be disregarded.

The most interesting and viable scenario is $0\leq \kappa_1 <\epsilon$. When $\kappa_1$ is sufficiently large $0.9~{\text{\AA}}^{-1}\leq \kappa_1\leq \epsilon$ the model only allows a single vacuum. That is there are only alpha helices connected by loops shorter than $18$~\AA. When $\kappa_1$ is moderate, $0.05~{\text{\AA}}^{-1}\leq \kappa_1\leq 0.9~{\text{\AA}}^{-1}$, the theory has two vacua, possible, but metastable, beta-strand-like configurations with $\kappa=0$, and longer loops, $10-40$~{\AA} connecting alpha helices. This resembles the situation of helical proteins like myoglobin. Finally, in the regime $0< \kappa_1\leq 0.05~{\text{\AA}}^{-1}$, the two vacua become nearly degenerate. The loops connecting alpha helices contain long, more than $50$~\AA, nearly linear ($\kappa\simeq 0$) pieces, which might themselves be perceived as beta strands. In fact, in in this situation it becomes harder to distinguish the linear part of stable kinks and pieces of the false vacua $\kappa=0$, which become relatively stable in the $\kappa_1\to 0$ limit. Hence this situation resembles the case of proteins with less obvious, probably short, helical structures and large number (sequences) of nearly straight pieces (strands). Thus $\kappa_1$ can be seen as an external parameter, like the chemical composition, which controls the abundance of the beta strands.

\paragraph{Acknowledgements}  The authors thank Greyson~J.~Coelho for collaboration at the initial stage of this project. DM would like to thank Antti Niemi and Ara Sedrakyan for many useful discussions on the application of effective field theory and topology to protein physics. DM is also grateful to Dionisio Bazeia, Sergei Brazovskii and the participants of the workshop ``Physics and Biology of Proteins'' held at the International Institute of Physics in Natal in June 2017 for interesting ideas and discussions. This work was supported by the grant No.~16-12-10344 of the Russian Science Foundation.

\end{document}